\documentclass[prx,superscriptaddress,twocolumn,aps,showpacs,amsmath,amssymb,floatfix,longbibliography]{revtex4-1}
\usepackage{comment}
\usepackage{bm}
\usepackage{hyperref}
\usepackage{graphics}
\usepackage{graphicx}
\usepackage{braket}
\usepackage{MnSymbol}
\usepackage{amssymb}
\hypersetup{colorlinks=true}
\usepackage{amsmath,empheq}
\usepackage{mathrsfs}
\usepackage{textgreek}
\usepackage{xcolor}
\usepackage{enumitem}
\usepackage{dcolumn}
\usepackage[normalem]{ulem}
\usepackage{amsfonts}
\usepackage[utf8]{inputenc}
\hypersetup{colorlinks,linkcolor=blue,urlcolor=blue,citecolor=blue}

\usepackage[export]{adjustbox} 

\DeclareUnicodeCharacter{2124}{$\mathbb{Z}$}

\newcommand{\uptau}{\boldsymbol{\tau}}
\newcommand{\sign}{\text{sign}}
\renewcommand{\v}[1]{\textbf{#1}}
\renewcommand{\H}{\mathcal{H}}
\begin{document}

\title{$\mathbb{Z}_N$ lattice gauge theories with matter fields}
\author{Kaustubh Roy}
\affiliation{Indian Institute of Science, CV Raman Road, Bengaluru 560012, India}
\affiliation{Max-Planck-Institut f\"{u}r Festk\"{o}rperforschung, Heisenbergstra{\ss}e 1, 70569 Stuttgart, Germany}
\author{Elio J. K\"{o}nig}
\affiliation{Max-Planck-Institut f\"{u}r Festk\"{o}rperforschung, Heisenbergstra{\ss}e 1, 70569 Stuttgart, Germany}

\date{\today}

\begin{abstract}
Motivated by the exotic phenomenology of certain quantum materials and recent advances in programmable quantum emulators, we here study fermions and bosons in $\mathbb Z_N$ lattice gauge theories. We introduce a family of exactly soluble models, and characterize their orthogonal (semi-)metallic ground states, the excitation spectrum, and the correlation functions. We further study integrability breaking perturbations using an appropriately derived set of Feynman diagrammatic rules and borrowing physics associated to Anderson's orthogonality catastrophe. In the context of the ground states, we revisit Luttinger's theorem following Oshikawa's flux insertion argument and furthermore demonstrate the existence of a Luttinger surface of zeros in the fermionic Green's function. Upon inclusion of perturbations, we address the transition from the orthogonal metal to the normal state by condensation of certain excitations in the gauge sectors, so-called ``$e$-particles''. We furthermore discuss the effect of dynamics in the dual ``$m$-particle'' excitations, which ultimately leads to the formation of charge-neutral hadronic $N$-particle bound states. We present analytical arguments for the most important phases and estimates for phase boundaries of the model. Specifically, and in sharp distinction to quasi-1D $\mathbb Z_N$ lattice gauge theories, renormalization group arguments imply that the phase diagram does not include an emergent deconfining $U(1)$ phase. Therefore, in regards to lattice QED problems, $\mathbb Z_N$ quantum emulators with $N<\infty$ can at best be used for approximate solutions at intermediate length scales.
\end{abstract}

\maketitle

\section{\label{sec:intro}Introduction}

Lattice gauge theories (LGTs) have enjoyed constant interest throughout the decades. Originally introduced in the context of statistical physics~\cite{Wegner1971}, they have since been of extraordinary relevance to understanding the strong coupling limit of the standard model of elementary particles~\cite{Wilson1974, Kogut1979, BrinoYoung2018} and to exotic phases of quantum materials~\cite{Sachdev2018}, including quantum spin liquids~\cite{SavaryBalents2016}. While theoretical activity is motivated by the fundamental connection of gauge theories to states with topological order~\cite{Wen1990} and to topological quantum computing~\cite{Kitaev2003}, very recent interest~\cite{CobaneraHassler2016,ZoharCirac2017,ZoharCirac2017b,EmontsZohar2020,CardarelliSantos2020,RobainaCirac2021} was boosted by the (prospect of the) realization of lattice gauge theories in quantum emulators~\cite{ZoharReznik2016,ErcolessiPepe2018}, in particular in the context of Rydberg atoms~\cite{CeliZoller2020,NotarnicolaNontagero2020,VerresenVishwanath2021,GiudiceGiudici2022} and arrays of superconducting qubits~\cite{DalmonteMontangero2016,BanulsZoller2020,SatzingerRoushan2021} -- including proposed implementations of discrete $\mathbb Z_N$ gauge theories with $N >2$.

From the perspective of quantum field theory, $\mathbb Z_N$ gauge theories coupled to fermionic matter are of considerable interest as toy models of hadron formation -- for example a finite string tension in $\mathbb Z_3$ gauge theories is expected~\cite{Pisarski2021} to favor baryonic gauge neutral fermionic trions, while in $\mathbb Z_2$ gauge theories meson-like (Cooper-) pairs are formed~\cite{GazitVishwanath2017}. While even the physics of non-Abelian gauge theories may dynamically emerge~\cite{GazitWang2018}, it is of interest to consider discrete $\mathbb Z_N$ LGTs with $N \gg 1$ as a proxy of $U(1)$ LGT~\cite{HaaseDellantonio2021}, which in contrast to quantum electrodynamics (QED) benefit from a finite local Hilbert space dimension which is more suitable to both classical numerical simulations and analog quantum  emulations.

From the statistical mechanics viewpoint, pure $\mathbb Z_N$ lattice gauge theories in $D = 2 + 1$ are obtained from $\mathbb Z_N$ quantum clock models by Kramers-Wannier duality transformation~\cite{Wegner1971, EinhornRabinovici1980}. This is in contrast to the case of $D = 1 + 1$ space-time dimensions, in which the quantum clock model maps back onto itself.  The differences between $D = 1+1$ and $D = 2 + 1$ persist to the nature of the (quantum) phase transition between ordered and disordered phases: While for quantum clock models for $N >4$ and in $D = 1 + 1$, the direct quantum phase transition is replaced by two Berezinskii-Kosterlitz-Thouless like phase transitions enclosing an intermediary gapless $U(1)$ phase~\cite{JoseNelson1977}, for the quantum clock model in $D = 2+1$, the quantum phase transition is direct, yet of $U(1)$ type~\cite{PatilSadvik2021, footnoteFirstOrder}. What, however, if the $\mathbb Z_N$ gauge theory in $D = 2+1$  is coupled to fermionic matter fields~\cite{AltshulerMillis1994,KimLee1994,HermeleWen2004,NaveLee2007,Lee2009,MrossSenthil2010}? Is an intermediate $U(1)$ phase emerging in the same way as in ladder systems~\cite{MagnificoBermudez2019,NyhegnBurello2021,PradhanErcolessi2022}, where $\mathbb Z_N$ symmetry breaking perturbations can be RG irrelevant? An affirmative answer would clearly facilitate numerical studies of QED$_3$. 

Finally, from the perspective of quantum materials, $\mathbb Z_N$ gauge theories with $N>2$ have been very recently proposed to explain strange and bad metal behavior~\cite{XuXu2022}. Quantum spin-liquid states with $\mathbb Z_N$ gauge group have also been studied over the years~\cite{Vaezi2014,BarkeshliQi2015,MaciejkoZhang2012}. In general, fractionalization approaches to correlated systems, including slave rotors~\cite{FlorensGeorges2004} which are related to the $N \rightarrow \infty$ limit of our theory, have been successfully employed (see, e.g.,~\cite{Coleman1984,KotliarRuckenstein1986,SachdevRead1991,YuSi2012,BonettiMetzner2022}). Of relevance for the present work are studies of $\mathbb Z_2$ fractionalized metals~\cite{GuWen2014,BorlaMoroz2022,EmontsMoroz2023} and of orthogonal (semi-)metals in two spatial dimensions~\cite{NandkishoreSenthil2012,ZhongLuo2013,GazitVishwanath2017,GazitWang2018,HohenadlerAssaad2019}. These states of matter can appear in $\mathbb Z_2$ slave spin constructions~\cite{RueggSigrist2010} and display a gap in the fermionic Green's function -- as if they were an insulator -- but sustain gapless 2-body correlators -- as if they were a metal.
Fascinatingly, the deconfining phases of $\mathbb Z_2$ LGTs allow to circumvent~\cite{ParamekantiVishwanath2004,ScheurerSachdev2018,KoenigTsvelik2020} the Luttinger-Oshikawa theorem~\cite{Luttinger1960,Oshikawa_2000}, i.e. the paradigm that in the absence of symmetry breaking the volume of the Fermi sea is determined by the fermion density alone, independently of the strength of interactions. Experimentally, this paradigm is challenged by the Fermi surface reconstruction in the cuprates including the putative direct transition into the normal Fermi liquid phase as observed in magnetotransport~\cite{ProustTaillefer2019}. A similar scenario of Fermi surface reconstruction in heavy-fermion materials~\cite{SiSteglich2010} is corroborated by quantum oscillation experiments. Another remarkable conundrum is the observation of quantum oscillations in electrical insulators, most notably YbAl$_{12}$~\cite{XiangLi2018} but potentially also in SmB$_6$~\cite{LiLi2014,TanSebastian2015} and RuCl$_3$~\cite{CzajkaOng2021}, which suggest the presence of a ``hidden'' Fermi surface. More precisely than noted above, Luttinger's theorem identifies the density to the volume where the fermionic Green's function is positive, which is enclosed by the area of poles \textit{or zeros} of the Green's function. Therefore, in addition to deconfining gauge theories, Luttinger's theorem may also be non-trivially satisfied in systems which stabilize a Luttinger surface of Green's function zeros~\cite{Dzyaloshinskii2003}. This avenue, which has attracted substantial renewed attention recently\cite{Fabrizio2023,WagnerSangiovanni2023,SettySi2023,BlasonFabrizio2023}, provokes the question if and when the LGT scenario and the scenario of Luttinger surfaces are two distinct descriptions of the same state of matter.  

Here we study the $\mathbb Z_N$ toric code~\cite{Kitaev2003,Wen2003,BullockBrennen2007,SchulzSchmidt2012,ZouHaah2016,PachosBook,WatanabeFuji2023} coupled to fermionic matter, which realizes a $\mathbb Z_N$ LGT in $D = 2 + 1$ space-time dimensions including both bosonic and fermionic matter fields. We argue that such a model may describe strongly coupled quantum materials in which a slave-clock fractionalization scheme is accurate and summarize the duality mapping between our model and $\mathbb Z_N$ LGTs with both fermionic and bosonic matter fields. Starting from the integrable limit describing the deconfining $\mathbb Z_N$ orthogonal (semi)~metallic~\cite{NandkishoreSenthil2012} phases we revisit Luttinger's theorem for arbitrary $N$ and the question of Fermi surface reconstruction without symmetry breaking. Upon inclusion of small perturbations, we demonstrate that $\mathbb Z_N$ orthogonal metals feature a Luttinger surface of Green's function zeros. We furthermore develop a diagrammatic perturbation theory allowing to resum perturbations and to determine the phases~\cite{FradkinShenker1979} of the $\mathbb Z_N$ LGT, their observable signatures, and the most important phase boundaries. We also present renormalization group calculations which imply that a putative deconfining intermediate phase of a $U(1)$ LGT with a Fermi surface is unstable towards $\mathbb Z_N$ perturbations and thus absent in the phase diagram. Instead, we find the formation of hadrons (charge neutral $N$ particle states) near the transition.

\begin{figure}[t]
    \centering
    \includegraphics[width = 0.6 \linewidth]{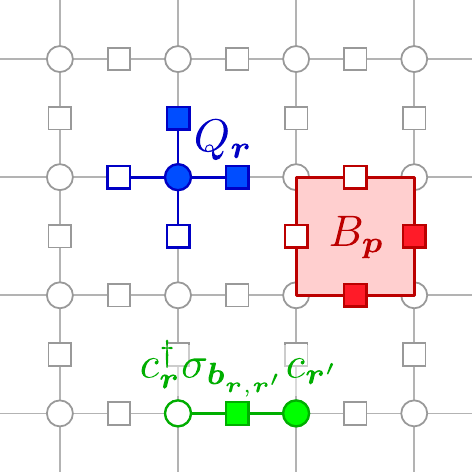}
    \caption{The $\mathbb Z_N$ toric code, with the star, plaquette and fermionic terms.}
    \label{fig:model}
\end{figure}

The remainder of the paper is structured as follows: In Sec.~\ref{sec:model} we introduce the model 
{in its integrable limit and motivate a $\mathbb{Z}_N$ fractionalization scheme. In Sec. \ref{sec:GS_Excitations} we find the ground state and excitations in the soluble limit. Sec. \ref{sec:LO} lays the modification to Luttinger's theorem obeyed by this ground state. In Sec. \ref{sec:Correlators_diagrammatics}, we develop Feynman rules and diagrammatics for the model. This allows perturbative addition of terms that destabilize the ground state and give rise to new phases, which are investigated in Sec. \ref{sec:Perturbations}. In Sec. \ref{sec:RG}, {we present} renormalization group {arguments for the absence} of a $U(1)$ phase. {We conclude with a summary and outlook and delegate important technical details to the appendices.}}

\section{Model and motivation}
\label{sec:model}

In this section, the integrable limit of the model is introduced  and motivated by a $\mathbb Z_3$ fractionalization of generalized Hubbard models. Integrability breaking perturbations are discussed later, in Sec.~\ref{sec:Perturbations}.

\subsection{Integrable limit}

In its integrable limit, the model under consideration is described by a generalization of Kitaev's toric code with mobile fermions on the lattice,
$
    \H = \H_K + \H_h + \H_w,
$
with
\begin{subequations} \label{eq:model}
    \begin{align}
        \H_K &= -\frac{K}{2}\sum_{\v p}e^{-i\phi}B_{\v p} + \text{H.c.}, \label{eq:HK}\\
       \H_h &= -\frac{h}{2}\sum_{\v r} {e^{- i \theta}} Q_{\v r} + \text{H.c.},\\
        \H_w &= -w\sum_{\v r}\sum_{\v e \in\{\hat{x},\hat{y}\}}c^{\dagger}_{r, \alpha}\sigma_{\v r+\frac{1}{2}\v e}c_{r+\v e, \alpha} + \text{H.c.}
        -\mu\sum_{\v r}c^{\dagger}_{\v r, \alpha}c_{\v r, \alpha}. \label{eq:Hw}
    \end{align}
\end{subequations}
Here and in what follows, $\v r$, $\v b$ and $\v p$ denote the positions of vertices, bonds (or edges) and plaquettes (or faces) of the square lattice, and $c_{\v r}$ are the fermionic annihilation operators in position space. In our notation, $K, h$ and $w$ are real positive quantities and $\theta, \phi$ are real angles. The quantities $B_{\v p}$ and $Q_{\v r}$ are the plaquette (``flux") and star (``charge") operators defined on the lattice as (see Fig.~\ref{fig:model})
\begin{subequations}
    \begin{align}
        B_{\v p} = \prod_{\v e \in\{\hat{x},-\hat{y}\}}&\sigma_{\v p + \frac{1}{2}\v e}\sigma^{\dagger}_{\v p - \frac{1}{2}\v e},\\
        Q_{\v r} = \omega^{\hat{\v n}_{\v r}} \prod_{\v e \in\{\hat{x},\hat{y}\}}&\tau_{\v r + \frac{1}{2}\v e}\tau^{\dagger}_{\v r - \frac{1}{2}\v e}.
    \end{align}
\end{subequations}
Here, $\hat n_{\v r} = c^\dagger_{\v r, \alpha}c_{\v r, \alpha}$ is the fermionic density. The operators $\sigma_{\v b}, \tau_{\,\v b'}$ represent the $\mathbb{Z}_N$ clock variables at the bonds, satisfying the algebra 
\begin{equation}
    \sigma_{\v b} \tau_{\,\v b'} = (\omega)^{\delta_{\v b,\v b'}}{\tau}_{\, \v b'} {\sigma}_{\v b},   \label{eq:ZNAlgebra}
\end{equation}
where $\omega \equiv \exp(\frac{2\pi i}{N})$ is the N-th root of unity.  
Fermions may or may not have an internal quantum number, e.g. a spin or flavor, $\alpha = 1, \dots, N_f$ (where Einstein summation is implied throughout the manuscript).

One may use a series of well-known steps of gauge fixing to demonstrate that Eq.~\eqref{eq:model} is a $\mathbb Z_N$ gauge theory containing both fermionic and bosonic matter fields (see App. A for details). In the limit $h \rightarrow \infty$, $\theta = 0$, this model becomes a theory containing only gauge fields and fermions, complemented with a local Gauss' law $Q_{\v r} = 1$ (no background $\mathbb{Z}_N$ charge).

\subsection{Motivation: $\mathbb{Z}_3$ fractionalization of the Hubbard model}
\label{sec:Motivation}

To heuristically motivate the $\mathbb Z_N$ gauge theory with fermions beyond the literature covered in the introduction, we here review the basic aspects of a $\mathbb Z_3$ slave spin fractionalization of a Hubbard-like single band model
\begin{equation}
    \H = -t\sum_{\langle \v r,\v r'\rangle}c^{\dagger}_{\v r,\alpha}c_{\v r'\!,\alpha} + \sum_{\v r} \frac{U}{2} n_{\v r}(n_{\v r}-1) - \mu n_{\v r} + \dots .
\end{equation}

The ellipses denote unknown additional terms which stabilize the fractionalized phase.

The local Coulomb repulsion splits the on-site energy into three distinct levels, with energies 0, $\mu$ and $U-2\mu$ and filling $n = 0, 1,$ and $2$, respectively. The $\mathbb Z_3$ slave spins act in the space of these three different charge states. This is similar to the well-established $\mathbb Z_2$ slave-spin approach~\cite{RueggSigrist2010}, which however is applicable only to the special case $\mu = U/2$ in which the two levels with even fermion parity become degenerate. Generalizing the established approach leads to a $\mathbb{Z}_3$ fractionalization of the physical fermion operator~\cite{NandkishoreSenthil2012} 
\begin{equation}
    c_{\v r,\beta}^{\dagger} = f_{\v r,\beta}^{\dagger} \varsigma_{\v r}.
\end{equation}
The $f_{\beta}$ carries the fermion charge and spin, while the $\varsigma$ operator acts like a raising operator on the fermion occupancy basis $\ket{n=0,1,2}$. The fractionalization has an emergent $\mathbb{Z}_3$ local symmetry  ${f}_{\v r, \beta}\rightarrow { e^{i \alpha_{\v r}} f}_{\v r, \beta}, {\varsigma}_{\v r}\rightarrow {e^{i \alpha_{\v r}} \varsigma}_{\v r}$ {with $\alpha_{\v r} \in \{0, 2\pi/3, -2\pi/3\}$. It is }generated by the unitary operator $Q_{\v r} \equiv \omega^{-\hat{n}_{\v r}}{\uptau}_r$, where $\omega \equiv e^{\frac{2\pi i}{3}}$ is the third root of unity and
\begin{equation}
    \varsigma_{\v r} = 
    \begin{pmatrix}
    0 &1 &0\\
    0 &0 &1\\
    1 &0 &0\\
    \end{pmatrix},\quad
   \uptau_{\v r} = 
    \begin{pmatrix}
    1 &0 &0\\
    0 &\omega &0\\
    0 &0 &\omega^2\\
    \end{pmatrix}.
\end{equation}
{The $\varsigma, \uptau$ matrices suffice the same clock algebra, Eq.~\eqref{eq:ZNAlgebra} as $\sigma, \tau$.} In the process of fractionalization, we have artificially enlarged the Hilbert space. To return to the physical subspace, we impose the constraint
\begin{equation}
    Q_{\v r} = 1.
\end{equation}

In this gauge sector, {and up to a constant}, the Hamiltonian takes the following form
\begin{align}
    \H  &= -t\sum_{\langle \v r, \v r' \rangle} f^{\dagger}_{\v r,\alpha} \varsigma_{\v r} \varsigma^{\dagger}_{\v r'}f_{\v r',\alpha} - \frac{h}{2}\sum_{\v r} (
    e^{- i \theta}\uptau_{\v r} + H.c.)+\dots ,
\end{align}
where $ he^{- i \theta} = \left [U-3 \mu -i \sqrt{3} (U-\mu )\right]/3$. Thus, for empty sites ($\mu \leq 0$), $\theta\in [-\pi/3,\pi/3]$,  for single occupation ($0\leq \mu\leq U$), $\theta \in [\pi/3, \pi]$ and for double occupation ($\mu > U$), $\theta \in [\pi, 5\pi/3]$.

At this point, the hopping term became an interaction term. Upon integration of high-energy modes in a renormalization group procedure, new operators may emerge. These include hopping of fermions 
\begin{equation}
    \H_w = -\frac{w}{2} \sum_{\v r, \v e \in \{\hat{x},\hat{y}\}} \bar{\sigma}_{\left(\v r + \frac{1}{2} \v e\right)} f^{\dagger}_{\v r,\alpha} f_{\v r + \v e,\alpha} + H.c.
\end{equation}
    and bosons (clock degrees of freedom)
\begin{equation}
        \H_J = -\frac{J}{2} \sum_{\v r, \v e \in \{\hat{x},\hat{y}\}} \bar{\sigma}_{\left(\v r + \frac{1}{2} \v e\right)}\varsigma^{\dagger}_{\v r}\varsigma_{\v r + \v e}.    
\end{equation}

Note that by gauge symmetry, the new coupling constants obtained by contraction of the type $t\langle \varsigma_{\v r} \varsigma^\dagger_{\v r'} \rangle_{\rm fast} = w \bar \sigma_{\v b_{\v r, \v r'}}/2$ retain a non-trivial transformation under gauge symmetry. To account for this, we introduce the emergent link degree of freedom (gauge potential) with non-trivial transformation $\bar \sigma_{\v b_{\v r, \v r'}} \rightarrow e^{i \alpha_{\v r}} \bar\sigma_{\v b_{\v r, \v r'}}e^{- i \alpha_{\v r'}}$ under $\mathbb Z_3$ gauge symmetry. In subsequent RG steps, terms accounting for the dynamics of gauge field variables emerge $\bar \sigma_{\v b}$ (both electric and magnetic terms).

In Appendix \ref{app:ClockToTC} we provide details on the mapping~\cite{Wegner1971, TupitsynStamp2010, KoenigTsvelik2020} of  the $\mathbb Z_N$ gauge theory of $f_{\v r}, \varsigma_{\v r}, \bar \sigma_{\v b}$ variables to Eq.~\eqref{eq:model} with additional perturbations. In particular, the Gauss law as well as appropriate gauge fixing allows to remove $\varsigma_{\v r}$ degrees of freedom from the Hamiltonian (1), which thereby takes the form of a toric code with integrability breaking extra-terms. Depending on the unspecified terms in the model indicated by $\dots$, an effective low-energy deconfining state may be stabilized. While the nature of such stabilizing unknown terms is beyond the scope of the analysis in this paper, the present goal is to understand the physics of all such models which lie in the basin of attraction of the deconfining phase, i.e. of the ground state of Hamiltonian, Eq.~\eqref{eq:model} discussed in the next section.

\section{Ground states and Excitations \label{sec:GS_Excitations}}

In this section, we determine the ground state of the model and the leading excitations in the model.

 As in the usual toric code, the plaquette and star operators mutually commute ensuring the integrability of the model. Moreover, they also commute with the fermionic term $\H_w$.

\subsection{Ground State}

In this paper, we will work in the asymptotic limit where the energy contribution of the fermionic part is negligible. We provide a mathematically concrete version of this qualitative statement at the end of the subsequent section, Sec.~\ref{sec:Excitations}.

\begin{figure}[t]
    \includegraphics[scale=1]{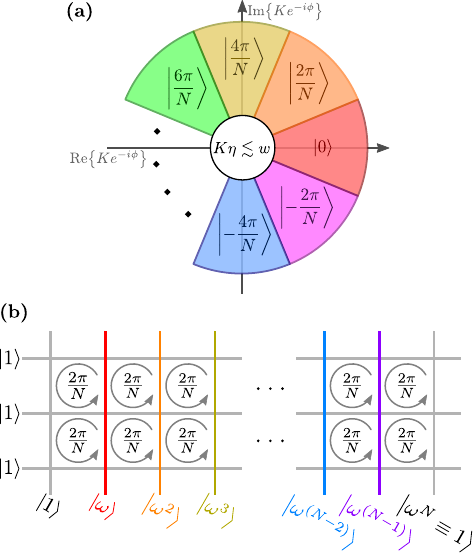}
    \caption{\textbf{(a)} Phase diagram illustrating the ground state of the system $\ket{\varphi}$ as a function of the phase of $K e^{-i\phi}$. {A similar phase diagram occurs in the complex plane of $h e^{-i \theta}$.} \textbf{(b)} Landau gauge choice for the system with $2\pi/N$ flux. The entry inside kets indicates the eigenvalue of clock operators, $\sigma_{\v b} \ket {\omega^n} =\omega^n  \ket {\omega^n }$, on the corresponding links. 
    }
    \label{fig:landau}
\end{figure}

Since they commute with the Hamiltonian, each $B_{\v p}$, $Q_{\v r}$ individually encodes an integral of motion, with eigenvalues 1, $\omega$, $\omega^2$, \dots, $\omega^{N-1}$. The ground state has equal flux through all plaquettes $B_{\v p}\ket{GS} = e^{i\varphi} \ket{GS}$ with $\varphi = \frac{2\pi n}{N}$ and equal charge on all sites $Q_{\v r}\ket{GS} = e^{i\vartheta} \ket{GS}$ with $\vartheta = \frac{2\pi n'}{N}$ 
and $ n, n' \in \{0,1,\dots,N-1\}$. The exact ground state flux/charge is determined by minimizing the phase differences $\phi - \varphi, \; \theta - \vartheta$ (on the unit circle, see Fig.~\ref{fig:landau} \textbf{(a)}). 

To construct the ground state, we first set $h=0$. We then choose a gauge configuration that satisfies the plaquette terms. For our analysis the ``Landau" gauge is utilized, Fig. \ref{fig:landau} \textbf{(b)}, {which corresponds to a wave function}
$ \ket{\varphi}_\sigma = \ket{{\sigma_{\v b, \text{horizontal}} = 1},\sigma_{\v b,{\text{vertical}}}= \omega^{n b_x}}.$
To satisfy the Hamiltonian for $h>0$, we project the constructed state onto the subspace where $Q_{\v r}= e^{i \vartheta}$ is satisfied ($P^{{(\vartheta)}}_{\v r} \equiv \frac{1}{N}\sum_{j=0}^{N-1} [Q_{\v r}e^{- i \vartheta}]^j$)

\begin{equation}
    \ket{GS} = \left[\prod_{\v r} P^{{(\vartheta)}}_{\v r}\right]\ket{\varphi}_\sigma \ket{FS}_c.
\end{equation}

\begin{figure}
    \centering
        \includegraphics[scale=1]{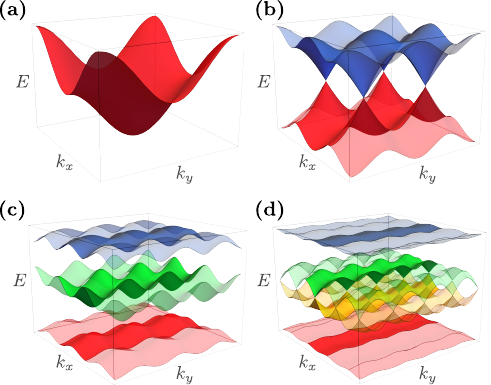}
    \caption{Band structures for different phases in the soluble limit of the model. The range of the plot is the full Brillouin zone, $BZ: -\pi \leq k_x, k_y < \pi$, with the darker section indicating the reduced Brillouin zone, $rBZ = BZ \cap \{k_x \in [-\pi/q, \pi/q)\}$ for the flux $\varphi = 2\pi p/q$. \textbf{(a)}: $\varphi=0$, \textbf{(b)}: $\varphi=\pi$, \textbf{(c)}: $\varphi=2\pi/3$, \textbf{(d)}: $\varphi=\pi/2$.} 
    \label{fig:bands}
\end{figure}

Here, $\ket{FS}_c$ represents the fermionic ground state for the two-dimensional {gauge-fixed} lattice {with Hamiltonian $\H_{\varphi} = \bra{\varphi}\H_w\ket{\varphi}$}. The projection operator in effect creates a symmetric combination of all allowed gauge configuration, also referred to as a ``quantum loop gas". 
The band structure in the fermionic sector consists of up to $N$ distinct bands (see Fig.~\ref{fig:bands}).
If $N$ is even, and depending on the value of the flux $\varphi$, bands may touch at half-filling, forming Dirac cones with a linear band dispersion. 
 It is important to stress that the ground state described above does not break any crystalline translation and rotation symmetries. 

\subsection{Excitations}\label{sec:Excitations}

The single-particle fermionic excitations of the system can be obtained by inserting the fermionic operators to the \textit{right} of the projectors while constructing the ground state.

\begin{subequations}
\begin{align}
    \ket{e;\v{k},i} &= \left[\prod_{\v r} P^{(\vartheta)}_{\v r}\right]c^{\dagger}_{\v k,i}\ket{\varphi}_\sigma \ket{FS}_c, \quad \v k \notin FS , \\
    \ket{h; \v{k},i} &= \left[\prod_{\v r} P^{(\vartheta)}_{\v r}\right]c_{\v k,i}\ket{\varphi}_\sigma \ket{FS}_c, \quad \v k \in FS .
\end{align}
\end{subequations}

Here, $i$ indicates the band index. Analogously, one can construct multi-particle excitations by applying multiple fermionic operators. If the fermionic operators were instead applied to the left of the projectors, then the excitations would no longer be gapless, since the fermionic operators do not commute with the star operators $Q_{\v r}$ because of the density term $\hat n_{\v r}$. These excitations are therefore gapped. In the simplest cases when $\theta  = \frac{2\pi n}{N}, \; n = 0, \dots N-1$
the associated cost in energy is $\eta h$, where $\eta = (1-\cos(\frac{2\pi}{N}))$. This corresponds to the creation of an ``$e$-particle" on the lattice, which can equivalently be created by electric string operators

\begin{equation}
    \mathcal{W}^{(e)} = \prod_{\v b_{\v r_1, \v r_2} \in \gamma} \sigma_{\v r_1, \v r_2}.
\end{equation}

\begin{figure}
    \centering
    \includegraphics[scale=1]{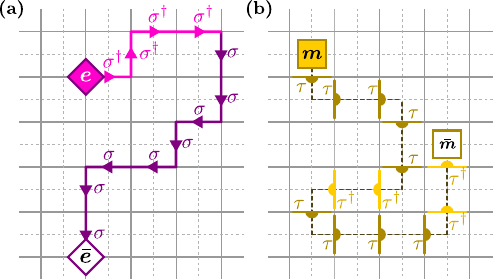}
    \caption{The two gapped excitations on the $\mathbb{Z}_N$ toric code. The direction of the arrow (half-moon) indicates the appropriate choice for $\sigma_{\v r_1, \v r_2} (\tau_{\v p_1, \v p_2})$, pointing from $\v r_1 (\v p_1)$ to $\v r_2 (\v p_2)$. \textbf{(a)}: An $e$ particle-antiparticle pair, joined by a contour on the lattice (solid gray lines), \textbf{(b)}: An $m$ particle-antiparticle pair, joined by a contour on the dual lattice (dotted gray lines).}
    \label{fig:excitations}
\end{figure}

The notation $\sigma_{\v r_1, \v r_2}$ is necessary to specify the correct orientation; it corresponds to $\sigma^{\dagger}$ when $(\v r_2 - \v r_1) = \{\hat{x}, \hat{y}\}$ and $\sigma$ otherwise, see Fig. \ref{fig:excitations} \textbf{(a)}.

A key distinction is that the electric string operator $\mathcal{W}^{(e)}$ produces two $e$-particles (for $N\neq 2$, a particle-antiparticle pair) at each end of the contour $\gamma$ with total excitation energy $2\eta h$, whereas the fermionic operator to the left of the projector creates just one $e$-particle coupled to the electron or hole. 

There is another gapped excitation on top of the toric code ground state, the $m$-particles, which reside on the dual lattice and are joined by $\tau$ insertions, analogous to the electric strings, Fig. \ref{fig:excitations} \textbf{(b)}

\begin{equation}
    \mathcal{W}^{(m)} = \prod_{\v b_{\v p_1, \v p_2} \in \gamma^*} \tau_{\v p_1, \v p_2}.
\end{equation}

Also analogously, $\tau_{\v p_1, \v p_2}$ is oriented, corresponding to $\tau$ when $(\v p_2 - \v p_1) = \{\hat{x}, -\hat{y}\}$ and $\tau^{\dagger}$ otherwise. 

In the regular toric code, both $e$- and $m$-particles excitations are static eigenstates, but in the fermionic toric code the $m$-particles are not static due to their coupling to the Fermi sea. This peculiarity will be explored in more detail in Section \ref{sec:m_particle}, but we anticipate here, that, in the absence of fermions, the energetic cost of $m$ particles in the simplest case $\phi = 2\pi n/N$, where $n = 0, 1, \dots, N-1$, is $\eta K$. The assumption introduced in the previous section, according to which the fermionic contribution to the ground state energy is negligible, is equivalent to $w \ll K \eta$. Contrary to this, in the opposite limit where $0<K \eta \lesssim w$, the ground states do break crystalline symmetries, even when represented projectively (i.e. symmetries modulo gauge transformations). The flux passing through the plaquettes is no longer homogeneous and is dominated by the fermionic sector. The fractional average flux is determined by the filling in the system - for a filling $\rho$, one expects the average flux density~\cite{HasegawaWiegmann1989} to be $\varphi = 2\pi \rho$. This has been corroborated by Monte Carlo simulations~\cite{ProskoMaciejko2017} as well as a semi-analytical study~\cite{KoenigTsvelik2020} in the special case $N=2$. Relaxing the condition $\phi = 2\pi n/N$, we furthermore expect that the gapless fermionic sector influences the phase transitions between different sectors of Fig.~\ref{fig:landau} \textbf{(a)} and leave detailed analyses of this regime for a general $N$ to future studies.

\section{Luttinger-Oshikawa theorem \label{sec:LO}} 

The phase $\phi$ of the plaquette term, Eq.~\eqref{eq:HK} alters the band structure of the fermionic sector and more importantly, changes the volume of the Fermi surface. This appears to violate 
Luttinger's theorem ~\cite{Luttinger1960}, according to which
the volume of the Fermi surface is only dependent on the fermionic filling $\rho$. For simplicity, and without loss of generality, in this section we concentrate on the spinless case $N_f$, in which $\rho$ coincides with the density. 
In its modern formulation due to Oshikawa, a momentum balance argument~\cite{Oshikawa_2000} leads to Luttinger's theorem for conventional Fermi liquids (see more details below). However, a key step in the proof is assuming that momentum can only be carried by the quasiparticles of a conventional Fermi liquid. In our model, and generally in deconfining $\mathbb Z_N$ gauge theories, $m$-particles are able to carry (angular) momentum,
which leads to a modification in Oshikawa's proof~\cite{ParamekantiVishwanath2004}. We here generalize the Oshikawa proof to $\mathbb Z_N$ gauge theories, and illustrate the proof for the simplest $\mathbb{Z}_3$ case in the $\ket{\varphi = 2\pi/3}$ state, Fig.~\ref{fig:bands} \textbf{(c)}. We mention in passing that the generalization of Oshikawa's proof to $SU(N)$ theories, of which $\mathbb {Z}_N$ is the center, was recently reported in Ref.~\cite{HazraColeman2021}.

\begin{figure*}
    \includegraphics[scale=1]{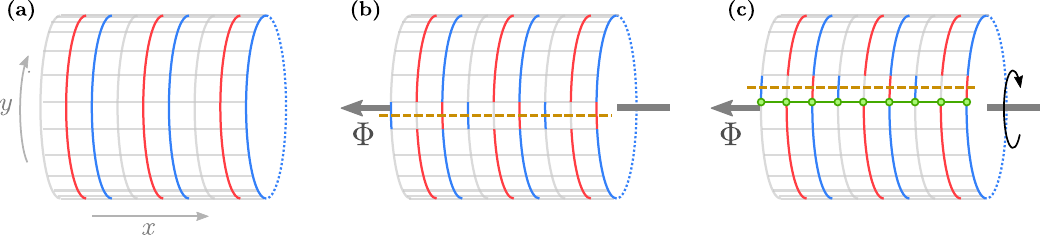}
    \caption{\textbf{(a)}: Modified Oshikawa's argument demonstrated for the $\ket{\varphi = 2\pi/3}$ flux state. \textbf{(b)}: An adiabatic flux $\Phi = 2\pi/3$ through the system absorbed into the gauge field as a vison threading (orange). \textbf{(c)}: The translation operator along the y-axis $T_y$ acting on the cylinder along the contour $\mathcal{C}$ (green).}
    \label{fig:Oshikawa}
\end{figure*}

Consider the model in a cylindrical setup with the specific gauge choice as shown in Fig. \ref{fig:Oshikawa} \textbf{(a)}, at half-filling. We now adiabatically thread a physical flux of $2\pi/3$ through the cylinder. This flux can be absorbed into the internal gauge field
by changing the bonds as shown in Fig. \ref{fig:Oshikawa} \textbf{(b)}. The ground state describing the new lattice is related to the original ground state by a unitary transform

\begin{equation}
    \ket{GS'} = \mathcal{W}^{(m)}\ket{GS},
\end{equation}

where $\mathcal{W}^{(m)}$ is the Wilson loop passing through the plaquettes. 

One can show that the operator $T_{y} = \prod_{r\in \mathcal{C}} \omega^{\hat{n}_{\v r}} Q^\dagger_{\v r}$ translates the lattice by one unit in the y- direction, Fig. \ref{fig:Oshikawa} \textbf{(c)}. Applying this to $\ket{GS'}$ gives

\begin{equation}\label{gauge_m}
\begin{aligned}
    T_y \ket{GS'} &= \prod_{\v r \in \mathcal{C}} \omega^{\hat{n}_{\v r}} Q^\dagger_{\v r} (\mathcal{W}^{(m)}\ket{GS}) \\
    &= \prod_{\v r \in \mathcal{C}} \exp\left(i\frac{2\pi }{3}\hat{n}_{\v r}\right) \mathcal{W}^{(m)} \ket{GS}\\
    &= \exp\left(i\frac{2\pi L_x}{3}\rho\right)  \ket{GS'},
\end{aligned}
\end{equation}

where we have assumed a homogeneous fermion filling $\rho = \langle \hat{n}_{\v r} \rangle_{\v r\in C}$.

Therefore, the momentum carried by the gauge field in this configuration, Eq.~\eqref{gauge_m}, is $\Delta P = -\frac{2\pi L_x}{3}\rho$. This is the exact momentum imparted to the cylinder by the adiabatic flux insertion $\Delta P_{tot} = \Phi L_x \rho$, where $\Phi = -2\pi/3$. The momentum balance is accounted for and does not affect the fermionic distribution.

We now present a generalized relationship between fermion density and size of the Fermi surface. 

Arguments analogous to above hold for a flux insertion of $4\pi/3$, where the gauge sector absorbs the flux with two Wilson loops instead of one ($(\mathcal{W}^{(m)})^2 \ket{GS}$). The gauge field can thus carry away momenta in steps of $2\pi/3$. Adding this condition to the standard momentum balance calculation done before leads to the modified result, Eq.~\eqref{LuttNew}:
\begin{equation}\label{LuttNew}
    \frac{V_{FS}}{(2\pi)^2} \pm \frac{1}{3} = \rho + p, 
\end{equation}

where $p\in \mathbb{Z}$. We see this is true at $\rho=1/2$, when the volume of the Fermi surface is $V_{FS} = (2\pi)^2/6$. Including the case where the flux is not threaded at all and entirely carried by the Fermi quasiparticles, the final relation is (in terms of a general N)

\begin{equation}
    \frac{V_{FS}}{(2\pi)^2} = \rho \left(\text{mod}\ \frac{1}{N}\right) . \label{eq:Luttinger}
\end{equation}

It is of note that we arrived at 
this result without any knowledge of the band structure or details about the Fermi surface.

\section{Correlators and Diagrammatics \label{sec:Correlators_diagrammatics}}

In this section, we first study the correlators of fermions as well as $\sigma$ insertions on the lattice and present a set of Feynman diagrams to graphically encode these correlators (see App. \ref{app:feynman_rules}). Subsequently, we discuss $\tau$ insertions.

\subsection{Feynman rules for fermions and e-particles}
\label{sec:Feynman}

To evaluate an arbitrary imaginary-time-ordered ground state correlator with $c_{\v r}, c^{\dagger}_{\v r}$ and $\sigma_{\v b}$ with the general form 
\begin{equation}
    C(\{\v r,\tau\}) = -\braket{\mathcal{T}\{\prod_n O^{(c)}_{\v r_n}(\tau_n) \prod_m O^{(\sigma)}_{\v r_m, \v r'_m}(\tau_m) \}}_{GS}. \label{eq:Correlator}
\end{equation}
Here, $O^{(c)}_{\v r_n}(\tau_n) \in \{ c_{\v r_n}(\tau_n), c^\dagger_{\v r_n}(\tau_n)\}$ and $ O^{(\sigma)}_{\v r_m, \v r'_m}(\tau_m)$ is defined for nearest-neighbor $\v r_m, \v r_m'$ as
\begin{equation}
       O^{(\sigma)}_{\v r_m, \v r'_m} = 
        \begin{cases}
        \sigma^{\dagger} & \text{for } \v r_m' - \v r_m = \hat{x}, \hat{y},\\
        \sigma & \text{for }\v r_m' -\v r_m = - \hat{x}, -\hat{y}.
        \end{cases}
\end{equation}

To facilitate clear notation, we hereafter consider the simplest case $\theta = 0$.
\begin{enumerate}
    \item Draw $\circ$ for $c_{\v r_n}(\tau_n)$, $\bullet$ for $c_{\v r_n}^{\dagger}(\tau_n)$ and \includegraphics[valign = c]{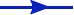} for $O^{(\sigma)}_{\v r_m, \v r'_m}(\tau_m)$
    , pointing from $\v r_m$ to $\v r_m'$. 
    Higher powers of $\sigma$ are indicated by multiple arrows, e.g.: \includegraphics[valign = c]{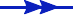} for $(O^{(\sigma)}_{\v r_m, \v r_m'})^2$.
    \item The ``charge" $q_i \in \{1,\dots,n\}$ is defined as the power of $\omega$ obtained from commuting $O_i \in \{O^{(c)}_{\v r}, O^{(\sigma)}_{\v r, \v r'}\}$ across $Q_{\v r}$. 
    \begin{equation}
        Q_{\v r}O_{i} = \omega^{q_i}O_{i}Q_{\v r}.
    \end{equation}
    For $n$ operators $O_1,\dots ,O_n$ at site $\v r$ with times $\tau_1,\dots,\tau_n$ respectively,
    \begin{itemize}
        \item The operators must satisfy charge neutrality modulo $N$, 
        \begin{equation}\label{eq:CN}
            \sum_{i=1}^n q_i = N\mathbb{Z}.
        \end{equation}
        Note that $O^{(\sigma)}_{\v r_1, \v r_2}$ contributes to both sites $\v r_1$ (with charge $+1$) and $\v r_2$(with charge $-1$).
        \item Connect operators on the same site by wavy lines. If there are more than two operators, encircle them. This corresponds to the exponential factor 
        \begin{equation}\label{eq:exp_e-particles}
            \exp\left[h\sum_{i=1}^n \tau_{p(i)} \Re \left\{\omega^{-\sum_{j=1}^i q_{p(j)}}\left(1-\omega^{q_{p(i)}}\right)\right\}\right],
        \end{equation}
        where $p(i): \{1,\dots,n\} \rightarrow \{1,\dots,n\}$ is the index permutation that orders the times appropriately
        \begin{equation}
            \tau_{p(1)}>\tau_{p(2)}>\dots>\tau_{p(n-1)}>\tau_{p(n)} .
        \end{equation}
        For only two operators, this takes the simple form $D(\tau_1,\tau_2) = e^{-\eta h|\tau_1 - \tau_2|}$, where we remind the reader of the excitation energy $\eta h = (1 - \cos(\frac{2\pi}{N}))h$ of an $e$-particle. 
    \end{itemize}
    \item Replace all \includegraphics[valign = c]{Figures/sigma.pdf} with their expectation value $\bra{\varphi} \sigma_{\v r_1, \v r_2} \ket{\varphi}$.
    \item Connect all $\circ, \bullet$ in all possible combinations in accordance with the standard rules for fermionic diagrammatics, with solid lines representing the fermionic Green's function,
    \begin{align*}
        G_{FS}(\v r_1,\v r_2;\tau) &= -\bra{FS} \mathcal{T}\left[\bar{c}_{\v r_1}(\tau)\bar{c}^\dagger_{\v r_2}(0)\right] \ket{FS} \\
        &= -\bra{FS} \mathcal{T}\left[e^{{\H_{\varphi}} \tau}c_{\v r_1}e^{-{\H_{\varphi}} \tau}c^\dagger_{\v r_2}\right] \ket{FS},
    \end{align*}
    where the bar above the operator indicates time evolution with respect to ${\H_{\varphi}}$ only.
\end{enumerate}

\begin{figure}[t]
    \centering
    \includegraphics[scale=1]{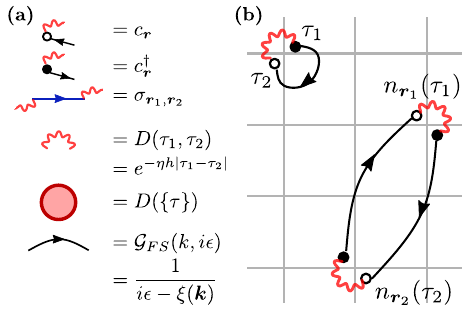}
    \caption{\textbf{(a)}: Feynman rules for calculating correlators. \textbf{(b)}: The onsite fermionic Green's function (top-left) and the polarization operator (right).}
    \label{fig:rules_and_correlators}
\end{figure}

The simplest correlator one can calculate is the two-point Green's function (Fig. \ref{fig:rules_and_correlators} (b)): $\mathcal{G}(\v r_1, \v r_2; \tau) = -\langle \mathcal{T} c^{\dagger}_{\v r_1}(\tau)c_{\v r_2}(0)\rangle = \delta_{\v r_1, \v r_2} e^{-\eta h|\tau|}\mathcal{G}_{FS}(\v r_1, \v r_1; \tau)$. The ultralocal correlator is a consequence of the fact that e-particles do not move in the integrable limit of the toric code. In the frequency representation,

\begin{equation}
    \mathcal{G}(z) = \int (dk) \frac{1}{z - \xi_0(\v k) - \sign[\xi_0(\v k)]\eta h}, \label{eq:SingleParticleGF}
\end{equation}
\vspace{5pt}

{where $\xi_0(\v k) = -2w(\cos(k_x)+\cos(k_y))$ is the dispersion relationship for the zero-flux state and $(dk) = d^2k/(2\pi)^2$}. The $e$-particles (cf. Section \ref{sec:Excitations}) effectively create a gapped density of states, represented by the above equation. However, two-particle correlators in the particle-hole channels display regular Fermi liquid behavior. For example, the polarization operator (Fig.~\ref{fig:rules_and_correlators} \textbf{(b)}) is gapless, since the $e$-particle propagator at each site evaluates to $D(\tau_i - \tau_i) = 1$. This drastically  different behavior of single and multiparticle correlators is characteristic of an orthogonal metallic phase~\cite{NandkishoreSenthil2012}. In contrast to the case $N = 2$, $\mathbb Z_N$ orthogonal metals with $N>2$ have gapped two-particle correlators in the particle-particle channel, effectively preventing a Cooper instability.
\\

\subsection{Insertion of $m$-particles} \label{sec:m_particle}

Unlike the insertion of the $\sigma$ operators which commute with the fermionic part of the Hamiltonian, Eq.~\eqref{eq:Hw}, the $\tau$ operators do not $
    \left[\tau_{\v b}, \mathcal{H}_w\right] \neq 0$. As a consequence, insertion of $\tau$ operators (i.e. insertion of $m$-particles) triggers an Anderson orthogonality catastrophe, which we shall now explore. We mention in passing, that a similar interplay of local spontaneous flux insertions with gapless fermionic modes is of fundamental importance in the physics of $U(1)$ and Kitaev quantum spin liquids~\cite{KnolleMoessner2014} as well as complex Kondo impurity systems~\cite{KoenigKomijani2021}.

We start by finding the $\tau$-$\tau$ correlator corresponding to placing $m$-particles at adjacent sites at time 0 and removing them at time $t$. (We switch to real time here momentarily to avoid confusion between $\tau_{\v b}$, the clock operator, and $\tau$, the imaginary time and to remain consistent with previous literature; we will return to imaginary time at the end).
One obtains an energy shift from the plaquettes $\mathcal{H}_K$, analogous to the $\sigma$ insertions, but one gets an additional term from the fermionic contribution,

\begin{subequations}
\begin{align}
    \mathcal{G}(\v b, t) &= -i\bra{0}\mathcal{T}\left\{\tau^{\dagger}_{\v b}(t)\tau_{\v b}(0)\right\}\ket{0} \notag \\
    &= \underbrace{-i\bra{FS}\mathcal{T}\exp\left\{-i\int_0^{|t|} dt'\, \hat{V}(t') \right\} \ket{FS}}_{\mathcal{G'}_{FS}(\v b; t)} \notag \\
    &\times {\exp\left(-2i\eta K|t|\right)},\\
    \hat{V}(t) &=\frac{w}{2} (1-\omega) c^{\dagger}_{\v r_1}(t) c_{\v r_2}(t) + \text{H.c.} ,\label{eq:HoppingChange}
\end{align}
\end{subequations}
where $\v r_1, \v r_2$ are the sites corresponding to the bond $\v b$. One must note that due to the translational and fourfold rotational symmetry of the 0-flux state, the choice of bond is irrelevant. The propagator $\mathcal{G'}_{FS}$ corresponds to scattering from a time-dependent perturbation $\hat{V}(t)$ of the bond $\v b$. This form is analogous to the X-ray edge problem~\cite{Anderson1967,GogolinTsvelikBook}, where the propagator gains not only an energy shift but also a transient part that results in a power-law decay at large times. Our propagator, however, differs in one key point: namely, the potential is not on-site but between neighboring sites.

Treating the potential perturbatively produces a set of connected and disconnected cycles, which can be summed using the linked cluster expansion~\cite{GogolinTsvelikBook} in two sets of complex vertices ($V\equiv \frac{w}{2}(1-\omega), \bar{V}$) and two types of propagators 
\begin{subequations}
\begin{alignat}{2}
    h_0(\tau_2-\tau_1) &= -i\langle\mathcal{T}c_{\v r_2}(\tau_2)c^{\dagger}_{\v r_1}(\tau_1)\rangle &&= -i\langle\mathcal{T}c_{\v r_1}(\tau_2)c^{\dagger}_{\v r_2}(\tau_1)\rangle,\\
    g_0(\tau_2-\tau_1) &= -i\langle\mathcal{T}c_{\v r_1}(\tau_2)c^{\dagger}_{\v r_1}(\tau_1)\rangle &&= -i\langle\mathcal{T}c_{\v r_2}(\tau_2)c^{\dagger}_{\v r_2}(\tau_1)\rangle.
\end{alignat}
\end{subequations}

\begin{figure}[t]
        \centering
        \includegraphics[scale=1]{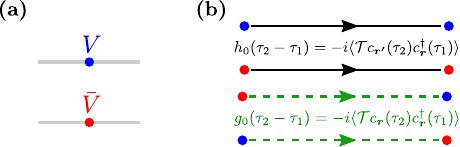}
    \caption{\textbf{(a)}: The two vertices in the perturbative expansion, $V = \frac{w}{2}(1-\omega)$ and $\bar{V} = \frac{w}{2}(1-\bar\omega)$. \textbf{(b)}: The two sets of propagators; $g_0$ joins two different vertices and $h_0$ joins vertices with the same colors.}
\end{figure}

The solution can also be obtained in a nonperturbative fashion, following the method of Nozieres and de Dominicis~\cite{NozieresdeDominicis1969,GogolinTsvelikBook}. 
The details of the calculation are relegated to Appendix \ref{app:nd}. The result for the propagator is

\begin{equation} \label{eq:ndd}
    \Re{\ln \mathcal{G}(t)} = - N_f \underbrace{\left\{\left(\frac{\delta_1}{\pi}\right)^2+\left(\frac{\delta_2}{\pi}\right)^2\right\}}_{\equiv \alpha} \ln(|t|)
\end{equation}
with $$\delta_{i} = \arctan\{\pi[\rho^h_0\Re{V} +(-1)^{i} \sqrt{|V|^2 (\rho_0)^2-(\Im{V}\rho_0^h)^2}]\}.$$ Here, we have returned to the general number $N_f$ of internal degrees of freedom of the fermions, e.g. $N_f = 2$ due to spin.

As one can see, the factors $\delta_1, \delta_2$ depend on the local density of states (DOS), each acquiring a value of $\frac{\pi}{2}$ as the DOS approaches infinity, which occurs in the orthogonal metal phase at half-filling (cf.~\cite{Gogolin1993} for the orthogonality catastrophe at a van Hove singularity). In contrast, the DOS vanishes for half filling in the OSM phase (even $N$) and the orthogonality catastrophe term disappears. One should note that the $\tau$-$\tau$ propagator acquires a phase shift from the energy shift due to the scattering potential, but an estimate of this requires non-analytical methods. However, we can conclude that it is of the order of the bandwidth $w$ since that is the only energy scale in the correlator, and since we work in the regime $K\eta \gg w$ we can ignore this shift safely.

\begin{figure}
    \centering
    \includegraphics{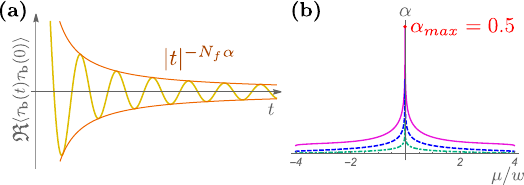}
    \caption{\textbf{(a)}: A schematic of the (real part of the) real-time $\tau$-$\tau$ correlator (yellow), with a power-law envelope (orange). \textbf{(b)}: Dependence of the power-law exponent $\alpha$ on the chemical potential $\mu$ for $N=3$ (purple, solid), $N=5$ (blue, dashed) and $N=7$ (green, dotdashed). At the van Hove singularity, all approach a maximum value of $0.5$.}
    \label{fig:aoc_exponent}
\end{figure}

The real-time propagator thus quickly oscillates on the scale $1/K\eta$ and its envelope slowly decays as $\mathcal{G}(t) \sim |t|^{-N_f \alpha}$, where the $N_f$ accounts for the flavor degeneracy of the fermions. The exponent $\alpha$ depends on the chemical potential in the OM state ($\mu$) as well as $N$, see Fig.~\ref{fig:aoc_exponent} \textbf{(b)}. In imaginary time, the oscillation is replaced by exponential decay, which overshadows the power-law {at scales beyond $(K \eta)^{-1}$.} 

\section{Perturbations and Instabilities \label{sec:Perturbations}}

In this section, we add various perturbations to the soluble limit which introduce a finite string tension and discuss the instabilities of the orthogonal metal phase. In analogy to the previous section, we first discuss the impact of spontaneous $\sigma$ insertions and subsequently of spontaneous $\tau$ insertions.

\subsection{Spontaneous $\sigma$ insertions: kinetics of $e$-particles, Luttinger surface and Higgs transition}
We now introduce a perturbation that delocalizes the $e$-particles (wavy lines), which were previously confined to one vertex. We do this by introducing $\sigma_{\v b}$ terms in the Hamiltonian

\begin{equation}
     \H_J = -\frac{J}{2}\sum_{\v b} (\sigma_{\v b} + \sigma^{\dagger}_{\v b}).
\end{equation}

We first consider the orthogonal metal phase (zero-flux state, $\bra{\varphi = 0}\sigma_{\v b}\ket{\varphi =0}$ = 1) The resummation of non-intersecting strings of \includegraphics[valign = c]{Figures/sigma.pdf} insertions is given by 
\begin{align}
    \v D(\v r_f, \v r_i;i\omega) &= D(i\omega)\delta_{\v r_f, \v r_i}\nonumber\\ &\hspace{-3em}+ \frac{J}{2} \sum_{\langle\v r,  \v r_f\rangle} \langle O^{(\sigma)}_{\v r, \v r_f} \rangle_0 \v D(\v r, \v r_f;i\omega) D(i\omega),
\end{align}
where $D(i\omega)=\mathcal F\{e^{-h|\tau|\eta}\}(\omega) = 2\eta h/(\omega^2 + (\eta h)^2)$ and $\mathcal F$ denotes Fourier transform. The object $\v D(\v r_f, \v r_i; i\omega)$ effectively describes the e-particle propagator.

In momentum space, this implies $\v D(\v q, i\omega) = D(i\omega) + {J}[\cos(q_x)+\cos(q_y)]D(i\omega)\v D(\v q, i\omega)$, giving us
\begin{equation}\label{eq:e_prop}
    \v D(\v q,i\omega) = \frac{2\eta h}{\omega^2 + \eta h(\eta h - {2}J[\cos(q_x) + \cos(q_y)])}.
\end{equation}

{Note that for $N=2$, the propagator gains an extra factor of 2 since $O^{(\sigma)}_{\v r_1, \v r_2} = O^{(\sigma)}_{\v r_2, \v r_1}$ and the resummation gains an extra term (cf.~\cite{KoenigTsvelik2020} for the $\mathbb{Z}_2$ propagator)}. One can add a second perturbation to this - a fermionic Hamiltonian with small nearest-neighbor hopping. It should be noted that this perturbation is added on top of the existing Fermi sea

\begin{equation}
    \H_t = -t\sum_{\langle\v r, \v r'\rangle} c^{\dagger}_{\v r} c_{\v r'}.
\end{equation}

The fermionic perturbation modifies the Green's function, with the resummation given by the equation 

\begin{align}
\v D_t(\v r_f, \v r_i;  i\omega) &= \v D(\v r_f, \v r_i;  i\omega) \nonumber\\
&\hspace{-5em}+ t \sum_{\v r} \sum_{\langle\v r', \v r \rangle} \v D(\v r, \v r_i; i\omega) \mathcal{G}_{FS}(\v r, \v r') \v D_t(\v r_f, \v r'; i\omega).
\end{align}

In momentum space,

\begin{equation}
    \v D_t(\v q, i\omega) = \v D(\v q, i\omega) + \bar{t} \, \v D(\v q,i\omega) \v D_t(\v q, i\omega),
\end{equation}

where $\bar{t} = 2t\; \mathcal{G}_{FS} (r,r+\hat{e}_x)$. The new propagator is, therefore, analogously to Eq.~\eqref{eq:e_prop} with the replacement $J \rightarrow (J+\bar{t})$.

\subsubsection{Luttinger surface of Green's function zeros}
\label{sec:zeros}

As outlined in the introduction, it is a question of current interest to understand the nature and occurrence of zeros in the electronic Green's function~\cite{Dzyaloshinskii2003} of strongly correlated systems. Specifically, these zeros, which can be understood as poles of the self-energy, contribute to the thermodynamics~\cite{Fabrizio2023} and topology~\cite{WagnerSangiovanni2023,SettySi2023,BlasonFabrizio2023} very much alike conventional quasi-particles in Fermi liquids and non-interacting electronic systems. At the same time, it is apparent that the low-energy thermodynamics of orthogonal (semi-)metals is equivalent to the thermodynamics of their non-orthogonal counterparts. In this section, we demonstrate a connection between orthogonal metals and the occurrence of zeros in the Green's function. However, the Luttinger surface of zeros is unrelated to the density count.

\begin{figure}
    \centering
    \includegraphics{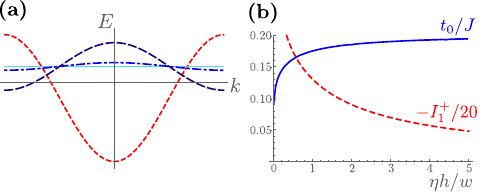}
    \caption{\textbf{(a)}: Schematic representation of the dispersion relation of orthogonal fermions (red, dotted), and Green's function zeros, Eq.~\eqref{eq:AllZerosMaintext}, for $t_0 = 0$ (light blue), $0<4t_0< \mu$ (blue, dot-dashed), $\mu<4t_0$ (dark blue, dashed). \textbf{(b)}: Dimensionless parameters entering $t_0$ and $\mathcal Z$, cf. Eq.~\eqref{eq:AllZerosMaintext}, as a function of $\eta h/w$.} 
    \label{fig:Zeros}
\end{figure}

To this end, we calculate the fermionic Green's function 

\begin{equation}
\mathcal G(\v p, i\epsilon)  = a^2 \int_{\rm BZ} (dq) \int (d \omega) \mathbf D(\v q, i \omega) G_{\rm FS}(\v p + \v q, i \epsilon + i \omega), \label{eq:FermionicG}
\end{equation}
perturbatively in $J$ and inside the gap of Eq.~\eqref{eq:SingleParticleGF}. Relegating details to App.~\ref{app:LuttingerSurface}, we find

\begin{subequations}
\begin{equation}
\mathcal G(\v p,z)  = -\frac{1}{\mathcal Z}  [z + \zeta_0( \v p) ], \label{eq:Zeros}
\end{equation}
with a dispersion of zeros of the form 
\begin{equation}
\zeta_0(\v p) = -2t_0 [\cos(p_x) + \cos(p_y)] - \mu,
\end{equation}
and
\begin{align}
t_0 & = - J \frac{ \eta h}{8 w} \frac{I_1^+ + 2\pi w/ \eta h}{I_1^+} >0,\\
\mathcal Z & = -(\eta h/\nu_0a^2) I_1^+ >0.
\end{align}
\label{eq:AllZerosMaintext}
\end{subequations}
Note that the ``weight of zeros'' $\mathcal Z$ has dimensions of energy squared. The dimensionless integral $I_1(\eta h/w)$ is plotted along with $t_0$ and a schematic representation of the dispersion of orthogonal fermions and of zeros in Fig.~\ref{fig:Zeros}. Crucially we observe that the Luttinger count enclosed by zeros is completely unrelated to the Fermi surface of orthogonal metals. Indeed, in the integrable limit $J = 0$, the zeros of the Green's functions are disperionless  and reside at energy $z = \mu$. Small $t_0< \mu/4$ does not lead to the appearance of a Luttinger surface. For larger $t_0$ the Luttinger surface appears, but, as mentioned, it is unrelated to the Fermi surface of fermions. 

In addition to the Luttinger surface being unrelated to fermion density, there are multiple distinctions between the zeros occurring in this context as compared to earlier discussions in the literature. The main reason is that the Luttinger surface encountered here originates from the dispersion of ``e'' particles (rather than, e.g., from a self-energy in the fermionic Green's function). Luttinger's theorem is fulfilled by the count of poles of orthogonal electrons, corrected for the effect of visons, Eq.~\eqref{eq:Luttinger}. While analogously to the situation discussed by Fabrizio~\cite{Fabrizio2023} the present fractionalized metal displays Fermi liquid thermodynamics, the distinction is that this thermodynamic response does not originate from the Luttinger surface but rather from the orthogonal fermions.

\subsubsection{Higgs transition}

The correlator $\v D(\v q = 0, i\nu = 0)$ represents the sum of open electric strings of all lengths. The singular behavior at {$\eta h = 4(J+\bar{t})$} 
indicates condensation of $e$-particles, corresponding to the confinement-deconfinement transition of the regular toric code. If $J+\bar{t}$ is increased further, the system undergoes a transition from an orthogonal metal to a regular Fermi liquid. At the critical point where confinement occurs, one can write down the continuum theory as a $\mathbb{Z}_N$ lattice gauge theory coupled to fermions (cf. ~\cite{NandkishoreSenthil2012,KoenigTsvelik2020} for the cases $N = 2,4$),

\begin{figure}
    \includegraphics[scale = 1]{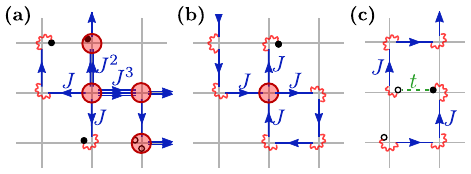}
    \label{fig:Vertices}
    \caption{Diagrams that contribute to the confinement-deconfinement transition. \textbf{(a)}: the $\mathbb Z_N$ term (illustrated specifically for $\mathbb{Z}_7$). \textbf{(b)}: the four-point self-interaction term. \textbf{(c)}: the fermion-string interaction term.}
\end{figure}

\begin{subequations}
    \begin{align}
        \mathcal S &= \mathcal{S}_f + \mathcal{S}_b + \mathcal{S}_{\rm IA},\\
        \mathcal{S}_b &= \int d\tau \, d^2x \;\bar \phi[-\partial_\tau^2 - v^2 \nabla^2  + r ] \phi  + u_0 \vert\phi\vert^4 \nonumber \label{eq:PhiAction} \\ 
        &\hspace{13em} + u  (\phi^N + \bar \phi^N),\\
        \mathcal{S}_f &=  \int d\tau \, d^2x \;\bar \psi [\partial_\tau + \epsilon(- i\nabla)] \psi,\\
        \mathcal{S}_{\rm{IA}} &= \gamma \int d\tau \, d^2x \;\bar \psi  \, [\cos(-i \partial_x) + \cos(-i \partial_y)] \, \psi \vert \phi \vert^2.
    \end{align}
    \label{eq:Criticality}
\end{subequations}

The complex field $\phi$ (Grassmann field $\psi$) describes the fluctuations of the electric strings (fermions) near criticality ($\v{D}(\v x,\tau) = 2\eta h a^2 \langle\bar\phi(\v x, \tau)\phi(0,0)\rangle, \mathcal{G}_{FS}(\v x, t) = a^2\langle\bar\psi(\v x, \tau)\psi(0,0)\rangle$ and $a$ is the lattice spacing). {A first order time derivative does not occur given the quadratic frequency dependence of the resummed propagator, Eq.~\eqref{eq:e_prop}.} Observing equation \eqref{eq:e_prop}, one can identify the constants $v^2 = \eta h[\eta h - {4}(J + \bar{t})]$ and {$r = \eta h (J+\bar{t})  a^2$}. Moreover, we have determined the four-point self-interaction vertex $u_0 \sim a^2 J^4/(\eta h)$ as well as the $\mathbb Z_N$ vertex $u \sim a^{N-2} J^N/h^{(\frac{N}{2} - 1)}$ (cf. Appendix \ref{app:Criticality}). We emphasize that the fermionic interactions are added on top of the Fermi sea, so $\gamma \sim t$.

\begin{figure}
    \centering
    \includegraphics[scale=1]{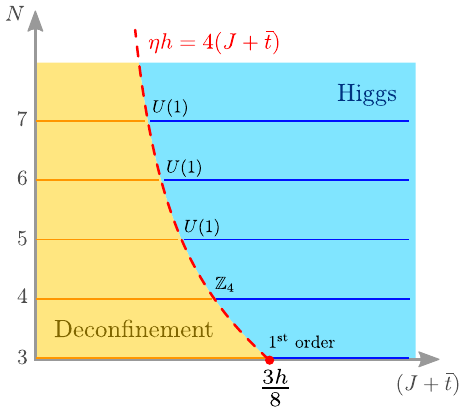}
    \caption{Phase diagram representing the confinement-deconfinement transition for $N>2$. The phase boundary (red) occurs at $\eta h = 4(J+\bar{t})$.
    }
    \label{fig:e_condensation}
\end{figure}

The nature of the phase transition hinges on the value of $N$, as shown in Fig. \ref{fig:e_condensation}. Note that for $N \geq 5$, the transition will be of $U(1)$ type. We also remark that, contrary to the quasi-1D case, the Higgs transition is not expected to be preempted by a $U(1)$ phase. This is because the Higgs transition stems from the toric code sector alone and intermediate $U(1)$ phases are not permitted in two spatial dimensions (see Sec.~\ref{sec:RG} for a review).

\subsection{Spontaneous $\tau$-insertions: Hadron formation}
\label{sec:Hadrons}

While in the previous section, we studied the effect of statistically appearing $e$-particles (obtained by $\sigma$ insertions), we here study the effect of $m$-particles. To this end we add

\begin{equation}
    \H_g = -\frac{g}{2}\sum_{\v b}(\tau_{\v b} + \tau^{\dagger}_{\v b})
\end{equation}
to the basic Hamiltonian, Eq.~\eqref{eq:model}, but set $J = t = 0$.

As we already saw, insertion of $\tau$ results in an Anderson orthogonality catastrophe, such that the $m$-particle dynamics in the presence of fermions is very different from the dynamics of $e-$particles, Eq.~\eqref{eq:e_prop}. Technically, this is reflected in a much less straightforward computation as the previous section. 

As a remedy to avoid the orthogonality catastrophe physics, it is physically favorable to bunch fermions into charge neutral $N$-particle bound states (``hadrons''). Here, we present details on the hadron formation under the assumption of $N_f = N$ fermion flavors. For simplicity, we concentrate on the limit of large $h$, $\phi = 0$ and low fermion filling (so that the Fermi surface is given by a circle and Umklapp effects generating order in the particle-hole channels~\cite{GazitVishwanath2017} is disabled.)

\begin{figure}[t]
    \includegraphics[scale=1]{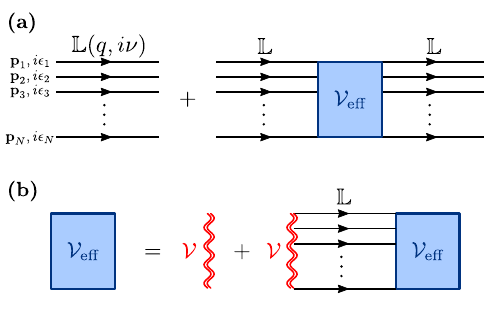}
    \caption{Ladder resummation of the $N$-particle propagator. \textbf{(a)}: The resummation in terms of the non-interacting ladder $\mathbb{L}(\v q = \sum_j \v p_j, i\nu = \sum_j i \epsilon_j)$ and the effective interaction vertex $\mathcal{V}_{\text{eff}}$. \textbf{(b)}: The Bethe-Salpeter equation defining the effective interaction vertex in terms of $\mathbb{L}$ and the bare interaction $\mathcal{V}$.}
    \label{fig:ladder}
\end{figure}

We first focus on the limit of low magnetic string tension, $g \ll \eta K$. Further, since we are already working in the limit $\eta K\gg w$, the orthogonal fermions are relatively slow as compared to the $m$-particles introduced with the perturbation. These $m$-particles induce an attractive interaction within the fermions, and for large enough $g$, create an instability in the OM phase. The instability can be identified as a singularity in the $N$-particle interaction channel, the leading contribution to which is obtained from a ``ladder" resummation, see Fig. \ref{fig:ladder}.  (for $N=2$, the calculation is equivalent to that of the Cooper instability). 

The bare interaction induced by the $m$-particles is characterized by $\mathcal{V}$. We can estimate $\mathcal{V}$ by finding the effective $N$-fermion hopping amplitude from perturbing around the OM ground state, treating $g$ as a small parameter in the partition function

\begin{equation}
    \frac{Z}{Z_0} = \text{Tr}\left[\mathcal{T} \exp\left(-\int d\tau \, \mathcal{H}_g(\tau)\right)\right].
\end{equation}

Ignoring higher-order contributions corresponding to $m$-particle insertions in loops, the first non-zero contribution to the hopping arises from $N$-th order in $g$, with an amplitude of $(gw)^N/(\Delta E_m)^{2N-1}$, $\Delta E_{m}$ being the energy gap of the $m$-particles. While at the integrable limit this is simply $\eta K$, it gets modified to $\sqrt{\eta K(\eta K - 4g)}$ as one approaches $m$-particle confinement (analogous to $e$-particles, cf. Eq.~\eqref{eq:e_prop}). 

Appendix \ref{app:Hadron} presents the details of the $N$-ladder correlator as well as the calculations of the critical temperature at which the hadron instability sets in, given by

\begin{equation}
    T_h(g \ll K) = {T_h^{(0)}} \left(1-\frac{1}{v\alpha_N(0)}\right),
\end{equation}
where $\alpha_N(0) = \int_0^1 \prod_i^N (d x_i) (\sum_i^N x_i)^{-1}$, is finite for all $N>2$ and behaves as $2/N$ in the large-$N$ limit. The dimensionless interaction $v$ is a function of the bare interaction $\mathcal{V}$ and the chemical potential $\mu$ (above the bottom of the band). 
The scaling of dimensionless $v$ as a function of $g$, $K$, $w$ as well as the chemical potential is

\begin{equation} 
    v(g/K) \sim \left(\frac{g}{\Delta E_m}\right)^N \left(\frac{w}{\Delta E_m}\right) \left(\frac{\mu'}{\Delta E_m}\right)^{N-2}. \label{eq:hadron_v_scaling}.
\end{equation}

\begin{figure}
    \centering
    \includegraphics{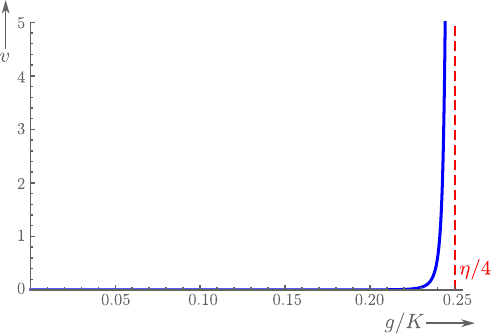}
    \caption{Estimate for the dimensionless hadron interaction $v$, Eq.~\eqref{eq:hadron_v_scaling}, varying with $g/K$ (blue). The plot has been constructed for $N=4$, $w/\eta K=0.3$ and $\mu/w=0.3$. $v$ has a vertical asymptote at $\eta K = 4 g$ (red), the point at which $m$-particle condensation occurs.}
    \label{fig:hadron-criticality-schematic}
\end{figure}

Here and henceforth the dashed quantity $\mu'$ indicates the chemical potential above zero filling, as opposed to the $\mu$ in Eq. \eqref{eq:model} which is zero at half-filling. As $g/K$ increases, one sees that the proposed transition always occurs before the condensation of $m$-particles, (see Fig. \ref{fig:hadron-criticality-schematic}), supporting the hypothesis that the $m$-particle dynamics result in a retarded attractive interaction and subsequent hadron instability. As $g$ approaches $\eta K/4$, $T_h$ approaches a maximum value $T_h^{(0)}\sim \mu'/N$.

In the opposite limit, deep in the confining phase for $m$-particles ($g\gg \eta K$) the $\mathbb{Z}_N$ clock variables on the lattice align in the $\tau$ basis which renders them disordered in the $\sigma$ basis. This results in an insulating phase for the fermions for low temperatures. One can carry out perturbation theory on the new ground state where $\tau_{\v b} = 1$, treating $w$ as a small parameter,

\begin{equation}
    \frac{Z}{Z_0} = \text{Tr}\left[\mathcal{T} \exp\left(-\int d \tau \, \H_w(\tau)\right)\right].
\end{equation}

Expanding this expression to $N$-th order in $w$ produces an effective nearest-neighbor hopping of $[c^\dagger_{\v r}]^N [c_{\v r'}]^N$, with an amplitude of the order $t_\text{eff} \sim w^N/g^{N-1}$. In this regime, assuming that the $N$-hadrons show bosonic statistics, i.e. $N \in 2 \mathbb N$, the condensation sets in as a Berezinskii-Kosterlitz-Thouless (BKT) transition (see~\cite{GazitVishwanath2017} for $N = 2$). One can then utilize BKT theory to estimate the transition temperature $T_h$ in this regime

\begin{equation}
    T_h(g \gg K) = T_{BKT} \sim t_{\text{eff}} \,n_b a^2,
\end{equation}

where $n_b$ is the boson density ($\sim (\mu/w)/N$) and $a$ the lattice spacing. This calculation suggests that the temperature $T_h$ develops a tail with a power-law decay ($\sim\frac{1}{g^{N-1}}$), tapering off for large $g/K$.  

\begin{figure}
    \centering
    \includegraphics{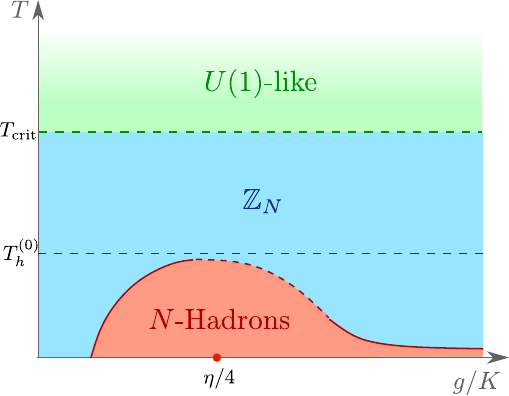}
    \caption{Schematic phase diagram of temperature $T$ vs. $g/K$ for small fermion filling showing the dome of hadron formation (red) buried in the $\mathbb{Z}_N$ phase (blue). The point $\eta K = 4g$ indicates the zero-temperature confinement-deconfinement transition for the $m$-particles. At higher temperatures $T>T_{\text{crit}}$, our RG-flow analysis suggests $U(1)$-like behavior (green).}
    \label{fig:master_phase_diag}
\end{figure}

In the regime between confinement and deconfinement of the magnetic sector ($g\sim \eta K$), one needs numerical tools to establish the exact phase boundary, but one expects to see a dome of $N$-hadron instability in the phase diagram as shown in Fig.~\ref{fig:master_phase_diag}, with the apex of the dome bounded above by the critical temperature $T_h^{(0)}$. A similar dome of superconductivity has been strongly evidenced for $N = 2$ in Ref.~\cite{GazitVishwanath2017} using quantum Monte Carlo techniques, yet for $N = 2$ the hadron formation occurs at infinitesimal $g/K$.

We leave a remark in passing here that interactions between these $N$-hadrons can lead to formation of higher order aggregates that are more energetically favourable, like the formation of $2N$-hadrons with bosonic statistics for odd $N$, which can form condensates unlike their counterparts with fermionic statistics. This requires further analysis and will be dealt with in future studies. 

\section{Absence of a $U(1)$ phase \label{sec:RG}}

In the previous section we have argued that the confinement-deconfinement transition of the $\mathbb Z_N$ gauge theory with fermions is buried under a dome of hadron formation (in the case $N=2$: a dome of superconductivity). Yet, one could have also expected that an intermediate $U(1)$ phase occurs between the two phases, or might at least be stabilized by additional interactions in the Hamiltonian. In this section, we provide arguments why this is not the case. Instead, we here show that in 2+1D, QED coupled to a Fermi surface is a renormalization group fixed point which is unstable to $\mathbb Z_N$ perturbations. According to these results, it is strictly speaking not possible to emulate $U(1)$ LGTs by finite Hilbert space $\mathbb Z_N$ approximants. However, we show that $U(1)$ physics may persist at intermediate system sizes smaller than the lengthscales at which N-hadron bound states may be observed.

We first summarize the situation in the absence of fermionic matter fields, where indeed an intermediate $U(1)$ phase occurs in one spatial dimension, but not in 2D. This can be understood as follows: While in $D = 1+1$ and large $N$, terms breaking $U(1)$ symmetry down to $\mathbb Z_N$ are renormalization group (RG) irrelevant near the transition~\cite{JoseNelson1977}, such a mechanism is disallowed for $D = 2+1$. Indeed, in the dual (gauge theory) language, the disorder - order transition corresponds to a deconfinement-confinement transition. At the same time, compact QED$_3$ without matter fields is always in a confining phase, which implies that any putative $U(1)$ phase would ultimately gap out at longest distances~\cite{Polyakov1987}. What, however, if the $\mathbb Z_N$ gauge theory in $D = 2+1$  is coupled to fermionic matter fields? Indeed, in contrast to the pure gauge theory, it is common knowledge that compact QED$_3$ coupled to a Fermi surface remains gapless~~\cite{AltshulerMillis1994,KimLee1994,HermeleWen2004,NaveLee2007,Lee2009,MrossSenthil2010}. Analogously to the logic applied in 1+1D, we here study the effect of terms breaking $U(1)$ symmetry down to $\mathbb Z_N$ in a renormalization group treatment.

\begin{figure}
    \centering
    \includegraphics{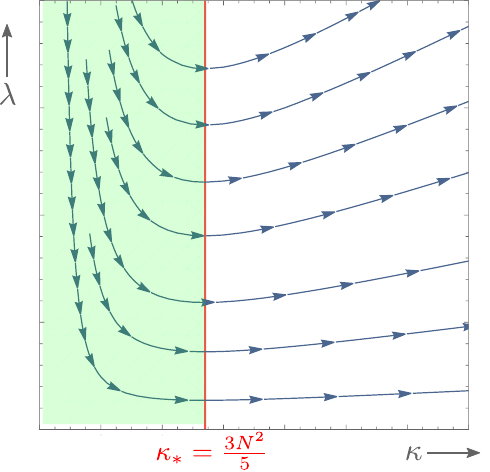}
    \caption{A schematic of the flow of of the constants $\kappa$, $\lambda$ under the renormalization group procedure. System scales corresponding to the green region can approximate $U(1)$ order, where $\kappa<\frac{3 N^2}{5}$.}
    \label{fig:RG_flow}
\end{figure}

We here study the renormalization group flow for a $\mathbb{Z}_N$ perturbation added to the non-Fermi liquid theory obtained by coupling a degenerate Fermi gas with $N_f$-fold flavor degeneracy to QED$_3$ gauge fluctuations (see Appendix \ref{app:U(1)_RG} for details). The flow equations are

\begin{subequations}
\begin{align} 
    \frac{d\kappa}{d \ln b} &= 6 \kappa, \label{eq:rg_flow_K}\\
    \frac{d \lambda}{d \ln b} &= \left(5 - \frac{3 N^2}{\kappa}\right) \lambda \label{eq:rg_flow_lambda},
\end{align}
\label{eq:RGMaintex}
\end{subequations}

where the constant $\lambda$ determines the strength of the $\mathbb{Z}_N$ perturbation locking the $U(1)$ gauge field to $N$ discrete angles and the constant $1/\kappa$ linearly enters the propagator of the $U(1)$ gauge field (analogously to a stiffness in Berezinskii-Kosterlitz-Thouless theory). One sees that since $\kappa$ is relevant, $\lambda$ is bound to flow to relevancy and thus the $\mathbb{Z}_N$ perturbation breaks $U(1)$ order in the system. Fig. \ref{fig:RG_flow} plots the flow equations Eq.~\eqref{eq:rg_flow_K}, \eqref{eq:rg_flow_lambda}. Although there is no $\mathbb{Z}_N$ order, one can see quasi-$U(1)$ behavior at intermediate system scales, corresponding to the green region of the phase space where $\lambda$ flows to irrelevancy. Using the  starting value $\kappa \sim 1/N_f^2$, we have calculated the critical length scale $L_{\text{crit}}$ up to which this behavior persists and its dependence on $N$ as well as the number of fermion flavors $N_f$
\begin{equation}
    L_{\text{crit}} \sim l \left(\frac{N}{N_f}\right)^{\frac{1}{3}},
\end{equation}

where $l$ is the UV length cutoff which can be estimated of the order of the Fermi wave-vector. The critical length scale endows a corresponding critical temperature $T_{\text{crit}} \sim \mu (N_f/N)$ above which the $U(1)$-like behaviour in the system persists, see Fig.~\ref{fig:master_phase_diag}. In the case $N_f = N$ studied in the previous section, it is ensured that the quasi-$U(1)$ behaviour persists only outside the $N$-hadron phase ($T_{\text{crit}}>T_h^{(0)}$)

{We emphasize that our calculations are performed within the random-phase approximation (RPA) of fermions coupled to emergent photons in 2D. It has been argued that a combination of large $N_f$ and an $\epsilon$ expansion justifies the existence of a phase described by this phenomenology~\cite{MrossSenthil2010}. However, in strictly two spatial dimensions, the infrared fixed point of QED$_3$ coupled to a Fermi surface is unknown. While we can not exclude a scenario in which $\lambda$ is relevant at the non-Fermi liquid RPA fixed point but irrelevant at the unknown physical infrared fixed point, such a scenario seems extremely unlikely.}

\section{Discussion and outlook}

In summary, we have coupled a $\mathbb{Z}_N$ generalization of Kitaev's toric code to fermions, obtaining a solvable model for fermions coupled to a $\mathbb{Z}_N$ lattice gauge theory in (2+1) dimensions. This model exhibits a {rich} phase diagram with exotic changes in the Fermi surface topology, switching from an orthogonal metal phase to an orthogonal semi-metal (for even $N$) or a band insulator (for odd $N$). The deconfining sector of the toric code admits momentum transfer through the gauge field, allowing violation of Luttinger's theorem without symmetry breaking. In the integrable limit, we have also developed a diagrammatic technique for correlators of fermions and ``$\sigma$'' insertions (corresponding to $e$-particles). We furthermore studied correlators of ``$\tau$'' insertions (corresponding to $m$-particles), which trigger an orthogonality catastrophe effect and a power-law decay at long times in the $m$-particle propagator. We have found a closed form for this power-law exponent systematically using linked cluster expansions.

We next included perturbations to the integrable limit which delocalize $e$ and $m$ particles of the toric code. In the deconfining phase, the term delocalizing $e$ particles generates a dispersive band of Green's function zeros.
We have also studied the Higgs transition corresponding to the condensation of $e$-particles. We predicted the position of the phase transition within mean-field theory and developed an effective field theory to investigate the critical behavior of the system at this transition beyond mean field. The terms dynamically delocalizing $m$-particles (which correspond to a finite string tension) tend to

 induce a hadron instability resulting in the formation of gauge-neutral hadrons consisting of $N$ particles. In the case of $N_f \geq N$ flavors of fermions we explicitly demonstrate that the hadron instability buries the confinement-deconfinement transition underneath a hadronic dome (superconductivity in the case $N = 2$). We finally address the question of the limit $N \rightarrow \infty$ to answer whether $\mathbb Z_N$ LGT are good emulators for QED. Concentrating on the case of a Fermi surface but far away from half-filling, we present RG arguments according to which QED is always unstable, i.e. $\mathbb{Z}_N$ LGTs do not have a $U(1)$ gapless phase in the infrared. However we argue that at intermediate energy scales $\mathbb Z_N$ LGTs with $N \gg 1$ can be good approximations to QED$_3$. 

In conclusion, both recent advances in quantum materials and the advent of new programmable quantum emulators have generated practical interest in topological phases of $\mathbb Z_N$ gauge theories. Here we studied fermionic fields in this context allowing to shed light on the physics of Luttinger surfaces of Green's function zeros and the problem of Fermi surface reconstruction without symmetry breaking. Interesting problems for the future involve the interplay of topological band structures with deconfining lattice gauge theories and the search for deconfining phases in more microscopic models of materials. At the same time, many physically relevant questions for quantum emulator implementations of deconfining gauge theories remain unanswered -- most importantly those addressing important aspects of experimental imperfections such as decoherence.

\acknowledgments
It is a pleasure to thank S.~Moroz, P.M.~Bonetti, M.S.~Scheurer, I.~Sodemann, U.~Seifert, D.~Vilardi and O.~Zilberberg for useful discussions. KR thanks the Max-Planck-Institute for Solid State Research for hospitality and the German Academic Exchange Service (DAAD) for support via the Working Internships in Science and Engineering (WISE) program. EJK thanks the Kavli Institute for Theoretical Physics for hospitality, where this work was partially completed. Thereby, this research was supported in part by the National Science Foundation under Grant No. NSF PHY-1748958. 

\appendix

\section{Quantum Clock Model to $\mathbb{Z}_N$ Toric Code}
\label{app:ClockToTC}

In this section we summarize the mapping of the $\mathbb{Z}_N$ gauge theory (with fermionic and bosonic matter fields) to the $\mathbb Z_N$ toric code model\cite{Wegner1971, TupitsynStamp2010, KoenigTsvelik2020}

\begin{equation}
    \mathcal{H} = \mathcal{H}_{\mathbb{Z}_N} + \mathcal{H}_{QC} + \mathcal{H}_f, \nonumber
\end{equation}
\begin{subequations}
    \begin{alignat}{1}
    \mathcal{H}_{\mathbb{Z}_N} &= - \frac{K}{2}\sum_{\v p}\prod_{\v e \in \{\hat{x},-\hat{y}\}} \bar{\sigma}_{\v r+ \frac{1}{2}\v e}  \bar{\sigma}^{\dagger}_{\v r - \frac{1}{2}\v e} \ -\frac{g}{2} \sum_{\v b} \bar{\tau}_{\, \v b} \ + \text{H.c.},\\
    \mathcal{H}_{QC} &= -\frac{J}{2} \sum_{\v r, \v e \in \{\hat{x},\hat{y}\}} \bar{\sigma}_{\left(\v r + \frac{1}{2} \v e\right)}\varsigma^{\dagger}_{\v r}\varsigma_{\v r + \v e} -\frac{h}{2} \sum_{\v r} \uptau_{\v r} \ + \text{H.c.},\\
    \mathcal{H}_f &= -\frac{w}{2} \sum_{\v r, \v e \in \{\hat{x},\hat{y}\}} \bar{\sigma}_{\left(\v r + \frac{1}{2} \v e\right)} f^{\dagger}_{\v r,\alpha} f_{\v r + \v e,\alpha} + \text{H.c.}.
    \end{alignat}
\end{subequations}
$\mathcal{H}_{QC}$ is the usual quantum clock Hamiltonian, with the interaction term modified with the addition of $\bar{\sigma_{\v b}}$ to couple it to the gauge field.

Both sets of operators obey the algebra
\begin{subequations}
    \begin{alignat}{1}
        \bar{\sigma}_{\v b} \bar{\tau}_{\,\v b'} &= (\omega)^{\delta_{\v b,\v b'}} \bar{\tau}_{\, \v b'} \bar{\sigma}_{\v b}, \\
        \varsigma_{\v r} \uptau_{\,\v r'} &= (\omega)^{\delta_{\v r,\v r'}} \uptau_{\, \v r'} \varsigma_{\v r} ,
    \end{alignat}
\end{subequations}
where $\omega = e^{2\pi i/N}$, the N-th root of unity. The model has a local $\mathbb{Z}_N$ symmetry at each site generated by the operator $\bar{Q}_{\v r}$,
\begin{equation}
    \bar{Q}_{\v r} = \underbrace{\omega^{\hat{n}_{\v r}}\uptau_{\v r}}_{\text{matter}} \underbrace{\prod_{\v e \in \{{\hat{x}}, {\hat{y}}\}} \bar{\tau}_{\,\v r+ \frac{1}{2}\v e} \bar{\tau}^{\dagger}_{\v r - \frac{1}{2}\v e}}_{\text{gauge}}.
\end{equation}

To fix the local charge, we impose Gauss' law on the physical subspace
\begin{equation}
    \bar{Q}_{\v r}\ket{\text{Phys}} = \ket{\text{Phys}}.
\end{equation}
We thus make the replacement $\uptau^{\dagger}_{\v r} \rightarrow \prod_{\v e \in \{\hat{x},\hat{y}\}} \omega^{\hat{n}_{\v r}} \bar{\tau}_{\,\v r+ \frac{1}{2}\v e} \bar{\tau}^{\dagger}_{\v r - \frac{1}{2}\v e} (\equiv Q_{\v r})$. We also enforce the unitary gauge on the physical subspace:
\begin{equation}
    \varsigma_{\v r}\ket{\text{Phys}} = \ket{\text{Phys}}.
\end{equation}
Finally, we replace the terms in the Hamiltonian with gauge-invariant quantities:
\begin{enumerate}
    \item $\mathcal{W}^{(e)}_{\gamma_{\v r, \v r'}} = \varsigma_{\v r}^{\dagger}\left[\prod_{\v r_i\in\gamma}\bar{\sigma}_{\v r_i, \v r_{i+1}}\right]\varsigma_{\v r'} \equiv \prod_{\v r_i\in\gamma}\sigma_{\v r_i, \v r_{i+1}}$
    \item $\bar{\tau}_{\v b} \equiv \tau_{\v b}$
    \item $c_{\v r,\alpha} = \varsigma^{\dagger}_{\v r} f_{\v r, \alpha}$
\end{enumerate}
giving us the gauge-fixed Hamiltonian:
\begin{equation}\label{eq:z_n}
\begin{aligned}
    \mathcal{H} = &- \frac{K}{2}\sum_{\v p}\prod_{\v e \in \{\hat{x},-\hat{y}\}} {\sigma}_{\v r+ \frac{1}{2}\v e}  {\sigma}^{\dagger}_{\v r - \frac{1}{2}\v e} \ -\frac{g}{2}\sum_{\v b} \tau_{\v b} + \text{H.c.}\\ 
    &-\frac{h}{2}\sum_{\v r} \underbrace{\omega^{\hat{n}_{\v r}}\prod_{\v e \in \{\hat{x},\hat{y}\}} {\tau}_{\,\v r+\frac{1}{2}\v e} {\tau}^{\dagger}_{\v r - \frac{1}{2}\v e}}_{Q_{\v r}} \ -\frac{J}{2}\sum_{\v r} \sigma_{\v r} + \text{H.c.}\\
    &-\frac{w}{2}\sum_{\v r, \v e \in \{\hat{x},\hat{y}\}} c^{\dagger}_{\v r,\alpha} \sigma_{\left(\v r + \frac{1}{2} \v e\right)} c_{\v r + \v e,\alpha} + \text{H.c.}.
    \end{aligned}
    \vspace{10pt}
\end{equation}

This concludes the argumentation exposed in Sec.~\ref{sec:Motivation} of the main text

\section{Derivation of Feynman rules \label{app:feynman_rules}}

In this Appendix we derive the Feynman rules presented in Sec.~\ref{sec:Feynman}. To this end, consider Eq.~\eqref{eq:Correlator} with the Heisenberg picture time evolution $O(\tau) = e^{H\tau} O e^{-H\tau}$.
\begin{itemize}
    \item Write $\ket{GS}$ as $\prod_{\v r} \hat P^{(0)}_{\v r} \ket{\varphi} \ket{FS}$. Commuting the projector on the left across to the other side, one sees that the charge $q_i$ on each site must be conserved modulo $N$ for the product of projectors to not vanish. This gives Feynman rule 2. This implies that two fermionic operators must be joined by a string of oriented $\sigma$ insertions, and equivalently that $\sigma$ insertions must either be closed loops or terminate in one (or one modulo $N$) fermionic operators.
    \item For an operator $O_{\v r}$ with charge $q_i$, the explicit time evolution is $O_{\v r}(\tau) = e^{\H_h\tau}e^{\H_w\tau} O_{\v r} e^{-\H_w\tau}e^{-\H_h\tau} = \bar{O}_{\v r}(\tau) \exp(\frac{h}{2}(1-\omega^{q})Q_{\v r}\tau + \text{H.c.})$, 
    where $q$ is the charge of the operator $O$ and $\bar{O}(\tau)$ indicates time evolution with respect to the fermionic part only.
    \item At site $\v r$, write the operators inside the correlator $C(\{\v r, \tau \}$ in time-ordered fashion. Replace all operators with the above time evolution expression: $O_{\v r}(\tau) = \bar{O}_{\v r}(\tau) \exp(\frac{h}{2}(1-\omega^{q})Q_{\v r}\tau + \text{H.c.})$. Commute the exponential terms to the right, making note of the charge of the subsequent operators ($Q_{\v r} O_{\v r'} = (\omega^{q})^{\delta_{\v r, \v r'}} O_{\v r'} Q_{\v r}$) and finally use $Q_{\v r}\ket{GS} = \ket{GS}$ coupled with Feynman rule 2 to obtain the $e$-particle interaction term, Eq.~\eqref{eq:exp_e-particles}. 
    \item The only dependence on {$\sigma$ variables} 
    now lies in $\bar{O}(\tau)$, specifically in $\H_w$, and the $\sigma$ insertions. The $\sigma$s can now be replaced with their expectation values on the gauge-fixed lattice ($\bra{\varphi}\sigma_{\v r}\ket{\varphi}$). This fixes the Hamiltonian $\H_w \rightarrow {\H_{\varphi}}$. 
    For example, in the zero-flux OM phase, $\langle\sigma_{\v r}\rangle_{0} = 1$ and $H_w$ reduces to the regular 2D tight-binding Hamiltonian.
    \item The correlator has now been brought to the form $C({\v r, \tau}) = \bra{FS}\mathcal{T}\{\bar{O}_{\v r_1}(\tau_1)\bar{O}_{\v r_2}(\tau_2)\dots\bar{O}_{\v r_n}(\tau_n)\}\ket{FS}$. Now Wick's theorem can be applied.
\end{itemize}

The real time Feynman rules are derived analogously, simply derived with a Wick rotation: $\tau \rightarrow it, |\tau_1-\tau_2| \rightarrow i|t_1 - t_2|$.

\section{Nozieres - de Dominicis solution} 
\label{app:nd}

In this appendix, we present details on the correlators of $\tau$ operators ($m$-particles) in the presence of fermions, thereby complementing Sec.~\ref{sec:m_particle}.

To understand the Nozieres - de Dominicis solution, we first lay out the perturbative method. The propagator $\mathcal{G'}_{FS}(t'-t) = -i\langle\mathcal{T} \exp{-i \int_t^{t'} d\tau \hat{V}(\tau)}\rangle$ can be expanded in $V\equiv\frac{w}{2}(1-\omega)$ and $\bar{V}$ into a series of connected and disconnected diagrams with two types of vertices. This series can be resummed to an exponential of just the connected diagrams using the linked cluster theorem
\begin{equation}
    \mathcal{G'}_{FS}(t) = -ie^{C(t)},
\end{equation}
where

\begin{equation}
    C(t) = \left\langle\mathcal{T} \exp{-i \int_0^{t} d\tau \hat{V}(\tau)}\right\rangle_{\text{connected}}.
\end{equation}

$C(t'-t)$ is the sum of all closed-loop diagrams, as shown below,
\begin{widetext}
\begingroup
\addtolength{\jot}{1em}
\begin{align*}
    \ln \mathcal{G}(t) &= \quad \includegraphics[width=1.2cm, valign=c]{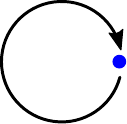} \quad + \quad \includegraphics[height=1.2cm, valign=c]{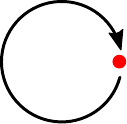}
    +\enspace\frac{1}{2} \left\{ \:\includegraphics[width=1.7cm, valign=c]{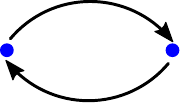} \quad + \quad 2\enspace\includegraphics[width=1.7cm, valign=c]{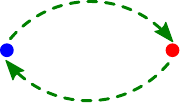} \quad + \quad \includegraphics[width=1.7cm, valign=c]{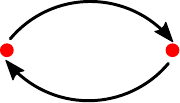}\:\right\}\\
    &+\enspace\frac{1}{3} \left\{ \: \quad \includegraphics[width=1.7cm, valign=c]{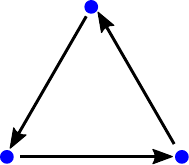} \quad + \quad 3\enspace\includegraphics[width=1.7cm, valign=c]{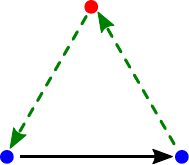} \quad + \quad 3\enspace\includegraphics[width=1.7cm, valign=c]{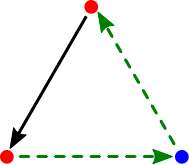} \quad + \quad \includegraphics[width=1.7cm, valign=c]{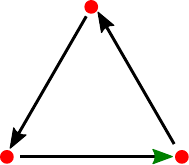}\:\right\}\\
    &+ \ \text{higher order clusters}.
\end{align*}
\endgroup
\end{widetext}

The diagrams above can be computed order by order, but only when the propagators in each diagram are non-singular. While we concentrate on the case away from half-filling, where the fermionic Green's function is $g_0(t) \sim 1/w t$, we mention in passing that at the van Hove singularity the long-time behavior of the onsite propagator acquires an additional logarithm ($g_0(t) \sim \frac{\ln (t)}{wt}$) and each term in the expansion is more singular than the last. This amplifies the orthogonality catastrophe. However, it was shown~\cite{Gogolin1993} for the case of an abruptly appearing onsite potential that the phase shift obtained at half-filling is the same as determined using the naive extrapolating of the final result (as obtained for finite density of states) to the van-Hove point. Based on the assumption that this observation also holds in the more complicated situation of an abrupt change in the hopping amplitude, cf.~Eq.~\eqref{eq:HoppingChange}, we will thus perform all calculations away from half-filling and simply extrapolate the final results when we discuss the half-filled situation.

To obtain the solution nonperturbatively, we need a closed equation for the sum of all linked clusters. We first split the diagrams into two sets of equivalent diagrams $C(t) = C_{red}(t) + C_{blue}(t)$, each of which has a vertex of the aforementioned type and generates the other propagator with the swap $V\rightarrow \bar{V}$ (One can always do this due to the form of the potential: for every connected diagram there exists a corresponding diagram with the vertices flipped).

\begin{widetext}
\begingroup
\addtolength{\jot}{1em}
\begin{align}
    C_{blue}(t) &= \quad \includegraphics[width=0.8cm, valign=c]{Figures/Linked_clusters/fd1b.pdf}
    +\enspace\frac{1}{2} \left\{ \:\includegraphics[width=1.2cm, valign=c]{Figures/Linked_clusters/fd2bb.pdf} + \includegraphics[width=1.2cm, valign=c]{Figures/Linked_clusters/fd2br.pdf}\:\right\}
    +\enspace\frac{1}{3} \left\{ \: \includegraphics[width=1.2cm, valign=c]{Figures/Linked_clusters/fd3bbb.pdf} + 2\enspace\includegraphics[width=1.2cm, valign=c]{Figures/Linked_clusters/fd3bbr.pdf} + \includegraphics[width=1.2cm, valign=c]{Figures/Linked_clusters/fd3brr.pdf}\:\right\} + \cdots,\\
    C_{red}(t) &= \quad \includegraphics[width=0.8cm, valign=c]{Figures/Linked_clusters/fd1r.pdf}
    +\enspace\frac{1}{2} \left\{ \:\includegraphics[width=1.2cm, valign=c]{Figures/Linked_clusters/fd2rr.pdf} + \includegraphics[width=1.2cm, valign=c]{Figures/Linked_clusters/fd2br.pdf}\:\right\}
    +\enspace\frac{1}{3} \left\{ \: \includegraphics[width=1.2cm, valign=c]{Figures/Linked_clusters/fd3rrr.pdf} + 2\enspace\includegraphics[width=1.2cm, valign=c]{Figures/Linked_clusters/fd3brr.pdf} + \includegraphics[width=1.2cm, valign=c]{Figures/Linked_clusters/fd3bbr.pdf}\:\right\} + \cdots.
\end{align}
\endgroup
\end{widetext}

We now take $C_{blue}(t)$ and isolate the blue vertex. The remaining parts of the propagator then form diagrams for the hopping propagator in the time dependent potential $V(\tau)$. We can resum these diagrams to obtain the fully normalized electron hopping propagator. The $1/n$ symmetry factor, which can be taken care of by multiplying each vertex with $\lambda$ and integrating with respect to $\lambda$ at the end ($\int_0^1 d\lambda (\lambda V)^{n} = V^n/n$). The transient hopping propagator and the transient onsite propagator is given by

\begin{align}
    h_{\lambda}(\tau) &= -i\frac{\left\langle\mathcal{T} c_{\v r_2}(\tau) c^{\dagger}_{\v r_1}(\tau')\exp\{-i\lambda\int dt' \hat{V}(t')\}\right\rangle}{\left\langle\mathcal{T}\exp\{-i\lambda\int dt' \hat{V}(t')\}\right\rangle},\label{eq:hop}\\
    g_{\lambda}(\tau) &= -i\frac{\left\langle\mathcal{T} c_{\v r_1}(\tau) c^{\dagger}_{\v r_1}(\tau')\exp\{-i\lambda\int dt' \hat{V}(t')\}\right\rangle}{\left\langle\mathcal{T}\exp\{-i\lambda\int dt' \hat{V}(t')\}\right\rangle}.\label{eq:onsite}
\end{align}

The expression we desire is obtained by reintroducing the removed vertex and integrating the normalized propagator $h_{\lambda}$,

\begin{equation}
    C_{blue}(t) = -V \int_0^t \! d\tau \int_0^1 \! d\lambda \, h_\lambda(\tau,\tau+|t).
\end{equation}

The two transient propagator equations \eqref{eq:hop}, \eqref{eq:onsite} satisfy the coupled Dyson-Schwinger equations:
\begin{align}
    h_{\lambda}(\tau,\tau'|t,t') =& h_0(\tau-\tau') \nonumber \\ 
    +& \lambda V \int_{t}^{t'} h_0(\tau-\tau'')h_{\lambda}(\tau'',\tau'|t,t')\nonumber\\
    +& \lambda \bar{V}\int_{t}^{t'} g_0(\tau-\tau'')g_{\lambda}(\tau'',\tau'|t,t'),\\
    g_{\lambda}(\tau,\tau'|t,t') =& g_0(\tau-\tau') \nonumber \\
    +& \lambda{V} \int_{t}^{t'} g_0(\tau-\tau'')h_{\lambda}(\tau'',\tau'|t,t') \nonumber \\
    +& \lambda \bar{V}\int_{t}^{t'} h_0(\tau-\tau'')g_{\lambda}(\tau'',\tau'|t,t').
\end{align}

One gets an equal number of corresponding diagrams $C_{red}$ which one can obtain by making the replacement $V \rightarrow \bar{V}$ in $C_{blue}$. This gives us the full propagator,
\begin{equation}
    C(t) = -V\int_0^{t}\!d\tau \int_0^1\! d\lambda \, h_{\lambda}(\tau,\tau\!+|t) + (V\mapsto \bar{V}).
\end{equation}

In the energy representation,
\begin{equation}
    g_0(\omega) = \int dE \frac{\rho(E)}{\omega - E + i\delta \sign{(E)}},
\end{equation}
where $\rho(E)$ is the density of states,
\begin{align*}
    \rho(E) &= \int \frac{d^2 k}{(2\pi)^2}\delta(E - \{-2w(\cos{k_x} + \cos{k_y})-\mu\})\\
    &= \frac{1}{4\pi^2 w} K\left\{\left[1-\left(\frac{E+\mu}{4w}\right)^2\right]^{\frac{1}{2}}\right\}
\end{align*}
and $K(x)$ is the modified Bessel function. 
Similarly,
\begin{equation}
    h_0(\omega) = \int dE \frac{\rho^{h}(E)}{\omega - E + i\delta \sign{(E)}}
\end{equation}
with the ``hopping" density of states $\rho^{h}$ defined as
\begin{equation}
\begin{aligned}
    \rho^{h}(E) &= \int \frac{d^2 k}{(2\pi)^2} e^{ik_y}\delta(E - \{-2w(\cos{k_x} + \cos{k_y})-\mu\})\\
    &= -\frac{1}{4w}(E+\mu)\rho(E).
\end{aligned}
\end{equation}
The integral can be evaluated in closed form due to the rotational symmetry of the 0-flux band dispersion. 
Away from half-filling, the small frequency behavior of the propagators can be approximated as 
\begin{subequations}
    \begin{align}
        g_0(\omega) &\approx A - i\pi \rho(0) \sign{(\omega)},\\
        h_0(\omega) &\approx B - i\pi \rho^{h}(0) \sign{(\omega)},
    \end{align}
\end{subequations}
where $A, B$ are real constants. It is apparent that the long time behavior of both propagators are $g_0(t), h_0(t) \sim 1/t$.

To solve for the normalized propagator, we first decouple the Dyson-Schwinger equations. Repressing all arguments except for the first since that is the only one that changes in the calculations
\begin{alignat}{1}
    h_{\lambda} &= \frac{- \rho_0^h}{\tau - \tau'}  + \lambda \int_0^t d\tau''  \frac{V \rho_0^h h_\lambda(\tau'')+\bar V \rho_0 g_\lambda(\tau'')}{\tau''- \tau},\\
    g_{\lambda} &= \frac{- \rho_0}{\tau - \tau'}  + \lambda \int_0^t d\tau''  \frac{V \rho_0 h_\lambda(\tau'')+\bar V \rho_0^h g_\lambda(\tau'')}{\tau''- \tau}.
\end{alignat}

To decouple these equations, one must compute the left eigenvectors and eigenvalues of the matrix

\begin{equation}
\v M = \lambda
\begin{pmatrix}
V\rho_0^{h} & \bar{V}\rho_0\\
V\rho_0 & \bar{V}\rho_0^h
\end{pmatrix}
.
\end{equation}
The eigenvalue equation for $\v M$ gives
\begin{align}
    \mu^2 - \lambda\rho_0^h(V+\bar{V})\mu + |\lambda V|^2((\rho^h_0)^2 - (\rho_0)^2) = 0,\\
    \mu = \lambda\rho^h_0\Re{V} \pm \lambda\sqrt{|V|^2 (\rho_0)^2-(\Im{V}\rho_0^h)^2}. 
\end{align}
and the left eigenvector $\v a \equiv (1,a)$ is defined by the characteristic equation
\begin{equation}\label{eq:lefteig}
    \lambda V \rho_0^{h} + a \lambda V \rho_0 = \mu_{1,2}
\end{equation}
with the solutions
\begin{equation}
    a_{1,2} = \frac{1}{V\rho_0}(\mu_{1,2}/\lambda-V\rho_0^h).
\end{equation}

The decoupled equations are
\begin{align}
    \{h + a_{(1,2)}g\}_\lambda(\tau) = &-\frac{\rho^{h}_0 +a_{(1,2)}\rho_0}{\tau - \tau'}\nonumber\\
    &+\mu_{1,2} \int_0^t \! d\tau'' \, \frac{\{h + a_{(1,2)} g\}_\lambda(\tau'')}{\tau''-\tau}.
\end{align}

This expression simplifies nicely due to Eq.~\eqref{eq:lefteig}, leaving us with
\begin{align}
    \{h + a_{(1,2)}g\}_\lambda(\tau) = &-\frac{\mu_{1,2}}{\lambda V(\tau - \tau')} \nonumber \\
    &+\mu_{1,2} \int_0^t \! d\tau'' \, \frac{\{h + a_{(1,2)} g\}_\lambda(\tau'')}{\tau''-\tau}.
\end{align}

Using the textbook~\cite{GogolinTsvelikBook}, one can directly read off the solution:

\begin{align}
    C_{blue}(t) = -V\int_0^{t}\!d\tau \int_0^1\! d\lambda \, &\left(\frac{a_2}{a_2-a_1}\right)\{h+a_1 g\}_{\lambda}(\tau,\tau\!+|t)\nonumber\\ - &\left(\frac{a_1}{a_2-a_1}\right)\{h+a_2 g\}_{\lambda}(\tau,\tau\!+|t). 
\end{align}

Ignoring the leading order energy shift, the transient part gives us

\begin{equation}
     C_{blue}(t) = -\left\{\left(\frac{a_2}{a_2-a_1}\right) \left(\frac{\delta_1}{\pi}\right)^2+\left(\frac{-a_1}{a_2-a_1}\right)\left(\frac{\delta_2}{\pi}\right)^2\right\}\ln(t),
\end{equation}
where $\delta_i = \arctan(\pi \mu_i/\lambda)$. Now we find $C_{red}$. Note that while $a_i$ is not invariant under the switch $V\mapsto\bar{V}$, $\mu_{1,2}$ are. 
This allows for a dramatic cancellation in the full propagator, leaving us with
\begin{equation}
C(t) = -\left\{\left(\frac{\delta_1}{\pi}\right)^2+\left(\frac{\delta_2}{\pi}\right)^2\right\} \ln(t).
\end{equation}

Our final equation will be the above multiplied by $N_f$ to account for internal flavor or spin degrees of freedom, giving equation \eqref{eq:ndd}. 

\section{Luttinger surface}
\label{app:LuttingerSurface}
In this appendix we present details for the calculation of the Luttinger surface in the orthogonal metal, Sec.~\ref{sec:zeros}, starting from Eq.~\eqref{eq:FermionicG} of the main text.

We here use the notation $M_{\v q} = \eta h \sqrt{1  - 2 J [\cos(q_x)+\cos(q_y)]} \simeq \eta h (1 - \delta M_{\v q})$, where $\delta M_{\v q} = J [\cos(q_x)+\cos(q_y)]/ (\eta h)$. Using the notation 
\begin{equation}
G_{\rm OM}(\v q, z) =\frac{1}{z - \xi_0(\v q) - \text{sign}(\xi_0(\v q)) \eta h}
\end{equation}
we thus obtain
\begin{widetext}
\begin{align}
\mathcal G(\v p, z)& = a^2\int_{\rm BZ} (dq) \frac{\eta h}{M_{\v q - \v p}} \frac{1}{z - \xi_0(\v q) - \text{sign}(\xi_0(\v q)) M_{\v q- \v p}} \notag\\
&\simeq  \underbrace{a^2\int_{\rm BZ} (dq) G_{\rm OM}(\v q, z) }_{\mathcal G^{(0)}(z)} +\underbrace{a^2\int_{\rm BZ} (dq) \delta M_{\v p-\v q} \left [ G_{\rm OM}(\v q, z) - \eta h \ \text{sign}(\xi_0(\v q)) G_{\rm OM}(\v q, z)^2 \right ]}_{\mathcal G^{(1)}(\v p, z)}.
\end{align} 
\end{widetext}
We next use a trigonometric identity to rewrite $\delta M_{\v p-\v q}$ as a product of functions containing $\v p$ and $\v q$, exploit that integrals over $\sin(q_{x,y})$ vanish by symmetry, exploit $C_4$ symmetry and write
\begin{equation}
\cos(q_x) + \cos(q_y) = (\xi_0(\v q) + \mu)/2w,
\end{equation}
so that, using $G_{\rm OM}(\xi, z) = [z - \xi - \text{sign}(\xi) \eta h]^{-1}$,
\begin{align}
\mathcal G^{(0)}(z) & = a^2\int d\xi \rho(\xi+\mu) G_{\rm OM}(\xi, z), \\ 
\mathcal G^{(1)}(\v p, z)  &= \frac{J (\xi_0(\v p) + \mu)}{8 \eta h w^2} a^2\int d\xi \rho(\xi+ \mu) (\xi + \mu)\notag \\
&\times\left [ G_{\rm OM}(\xi, z) - \eta h \text{sign}(\xi) G_{\rm OM}(\xi, z)^2 \right ] .
\end{align}

We next define the following dimensionless integrals
\begin{equation}
I_n^{\pm}  = \int_0^\infty d \bar \xi \frac{\bar \xi^{n - 1}}{\nu_0} \frac{\rho(\eta h \bar \xi + \mu) \pm \rho(\eta h \bar \xi - \mu)}{(z/\eta h)^2 - (\bar \xi + 1)^2},
\end{equation}
where $\nu_0 = 1/(2\pi w a^2)$ and 
\begin{equation}
\rho(\xi) = \nu_0 \theta((2w)^2- \xi^2) \frac{4 w K(1 - (2w/\xi)^2)}{\pi \vert \xi \vert}
\end{equation}
is the particle-hole symmeric density of states of the square lattice centered around zero and $K(x)$ denotes a Bessel function of the second kind.

Using this notation we find
\begin{align}
\mathcal G^{(0)}(z) & = \frac{z}{\eta h}a^2 \nu_0 I_1^+ + \eta h \nu_0(I_2^- + I_3^-) \\ 
\mathcal G^{(1)}(\v p, z)  &= (\xi_0(\v p) + \mu) \frac{J }{8 w^2} a^2\nu_0 \times\notag \\
& \times \Big \lbrace I_2^+ + I_3^+ + \eta h \partial_z (I_2^- + I_3^-)\notag \\
& + \frac{\mu}{\eta h} [I_1^- + I_2^- + \eta h \partial_z (I_1^+ + I_2^+)] \notag\\
& +\frac{z}{\eta h} I_2^{-} + \partial_z (z I_2^+)\notag\\
& + \frac{\mu z}{(\eta h)^2} I_1^+ + \frac{\mu}{\eta h} \partial_z (z I_1^-)\Big \rbrace.
\end{align}

We concentrate on the vicinity of half filling, i.e. $\mu = 0$. We thus assume also $z$ to be small and drop all orders $z^2, \mu^2, z\mu, zJ$ etc.. In this limit, we only need the following combination of intgrals (in addition to $I_1^+ <0$) which are moreover related to each other by partial integration
\begin{align}
I_2^- +I_3^- &\simeq - (\mu/\eta h) I_1^+ ,\\
2I_2^+ + I_3^+ &\simeq - I_1^+ - \frac{2}{\nu_0 \eta h} <0.
\end{align}
This leads to 
\begin{align}
\mathcal G^{(0)}(z) & = \frac{z-\mu}{\eta h} a^2\nu_0 I_1^+ ,\\ 
\mathcal G^{(1)}(\v p, z)  &= -(\xi_0(\v p) + \mu) \frac{J }{8 w^2} a^2\nu_0 (I_1^+ + \frac{2 \pi w}{\eta h}),
\end{align}

which can be rewritten as
\begin{align}
\mathcal G(\v p, z) & \simeq a^2\frac{\nu_0 I_1^+}{\eta h} \left [z - \mu  + \frac{J \eta h}{4 w} \frac{I_1^+ + 2\pi w/ \eta h}{I_1^+} \sum_{\mu = x, y} \cos(q_\mu) \right ] \notag \\
&  =: -\frac{1}{\mathcal Z}  [z + \zeta_0( \v p) ],
\end{align}
which is the result quoted in Eq.~\eqref{eq:AllZerosMaintext} of the main text.
 
\section{Criticality of electric strings in the OM phase} \label{app:Criticality}

In this section we derive Eq.~\eqref{eq:Criticality}. Both coefficients $\lambda$ and $\lambda_0$ arise from the self-interaction of the critical electric strings. The four-point correlation function corresponding to $\lambda_0$ arises from vertices with 2 $e$-particles and 2 $e$-antiparticles interacting. Since we only require the connected part to calculate the non-linearity, we must subtract from Eq.~\eqref{eq:exp_e-particles} the pairwise distinct interaction of $e$-particle/antiparticle pairs, i.e., noninteracting electric strings,

\begin{align}\label{eq:4-vertex}
    V_0(\{\v r, \tau\}) &= J^4 \prod_{n=1}^4 \sum_{\v r} \delta_{\langle \v r_n, \v r \rangle}\nonumber \\
    & \bigg[\exp\left\{h\sum_{i=1}^n \tau_{p(i)} \Re \left[\omega^{-\sum_{j=1}^i q_{p(j)}}\left(1-\omega^{q_{p(i)}}\right)\right]\right\} \nonumber \\
    & - D(\tau_1,\tau_2)D(\tau_3,\tau_4)- D(\tau_1,\tau_3)D(\tau_2,\tau_4) \nonumber\\
    & - D(\tau_1,\tau_4)D(\tau_2,\tau_3)  \bigg].
\end{align}

Except for the special case $N=2$, one of the three terms in the pairwise interactions always evaluates to 0 due to Feynman rule 2. To find the vertex for the continuum field theory, we evaluate equation \eqref{eq:4-vertex} at zero frequency, taking $\beta \equiv 1/T $ as the infrared cutoff for the theory

\begin{align}
    \int_0^{\beta} \prod_{n=1}^N (d\tau_n) \, V_0(\{\v r, \tau\}) = &J^4 \sum_{\v r} \delta_{\langle \v r_i, \v r \rangle}\nonumber \\
    & \left[16\left(\frac{3}{(\eta h)^4} - \frac{2\beta}{(\eta h)^3}  + \frac{\beta^2}{2(\eta h)^2}\right)\right.\nonumber\\
    + &\left.\frac{8}{(\eta-2)^2}\left(\frac{4\eta-9}{(\eta h)^2} - 
    \beta\frac{2-\eta}{2(\eta h)^3}\right)\right]\nonumber \\
    - & 2 \frac{4J^4}{h^2\eta^2} \left(\beta - \frac{1}{\eta h}\right)^2 \nonumber \\
    \stackrel{O(\beta)}{=}& -\beta\frac{J^4}{(\eta h)^3}\left(16 + \frac{4}{\eta-2}\right)\sum_{\v r}\delta_{\langle \v r_i, \v r \rangle}.
\end{align}

The $\mathbb Z_N$ vertex has two terms, each of which corresponding to the creation ($\phi^N$) and annihilation ($\bar{\phi}^N$) of $N$ $e$-particles respectively. Unlike the 4-vertex, there are no lower order contributions that need to be subtracted from the connected part of the propagator. Since all charges are $1$ (for the creation term, $-1$ for the annihilation term), equation \eqref{eq:exp_e-particles} can be simplified to

\begin{align}
    V(\{\v r, \tau\}) &= J^N \prod_{n=1}^N \sum_{\v r} \delta_{\langle \v r_n, \v r \rangle}\nonumber\\
    &\exp\left\{h\sum_{=1}^N \tau_{p(i)} \Re \left[\omega^{-i}\left(1-\omega\right)\right]\right\}.
\end{align}
 
The corresponding imaginary-time integral with the IR cutoff at $\beta$ has no closed form in terms of $N$, but dimensionally, one can observe that the exponent is of the order of $J^N/h^{N-1}$.

The addition of the nearest-neighbor hopping perturbation $t$ introduces new operators that in the critical limit give rise to interactions between strings and fermionic excitations, captured by the action $\mathcal{S}_{\text{IA}}$. The interaction (local in time) takes the form
\begin{equation}
    V(\v r_c, \v r_{c^\dagger}; \v r_{D_1}, \v r_{\bar{D_2}}) = t\,\delta_{\v r_c, \v r_{D_1}} \delta_{\v r_{c^\dagger}, \v r_{\bar{D_2}}}\delta_{\langle \v r_{c^\dagger} , \v r_c \rangle}.
\end{equation}
\\
Upon rescaling the fields, we arrive at the coupling constant $\gamma \sim \eta h t a^2 [\cos(k_x) + \cos(k_y)]$ in the long-wavelength limit of the $\phi$ fields.

\section{Hadron formation in large $K\eta/g$ limit: ladder resummation}
\label{app:Hadron}

In this appendix, we present details on the formation of hadrons. We concentrate on the large $K\eta/g$ limit, in which we calculate the temperature at which $N-$particle propagators form bound states.

We now evaluate the $N-$ particle Green's function, see Fig.~\ref{fig:ladder}, and consider first the non-interacting limit:

\begin{align}
    \mathbb{L}(\v q,i\nu) &= T^{N-1} \prod_{i=1}^N \left[\int  (d \v p_i) \sum_{\epsilon_i} \frac{1}{i \epsilon_i - \xi(\v p_i)} \right.\nonumber\\
    & \hspace{6em} \times \left.\delta\left(\sum \v p_i - \v q\right) \delta_{\left(\sum \epsilon_i\right),\nu} \vphantom{\int}\right] \notag\\
    &=  \int \prod_{i=1}^N (d \v p_i) \int_0^\beta d\tau \,  e^{-i\nu\tau}\delta^{(2)}\left(\sum \v p_i - \v q\right) \nonumber\\
    & \hspace{6em} \times \left(T\sum_{\epsilon_i} \frac{e^{i\epsilon_i \tau}}{i\epsilon_i - \xi(\v p_i)}\right).
\end{align}

In the second step, we have used the identity $\int_0^\beta (d\tau) e^{i\omega_n\tau} = \beta\delta_{\omega_n,0}$ for a bosonic Matsubara frequency $\omega_n$. This necessitates that the Matsubara frequency $\nu$ is bosonic when N is even and fermionic when N is odd, capturing the statistic of the composite excitation. To calculate the sum over fermionic Matsubara frequencies $\epsilon_i$, we employ the usual contour integration

\begin{align}
    T\sum_{\epsilon_i} \frac{e^{i\epsilon_i  \tau}}{i\epsilon_i - \xi(\v p_i)} &= -\frac{1}{2\pi i} \oint dz \, \frac{e^{z \tau}}{z-\xi} \times\frac{1}{e^{\beta z}+1} \notag\\
    &= n_{\text{f}}(\xi) e^{\xi \tau}.
\end{align}

The integral becomes

\begin{align}
    \mathbb{L} &= \int_0^\beta d\tau \int \prod_{i} (dp_i) \, e^{-i\nu\tau}\left(n_{\text{f}}(\xi_i)e^{\xi_i \tau}\right) \delta^{(2)}\left(\sum_i \v p_i - \v q\right) \notag\\
    &= \int \prod_{i} (dp_i) \, \delta^{(2)}\left(\sum_i \v p_i - \v q\right) \, n_{f}(\xi_i) \,  \frac{e^{-i\nu\beta} e^{\sum_i \xi_i} - 1}{\sum_i \xi_i - i\nu}. \label{eq:Ladder}
\end{align}

For now, we assume even N. One would like to examine the behavior of this ladder propagator at $\mathbb{L}(\v q=0,i\nu=0)$; $\mathbb{L}_0$. For $N=2$, one recovers the standard result for the Cooper instability in a Fermi liquid. For a general $N$, the propagator can be rewritten as

\begin{equation}
    \mathbb{L}_0 = \int \prod_i (d\v p_i) \,\delta^{(2)}\left(\sum_i \v p_i\right) \left[ \prod_i n_{f}(-\xi_i) - \prod_i n_f(\xi_i)\right].
\end{equation}

The term in square brackets is non-zero only when all excitations are electrons ($\xi > 0$) or holes ($\xi < 0$). In the limit of low filling, we can assume a parabolic (and thereby radially symmetric) dispersion relation ($p_i = \sqrt{2m(\mu+\xi_i)}$). One then replaces the momentum-conserving Dirac delta by averaging it out over the $N$ momenta

\begin{equation}
    \left\langle \delta^{(2)} \left(\sum_i \v p_i \right)\right\rangle = 2\pi\int dr \; r \left(\prod_i J_0 (p_i r)\right).
\end{equation}

Performing a change of variables $\v p_i \rightarrow \xi_i, d\v p_i\rightarrow\rho d\xi_i$, where $\rho$ is the density of states,

\begin{align}
    \mathbb{L}_0 =& \rho^N \int_{-\mu} \prod_i d \xi_i \int dr (2\pi r) \frac{\prod_i J_0(p_i(\xi_i)r)}{\sum_i \xi_i} \nonumber \\
    &\left[ \prod_i n_{f}(-\xi_i) - \prod_i n_f(\xi_i)\right].
\end{align}

If the effective interaction between the fermions occurs in a range of energies around the Fermi surface: $\xi_i \in [-\omega_0,\omega_0]$ and $\omega_0 \ll \mu$, the integral over the Bessel functions can be estimated by setting the momentum as the Fermi momentum

\begin{equation}
    2\pi \int r \, dr \, (J_0 (p_f r))^N = \frac{4\pi}{p_f^2}c_N.
\end{equation}
While $c_2$ diverges (this is a manifestation of the Cooper logarithm), $c_N$ with $N \geq 3$ are all finite and fall off asymptotically as $1/N$ for large $N$. We highlight that the assumption of a retarded interaction $\omega_0 \ll \mu$ is not valid in the case of $g$ perturbations, instead all states up to the band edge are involved $\omega_0 \sim \mu$. Therefore, while our calculations can be expected to be parametrically correct, non-universal constants from ultraviolet processes may be missed.

At finite temperature, approximating the Fermi distribution function as a cutoff to the integral, one gets

\begin{equation}
    \mathbb{L}_0 \simeq -2\frac{\rho^N}{p_f^2} 4\pi c_N \underbrace{\int_T^{\omega_0}\prod_i d \xi_i \frac{1}{\sum_i \xi_i}}_{\omega_0^{N-1} \alpha_N(T/\omega_0)}. \label{eq:ladder_integral}
\end{equation}

For small $T$, the integral can be further approximated by a Taylor expansion to first order about 0: $\alpha_N(T/\omega_0) \approx \alpha_N(0) - N\frac{T}{\omega_0} \alpha_{N-1}(0)$. 

We expect that the fermions have an attractive interaction induced due to $H_g$ whose strength is $\mathcal{V}$. The effective interaction vertex in the $N$-particle channel, $\mathcal{V}_{\text{eff}}$ is therefore

\begin{equation}
    \mathcal{V}_{\text{eff}} = \frac{\mathcal{V}}{1-\mathcal{V}\mathbb{L}_0}.
\end{equation}

The onset of hadron instability is characterized by the development of a singularity in the vertex, which occurs when 
\begin{equation}\label{eq:hadron_singular}
    \mathbb{L}_0=\mathcal{V}^{-1}.
\end{equation} 
Defining the scaled interaction $v$ as

\begin{equation} \label{eq:scaled_v}
    v = -2\mathcal{V}\frac{\rho^N}{p_F^2} 4\pi c_N \omega_0^{N-1},
\end{equation}

see Fig.~\ref{fig:hadron-criticality-schematic}, the critical temperature $T_h$ where Eq.~\eqref{eq:hadron_singular} holds is 

\begin{equation}
    T_h = \underbrace{\frac{\omega_0 \alpha_N(0)}{N \alpha_{N-1}(0)}}_{T_h^{(0)}} \left(1-\frac{1}{v\alpha_N(0)}\right).
\end{equation}

For odd $N$, the ladder propagator must be evaluated at $\nu = \pm \pi\beta$, the minimum fermionic Matsubara frequency. This changes the integrand in Eq.~\eqref{eq:ladder_integral} to $[\sum_i \xi_i + (\pi T)^2/(\sum_i \xi_i)]^{-1}$. Up to first order in $T$, this change does not significantly alter the integral and one can safely proceed with the result for even $N$.

As mentioned in the main text, the bare interaction $\mathcal{V}$ is expected to scale as $(wg)^N/(\Delta E)^{2N-1}$. By substituting into Eq.~\eqref{eq:scaled_v} and assuming low filling, one can show that in terms of $w, g$ and the filling $\mu'$ (above zero filling), $v$ scales as

\begin{equation} 
    v \sim \left(\frac{g}{\Delta E_m}\right)^N \left(\frac{w}{\Delta E_m}\right) \left(\frac{\mu}{\Delta E_m}\right)^{N-2}. 
\end{equation}

\section{Connection to QED$_3$}

This appendix is devoted to making a connection between our $\mathbb Z_N$ model and QED$_3$. We first discuss the limit $N \rightarrow \infty$ of our model, in order to identify the coupling constants of both theories. Next, starting from a deconfining state of QED$_3$ in the presence of a Fermi surface, we study the RG relevance of $\mathbb Z_N$ perturbations and thereby provide technical details for Sec.~\ref{sec:RG} of the main text.

\subsection{$\mathbb{Z}_N$ to QED$_3$}

In this appendix we present the connection between the Hamiltonians of $\mathbb Z_N$ and $U(1)$ gauge theories in the limit  of large $h$. 

Concentrating on the limit $\phi = 0$ the gauge theory part of our model is 
\begin{align}
    \H &= -\frac{g}{2}\sum_{\v r}\sum_{\hat e\in\{x,y\}} (\tau_{\v r,\hat e}+\tau_{\v r,\hat e}^{\dagger}) - \frac{K}{2} \sum_{\v p}(B_{\v p} + B_{\v p}^{\dagger}).
\end{align}
We now consider the limit $N \rightarrow \infty$ and make connection to a compact QED$_3$ with gauge potential $a_{\v r, \hat e}$ on each link. Note the slight difference in notation to the main text, in which $\v b$ denotes a link, rather than the multi-index $(\v r,\hat e)$. This is to facilitate the interpretation as a line integral over the $U(1)$ potential.
By imposing the same algebra as Eq.~\eqref{eq:ZNAlgebra} we can identify

\begin{align}
\tau_{\v r,\hat e}+\tau_{\v r,\hat e}^{\dagger} & = \cos\left(\frac{2\pi}{N}i\frac{\partial}{\partial a_{\v r, \hat e}} \right ),\\
B_{\v p} + B_{\v p}^{\dagger} &= 2\cos(\Phi_{\v p}),
\end{align}

where $\Phi_{\v p}$ is the flux passing through the plaquette $\v p$, i.e., the directed sum of vector potentials $a_{\v r, \hat e}$ defined on each link. Thus the Hamiltonian becomes
\begin{align}
H &= -g\sum_{\v r} \sum_{\hat e\in\{x,y\}} \cos\left(\frac{2\pi}{N}i\frac{\partial}{\partial a_{\v r, \hat e}}\right) - K \sum_{\v p}\cos(\Phi_{\v p}).
\end{align}
 The first term contain the conjugate ``electric" field $E_{\v r, \hat e} \equiv i\frac{\partial}{\partial a_{\v r, \hat e}}$ endowed with the commutation relation $[E_{\v r, \hat e},a_{\v r', \hat e'}] = i\delta_{\v r, \v r'}\delta_{\hat e,\hat e'}$. We carefully take the continuum limit and expand the Hamiltonian about its minima. After sorting out the factors of the lattice constant inside the vector potential, we obtain the Hamiltonian

\begin{equation}
    H = \int\! d^2x \, \frac{g}{2} \left(\frac{2\pi}{N}\right)^2 (E_x^2 + E_y^2) + \frac{K a^2}{2} B^2.
\end{equation}

Given this Hamiltonian, the corresponding QED$_3$ action is

\begin{equation}
    S = \int \! dt \, d^2x \, \left[\frac{N^2}{2g(2\pi)^2} \dot{{A}}^2 - \frac{K a^2}{2} (\nabla\times{A})^2\right],
\end{equation}
\begin{equation}
    S = \frac{K a^2}{2} \int \! dt \, d^2x \, \sum_{\mu = 1}^3 {A}_\mu \left(\frac{N^2}{gK(2\pi a)^2}\partial^2_t - \nabla^2 \right) {A}_\mu,
\end{equation}

with the speed of light $c = \sqrt{gK}(2\pi a/N)$. The fermionic part of the Hamiltonian is coupled to this $\mathbb{Z}_N$ gauge field as:

\begin{equation}
    \H_f = -w \sum_{\v r} \sum_{\hat e \in \{x,y\}} c^{\dagger}_{\v r} e^{-ia_{\v r,\hat e}} c_{\v r +\hat e} + \text{H.c.}
\end{equation}
{
This corresponds to a Peierls substitution where the line integral of the $U(1)$ gauge potential ($a_{\v r, \hat e} = \int_{\v r}^{\v r+\hat e} A(\v r')\cdot d\v r'$) modifies the hopping between sites. Taking the continuum limit recovers the Hamiltonian of fermions coupled to a $U(1)$ gauge field

\begin{equation}
    \H_w = \frac{w a^2}{2} \int \! d^2 x \, \psi^{\dagger}_{\v x} (-i\nabla - {A})^2 \psi_{\v x} .
\end{equation}
}

\subsection{RG flow calculations for $\mathbb{Z}_N$ perturbation in $\text{QED}_3$ \label{app:U(1)_RG}}

We now determine the regime in which the $\mathbb{Z}_N$ physics dominates the QED$_3$ action. Namely, we ask when is the action 
\begin{equation}
    S = \int\! d\tau \, d^2 x \sum_{\mu = 1}^3 \, A_\mu \left[-\nabla^2 - \partial_\tau^2 \right] A_\mu+ S_{\text{fermions}}\\
\end{equation}
stable to the $\mathbb{Z}_N$ perturbation 
\begin{equation}
\delta S=\int\! d\tau \, d^2 x \, \lambda \left[ \cos(N A_x) + \cos(N A_y)\right].
\end{equation} 
We work in the Coulomb gauge ($\nabla\cdot\v A=0$), where the propagator takes the form~\cite{NaveLee2007}
\begin{equation}
    \mathbb{D}_T(Q\equiv(\nu,\v q)) = \frac{1}{\gamma\frac{|\nu|}{q}+q^2 \chi_D},
\end{equation}
where on the bare level $\gamma \sim N_f k_F$ and $\chi_D \sim N_f/m$.

Formally, this propagator is obtained by introducing N$_f$ flavors of fermions followed by a large-$N_f$ {(and~\cite{MrossSenthil2010} an $\epsilon$-)} expansion. The action takes the form

\begin{equation}
    S_{T} = \int\frac{d\nu \, d^2 \v q}{(2\pi)^3} \left|A_T(Q)\right|^2 \left[\gamma \frac{|\nu|}{q} + q^2 \chi_D \right]/a^2.
\end{equation}

Here, $A_{T}(Q)$ is the transverse part of the gauge-fixed vector potential ($\v A(Q) = A_T(Q)[\hat{z}\times\v{q}]$) and $a$ is the lattice spacing. Our RG procedure now splits the phase space into two parts, the slow and fast fluctuations:

\begin{align}
    \sqrt{(\gamma\nu)^2 + (\chi_D q^3)^2} &\in \left[0, \frac{\Lambda}{b_\nu}\right] \quad \text{(slow)}\\
     \sqrt{(\gamma\nu)^2 + (\chi_D q^3)^2} &\in \left(\frac{\Lambda}{b_\nu}, \Lambda\right]  \quad \text{(fast)}
\end{align}

where $\Lambda$ is the UV cutoff. The quadratic action decouples in the slow and fast fluctuations. We discard the fast fluctuations and rescale the remaining action,

\begin{equation}
    \nu' \rightarrow b_\nu \nu, \quad \v{q'} \rightarrow b \v{q}.
\end{equation}

One can immediately see that for the propagator to retain its standard form, $b_\nu = b^3$, which motivated the choice for the momentum shells. 
Assuming that the $U(1)$ gauge field scales as

\begin{equation}
    \left|A_T^<\left(\frac{\nu}{b^3}, \frac{\v q}{b}\right)\right| = b^5 \left|A_T(\nu, \v{q})\right|,
\end{equation}

the action thus scales as

\begin{equation}
    S_T^< = b^3 \int_{[0,\Lambda]} \frac{d\nu \, d^2 \v{q}}{(2\pi)^3} \left|A_T(Q)\right|^2 \left[\gamma \frac{|\nu|}{q} + q^2 \chi_D \right]/a^2.
\end{equation}

Under this momentum-shell RG procedure, the $\mathbb{Z}_N$ perturbation scaling is
\begin{align}
    \delta S^< &= -\lambda \int d\tau\, d^2\v x \;\sum_{i=\{x,y\}} \frac{1}{2}\left\langle e^{iN(A_i^> + A_i^<)}+e^{-iN(A_i^> + A_i^<)}\right\rangle_> \notag\\
    &= -\lambda \int d\tau\, d^2\v x \; \sum_{i=\{x,y\}} \cos\left({N}A_i^<\right)e^{-\frac{N^2}{2}\langle (A_i^>)^2 \rangle}
\end{align}

\begin{align}
    \langle (A_i^>)^2 \rangle &= \frac{1}{(2\pi)^3}\int_{\text{fast}} d\nu\, d^2\v{q}\; a^2 \frac{[\hat{z}\times \v q]_i^2}{\gamma\frac{|\nu|}{q} + \chi_D q^2}\notag\\
    &= \frac{a^2}{2(2\pi)^3} \int_{\text{fast}} d\nu\, d^2\v{q}\; \frac{1}{\gamma\frac{|\nu|}{q} + \chi_D q^2}\notag\\
    &= \frac{a^2}{(2\pi)^3} \int_{\text{fast, }\nu>0} d\nu\, d^2\v{q} \;\frac{1}{\gamma\frac{\nu}{q} + \chi_D q^2}.
\end{align}

Switching to polar coordinates and making the substitution

\begin{equation}
    \bar{\nu} = \gamma \nu,\quad \bar{q} = (\chi_D)^{1/3} q,
\end{equation}

the integral is simplified to
\begin{align}
    \langle (A_i^>)^2 \rangle &= \frac{a^2}{(2\pi)^2}\frac{1}{\gamma\chi_D}\int_{\text{fast}, \bar\nu,\bar{q} > 0} d\bar\nu\, d\bar q\; \frac{\bar{q}^2}{\bar\nu + \bar{q}^3}\notag\\
    &= \frac{a^2}{12\pi^2}\frac{1}{\gamma\chi_D}\int_{\substack{\bar{\nu},\bar{x} > 0 \\ \sqrt{\bar\nu^2 + \bar{x}^2}\in{(\Lambda/b^3, \Lambda]}}}\hspace{-1em} d\bar\nu\, d\bar x\; \frac{1}{\bar\nu + \bar{x}}.
\end{align}
where $\bar x = \bar {q}^3$
Making another polar transformation ($\bar{\nu} = r \sin(\theta), \bar{x} = r \cos(\theta)$),
\begin{align}
    \langle (A_i^>)^2 \rangle &= \frac{a^2}{12\pi^2}\frac{1}{\gamma\chi_D}\int_{\Lambda/b^3}^{\Lambda} {dr} \int_0^{\frac{\pi}{2}}\frac{d\theta}{\sin(\theta) + \cos(\theta)}\notag\\
    &= \frac{\tilde{\Lambda}}{{\chi_D\gamma}} (1-b^{-3}),\label{eq:a_RG}
\end{align}
where $\tilde \Lambda \propto \Lambda$ is a velocity corresponding to the UV cut-off and has hence the same dimensions as $\gamma \chi_D$.

From Eq.~\eqref{eq:a_RG}, the $\mathbb{Z}_N$ perturbation scales as

\begin{align}
     \delta S^{<} = &-\lambda \exp\left\{\frac{N^2}{2}\frac{\tilde{\Lambda}}{{\chi_D\gamma}} (1-b^{-3})\right\} \nonumber\\
     &\hspace{5em}\int d\tau\, d^2\v x \sum_{i=\{x,y\}} \cos\left({N}A_i^<\right) \label{eq:lambda_RG}.
\end{align}

Examining Eq.~\eqref{eq:lambda_RG}, $\lambda$ scales under the RG procedure as

\begin{equation}
    \lambda \rightarrow b^5 \exp\left\{-\frac{N^2}{\tilde\chi \tilde\gamma} (1-b^{-3})\right\}\, \lambda 
\end{equation}

where the cutoff has been absorbed into a redefinition of couplings denoted $\tilde \chi$ and $\tilde \gamma$. The tree-level RG flow equations are

\begin{align}
        \frac{d\tilde\chi}{d \ln b} &= 3\tilde\chi, \label{eq:chi_flow}\\ 
        \frac{d\tilde\gamma}{d \ln b} &= 3\tilde\gamma, \label{eq:gamma_flow}\\
        \frac{d \lambda}{d \ln b} &= \left(5 - 3\frac{N^2}{\tilde\chi \tilde\gamma}\right)\lambda.\label{eq:lambda_flow}
\end{align}

Introducing the dimensionless coupling $\kappa = \tilde\gamma \tilde\chi \stackrel{\text{bare value}}{=} N_f^2 v_F/\tilde \Lambda$, we obtain the RG equations presented in Eq.~\eqref{eq:RGMaintex} of the main text. 
One can see that for however large $N$, since $\kappa$ is a relevant parameter under RG scaling, the $\mathbb{Z}_N$ parameter $\lambda$ always becomes relevant. Thus, there is no $U(1)$ phase in the system.

We can, however, estimate the regime in which the system approximately behaves as QED$_3$. From the tree-level flow equations \eqref{eq:chi_flow}, \eqref{eq:gamma_flow},

\begin{equation}
    \frac{d\kappa}{d \ln(L/l)} = 6\kappa \Rightarrow \kappa = \kappa_0 \left(\frac{L}{l}\right)^6.
\end{equation}

The $\mathbb{Z}_N$ perturbation scaling $\lambda$ does not grow as long as $\kappa = \frac{3N^2}{5}$. This endows a critical size to the system $L_{\text{crit}} = l\left(\frac{3 N^2}{5 \kappa_0}\right)^{1/6}$. For $L < L_{\text{crit}}$, a $U(1)$ phase can be well-approximated.

It is natural to assume $\tilde \Lambda \sim v_F$, so that the bare $\kappa = N_f^2$ (up to a coefficient of order one) and to set $l = 1/k_F$.

\bibliography{GaugeBibliography}

\begin{thebibliography}{91}%
\makeatletter
\providecommand \@ifxundefined [1]{%
 \@ifx{#1\undefined}
}%
\providecommand \@ifnum [1]{%
 \ifnum #1\expandafter \@firstoftwo
 \else \expandafter \@secondoftwo
 \fi
}%
\providecommand \@ifx [1]{%
 \ifx #1\expandafter \@firstoftwo
 \else \expandafter \@secondoftwo
 \fi
}%
\providecommand \natexlab [1]{#1}%
\providecommand \enquote  [1]{``#1''}%
\providecommand \bibnamefont  [1]{#1}%
\providecommand \bibfnamefont [1]{#1}%
\providecommand \citenamefont [1]{#1}%
\providecommand \href@noop [0]{\@secondoftwo}%
\providecommand \href [0]{\begingroup \@sanitize@url \@href}%
\providecommand \@href[1]{\@@startlink{#1}\@@href}%
\providecommand \@@href[1]{\endgroup#1\@@endlink}%
\providecommand \@sanitize@url [0]{\catcode `\\12\catcode `\$12\catcode
  `\&12\catcode `\#12\catcode `\^12\catcode `\_12\catcode `\%12\relax}%
\providecommand \@@startlink[1]{}%
\providecommand \@@endlink[0]{}%
\providecommand \url  [0]{\begingroup\@sanitize@url \@url }%
\providecommand \@url [1]{\endgroup\@href {#1}{\urlprefix }}%
\providecommand \urlprefix  [0]{URL }%
\providecommand \Eprint [0]{\href }%
\providecommand \doibase [0]{http://dx.doi.org/}%
\providecommand \selectlanguage [0]{\@gobble}%
\providecommand \bibinfo  [0]{\@secondoftwo}%
\providecommand \bibfield  [0]{\@secondoftwo}%
\providecommand \translation [1]{[#1]}%
\providecommand \BibitemOpen [0]{}%
\providecommand \bibitemStop [0]{}%
\providecommand \bibitemNoStop [0]{.\EOS\space}%
\providecommand \EOS [0]{\spacefactor3000\relax}%
\providecommand \BibitemShut  [1]{\csname bibitem#1\endcsname}%
\let\auto@bib@innerbib\@empty
\bibitem [{\citenamefont {Wegner}(1971)}]{Wegner1971}%
  \BibitemOpen
  \bibfield  {author} {\bibinfo {author} {\bibfnamefont {Franz~J}\ \bibnamefont
  {Wegner}},\ }\bibfield  {title} {\enquote {\bibinfo {title} {Duality in
  generalized {I}sing models and phase transitions without local order
  parameters},}\ }\href@noop {} {\bibfield  {journal} {\bibinfo  {journal}
  {Journal of Mathematical Physics}\ }\textbf {\bibinfo {volume} {12}},\
  \bibinfo {pages} {2259--2272} (\bibinfo {year} {1971})}\BibitemShut {NoStop}%
\bibitem [{\citenamefont {Wilson}(1974)}]{Wilson1974}%
  \BibitemOpen
  \bibfield  {author} {\bibinfo {author} {\bibfnamefont {Kenneth~G.}\
  \bibnamefont {Wilson}},\ }\bibfield  {title} {\enquote {\bibinfo {title}
  {Confinement of quarks},}\ }\href {\doibase 10.1103/PhysRevD.10.2445}
  {\bibfield  {journal} {\bibinfo  {journal} {Phys. Rev. D}\ }\textbf {\bibinfo
  {volume} {10}},\ \bibinfo {pages} {2445--2459} (\bibinfo {year}
  {1974})}\BibitemShut {NoStop}%
\bibitem [{\citenamefont {Kogut}(1979)}]{Kogut1979}%
  \BibitemOpen
  \bibfield  {author} {\bibinfo {author} {\bibfnamefont {John~B.}\ \bibnamefont
  {Kogut}},\ }\bibfield  {title} {\enquote {\bibinfo {title} {An introduction
  to lattice gauge theory and spin systems},}\ }\href {\doibase
  10.1103/RevModPhys.51.659} {\bibfield  {journal} {\bibinfo  {journal} {Rev.
  Mod. Phys.}\ }\textbf {\bibinfo {volume} {51}},\ \bibinfo {pages} {659--713}
  (\bibinfo {year} {1979})}\BibitemShut {NoStop}%
\bibitem [{\citenamefont {Brice\~no}\ \emph {et~al.}(2018)\citenamefont
  {Brice\~no}, \citenamefont {Dudek},\ and\ \citenamefont
  {Young}}]{BrinoYoung2018}%
  \BibitemOpen
  \bibfield  {author} {\bibinfo {author} {\bibfnamefont {Ra\'ul~A.}\
  \bibnamefont {Brice\~no}}, \bibinfo {author} {\bibfnamefont {Jozef~J.}\
  \bibnamefont {Dudek}}, \ and\ \bibinfo {author} {\bibfnamefont {Ross~D.}\
  \bibnamefont {Young}},\ }\bibfield  {title} {\enquote {\bibinfo {title}
  {Scattering processes and resonances from lattice {QCD}},}\ }\href {\doibase
  10.1103/RevModPhys.90.025001} {\bibfield  {journal} {\bibinfo  {journal}
  {Rev. Mod. Phys.}\ }\textbf {\bibinfo {volume} {90}},\ \bibinfo {pages}
  {025001} (\bibinfo {year} {2018})}\BibitemShut {NoStop}%
\bibitem [{\citenamefont {Sachdev}(2018)}]{Sachdev2018}%
  \BibitemOpen
  \bibfield  {author} {\bibinfo {author} {\bibfnamefont {Subir}\ \bibnamefont
  {Sachdev}},\ }\bibfield  {title} {\enquote {\bibinfo {title} {Topological
  order, emergent gauge fields, and {F}ermi surface reconstruction},}\
  }\href@noop {} {\bibfield  {journal} {\bibinfo  {journal} {Reports on
  Progress in Physics}\ }\textbf {\bibinfo {volume} {82}},\ \bibinfo {pages}
  {014001} (\bibinfo {year} {2018})}\BibitemShut {NoStop}%
\bibitem [{\citenamefont {Savary}\ and\ \citenamefont
  {Balents}(2016)}]{SavaryBalents2016}%
  \BibitemOpen
  \bibfield  {author} {\bibinfo {author} {\bibfnamefont {Lucile}\ \bibnamefont
  {Savary}}\ and\ \bibinfo {author} {\bibfnamefont {Leon}\ \bibnamefont
  {Balents}},\ }\bibfield  {title} {\enquote {\bibinfo {title} {Quantum spin
  liquids: a review},}\ }\href@noop {} {\bibfield  {journal} {\bibinfo
  {journal} {Reports on Progress in Physics}\ }\textbf {\bibinfo {volume}
  {80}},\ \bibinfo {pages} {016502} (\bibinfo {year} {2016})}\BibitemShut
  {NoStop}%
\bibitem [{\citenamefont {Wen}(1990)}]{Wen1990}%
  \BibitemOpen
  \bibfield  {author} {\bibinfo {author} {\bibfnamefont {Xiao-Gang}\
  \bibnamefont {Wen}},\ }\bibfield  {title} {\enquote {\bibinfo {title}
  {Topological orders in rigid states},}\ }\href@noop {} {\bibfield  {journal}
  {\bibinfo  {journal} {International Journal of Modern Physics B}\ }\textbf
  {\bibinfo {volume} {4}},\ \bibinfo {pages} {239--271} (\bibinfo {year}
  {1990})}\BibitemShut {NoStop}%
\bibitem [{\citenamefont {Kitaev}(2003)}]{Kitaev2003}%
  \BibitemOpen
  \bibfield  {author} {\bibinfo {author} {\bibfnamefont {A~Yu}\ \bibnamefont
  {Kitaev}},\ }\bibfield  {title} {\enquote {\bibinfo {title} {Fault-tolerant
  quantum computation by anyons},}\ }\href@noop {} {\bibfield  {journal}
  {\bibinfo  {journal} {Annals of Physics}\ }\textbf {\bibinfo {volume}
  {303}},\ \bibinfo {pages} {2--30} (\bibinfo {year} {2003})}\BibitemShut
  {NoStop}%
\bibitem [{\citenamefont {Cobanera}\ \emph {et~al.}(2016)\citenamefont
  {Cobanera}, \citenamefont {Ulrich},\ and\ \citenamefont
  {Hassler}}]{CobaneraHassler2016}%
  \BibitemOpen
  \bibfield  {author} {\bibinfo {author} {\bibfnamefont {Emilio}\ \bibnamefont
  {Cobanera}}, \bibinfo {author} {\bibfnamefont {Jascha}\ \bibnamefont
  {Ulrich}}, \ and\ \bibinfo {author} {\bibfnamefont {Fabian}\ \bibnamefont
  {Hassler}},\ }\bibfield  {title} {\enquote {\bibinfo {title} {Changing
  anyonic ground degeneracy with engineered gauge fields},}\ }\href {\doibase
  10.1103/PhysRevB.94.125434} {\bibfield  {journal} {\bibinfo  {journal} {Phys.
  Rev. B}\ }\textbf {\bibinfo {volume} {94}},\ \bibinfo {pages} {125434}
  (\bibinfo {year} {2016})}\BibitemShut {NoStop}%
\bibitem [{\citenamefont {Zohar}\ \emph
  {et~al.}(2017{\natexlab{a}})\citenamefont {Zohar}, \citenamefont {Farace},
  \citenamefont {Reznik},\ and\ \citenamefont {Cirac}}]{ZoharCirac2017}%
  \BibitemOpen
  \bibfield  {author} {\bibinfo {author} {\bibfnamefont {Erez}\ \bibnamefont
  {Zohar}}, \bibinfo {author} {\bibfnamefont {Alessandro}\ \bibnamefont
  {Farace}}, \bibinfo {author} {\bibfnamefont {Benni}\ \bibnamefont {Reznik}},
  \ and\ \bibinfo {author} {\bibfnamefont {J.~Ignacio}\ \bibnamefont {Cirac}},\
  }\bibfield  {title} {\enquote {\bibinfo {title} {Digital lattice gauge
  theories},}\ }\href {\doibase 10.1103/PhysRevA.95.023604} {\bibfield
  {journal} {\bibinfo  {journal} {Phys. Rev. A}\ }\textbf {\bibinfo {volume}
  {95}},\ \bibinfo {pages} {023604} (\bibinfo {year}
  {2017}{\natexlab{a}})}\BibitemShut {NoStop}%
\bibitem [{\citenamefont {Zohar}\ \emph
  {et~al.}(2017{\natexlab{b}})\citenamefont {Zohar}, \citenamefont {Farace},
  \citenamefont {Reznik},\ and\ \citenamefont {Cirac}}]{ZoharCirac2017b}%
  \BibitemOpen
  \bibfield  {author} {\bibinfo {author} {\bibfnamefont {Erez}\ \bibnamefont
  {Zohar}}, \bibinfo {author} {\bibfnamefont {Alessandro}\ \bibnamefont
  {Farace}}, \bibinfo {author} {\bibfnamefont {Benni}\ \bibnamefont {Reznik}},
  \ and\ \bibinfo {author} {\bibfnamefont {J.~Ignacio}\ \bibnamefont {Cirac}},\
  }\bibfield  {title} {\enquote {\bibinfo {title} {Digital quantum simulation
  of $\mathbb{Z}_{2}$ lattice gauge theories with dynamical fermionic
  matter},}\ }\href {\doibase 10.1103/PhysRevLett.118.070501} {\bibfield
  {journal} {\bibinfo  {journal} {Phys. Rev. Lett.}\ }\textbf {\bibinfo
  {volume} {118}},\ \bibinfo {pages} {070501} (\bibinfo {year}
  {2017}{\natexlab{b}})}\BibitemShut {NoStop}%
\bibitem [{\citenamefont {Emonts}\ \emph {et~al.}(2020)\citenamefont {Emonts},
  \citenamefont {Ba\~nuls}, \citenamefont {Cirac},\ and\ \citenamefont
  {Zohar}}]{EmontsZohar2020}%
  \BibitemOpen
  \bibfield  {author} {\bibinfo {author} {\bibfnamefont {Patrick}\ \bibnamefont
  {Emonts}}, \bibinfo {author} {\bibfnamefont {Mari~Carmen}\ \bibnamefont
  {Ba\~nuls}}, \bibinfo {author} {\bibfnamefont {Ignacio}\ \bibnamefont
  {Cirac}}, \ and\ \bibinfo {author} {\bibfnamefont {Erez}\ \bibnamefont
  {Zohar}},\ }\bibfield  {title} {\enquote {\bibinfo {title} {Variational
  {M}onte {C}arlo simulation with tensor networks of a pure $\mathbb{Z}_{3}$
  gauge theory in $(2+1)\mathrm{D}$},}\ }\href {\doibase
  10.1103/PhysRevD.102.074501} {\bibfield  {journal} {\bibinfo  {journal}
  {Phys. Rev. D}\ }\textbf {\bibinfo {volume} {102}},\ \bibinfo {pages}
  {074501} (\bibinfo {year} {2020})}\BibitemShut {NoStop}%
\bibitem [{\citenamefont {Cardarelli}\ \emph {et~al.}(2020)\citenamefont
  {Cardarelli}, \citenamefont {Greschner},\ and\ \citenamefont
  {Santos}}]{CardarelliSantos2020}%
  \BibitemOpen
  \bibfield  {author} {\bibinfo {author} {\bibfnamefont {Lorenzo}\ \bibnamefont
  {Cardarelli}}, \bibinfo {author} {\bibfnamefont {Sebastian}\ \bibnamefont
  {Greschner}}, \ and\ \bibinfo {author} {\bibfnamefont {Luis}\ \bibnamefont
  {Santos}},\ }\bibfield  {title} {\enquote {\bibinfo {title} {Deconfining
  disordered phase in two-dimensional quantum link models},}\ }\href {\doibase
  10.1103/PhysRevLett.124.123601} {\bibfield  {journal} {\bibinfo  {journal}
  {Phys. Rev. Lett.}\ }\textbf {\bibinfo {volume} {124}},\ \bibinfo {pages}
  {123601} (\bibinfo {year} {2020})}\BibitemShut {NoStop}%
\bibitem [{\citenamefont {Robaina}\ \emph {et~al.}(2021)\citenamefont
  {Robaina}, \citenamefont {Ba\~nuls},\ and\ \citenamefont
  {Cirac}}]{RobainaCirac2021}%
  \BibitemOpen
  \bibfield  {author} {\bibinfo {author} {\bibfnamefont {Daniel}\ \bibnamefont
  {Robaina}}, \bibinfo {author} {\bibfnamefont {Mari~Carmen}\ \bibnamefont
  {Ba\~nuls}}, \ and\ \bibinfo {author} {\bibfnamefont {J.~Ignacio}\
  \bibnamefont {Cirac}},\ }\bibfield  {title} {\enquote {\bibinfo {title}
  {Simulating $2+1\mathrm{D}$ ${Z}_{3}$ lattice gauge theory with an infinite
  projected entangled-pair state},}\ }\href {\doibase
  10.1103/PhysRevLett.126.050401} {\bibfield  {journal} {\bibinfo  {journal}
  {Phys. Rev. Lett.}\ }\textbf {\bibinfo {volume} {126}},\ \bibinfo {pages}
  {050401} (\bibinfo {year} {2021})}\BibitemShut {NoStop}%
\bibitem [{\citenamefont {Zohar}\ \emph {et~al.}(2015)\citenamefont {Zohar},
  \citenamefont {Cirac},\ and\ \citenamefont {Reznik}}]{ZoharReznik2016}%
  \BibitemOpen
  \bibfield  {author} {\bibinfo {author} {\bibfnamefont {Erez}\ \bibnamefont
  {Zohar}}, \bibinfo {author} {\bibfnamefont {J~Ignacio}\ \bibnamefont
  {Cirac}}, \ and\ \bibinfo {author} {\bibfnamefont {Benni}\ \bibnamefont
  {Reznik}},\ }\bibfield  {title} {\enquote {\bibinfo {title} {Quantum
  simulations of lattice gauge theories using ultracold atoms in optical
  lattices},}\ }\href {\doibase 10.1088/0034-4885/79/1/014401} {\bibfield
  {journal} {\bibinfo  {journal} {Reports on Progress in Physics}\ }\textbf
  {\bibinfo {volume} {79}},\ \bibinfo {pages} {014401} (\bibinfo {year}
  {2015})}\BibitemShut {NoStop}%
\bibitem [{\citenamefont {Ercolessi}\ \emph {et~al.}(2018)\citenamefont
  {Ercolessi}, \citenamefont {Facchi}, \citenamefont {Magnifico}, \citenamefont
  {Pascazio},\ and\ \citenamefont {Pepe}}]{ErcolessiPepe2018}%
  \BibitemOpen
  \bibfield  {author} {\bibinfo {author} {\bibfnamefont {Elisa}\ \bibnamefont
  {Ercolessi}}, \bibinfo {author} {\bibfnamefont {Paolo}\ \bibnamefont
  {Facchi}}, \bibinfo {author} {\bibfnamefont {Giuseppe}\ \bibnamefont
  {Magnifico}}, \bibinfo {author} {\bibfnamefont {Saverio}\ \bibnamefont
  {Pascazio}}, \ and\ \bibinfo {author} {\bibfnamefont {Francesco~V.}\
  \bibnamefont {Pepe}},\ }\bibfield  {title} {\enquote {\bibinfo {title} {Phase
  transitions in ${Z}_{n}$ gauge models: Towards quantum simulations of the
  {S}chwinger-{W}eyl {QED}},}\ }\href {\doibase 10.1103/PhysRevD.98.074503}
  {\bibfield  {journal} {\bibinfo  {journal} {Phys. Rev. D}\ }\textbf {\bibinfo
  {volume} {98}},\ \bibinfo {pages} {074503} (\bibinfo {year}
  {2018})}\BibitemShut {NoStop}%
\bibitem [{\citenamefont {Celi}\ \emph {et~al.}(2020)\citenamefont {Celi},
  \citenamefont {Vermersch}, \citenamefont {Viyuela}, \citenamefont {Pichler},
  \citenamefont {Lukin},\ and\ \citenamefont {Zoller}}]{CeliZoller2020}%
  \BibitemOpen
  \bibfield  {author} {\bibinfo {author} {\bibfnamefont {Alessio}\ \bibnamefont
  {Celi}}, \bibinfo {author} {\bibfnamefont {Beno\^{\i}t}\ \bibnamefont
  {Vermersch}}, \bibinfo {author} {\bibfnamefont {Oscar}\ \bibnamefont
  {Viyuela}}, \bibinfo {author} {\bibfnamefont {Hannes}\ \bibnamefont
  {Pichler}}, \bibinfo {author} {\bibfnamefont {Mikhail~D.}\ \bibnamefont
  {Lukin}}, \ and\ \bibinfo {author} {\bibfnamefont {Peter}\ \bibnamefont
  {Zoller}},\ }\bibfield  {title} {\enquote {\bibinfo {title} {Emerging
  two-dimensional gauge theories in {R}ydberg configurable arrays},}\ }\href
  {\doibase 10.1103/PhysRevX.10.021057} {\bibfield  {journal} {\bibinfo
  {journal} {Phys. Rev. X}\ }\textbf {\bibinfo {volume} {10}},\ \bibinfo
  {pages} {021057} (\bibinfo {year} {2020})}\BibitemShut {NoStop}%
\bibitem [{\citenamefont {Notarnicola}\ \emph {et~al.}(2020)\citenamefont
  {Notarnicola}, \citenamefont {Collura},\ and\ \citenamefont
  {Montangero}}]{NotarnicolaNontagero2020}%
  \BibitemOpen
  \bibfield  {author} {\bibinfo {author} {\bibfnamefont {Simone}\ \bibnamefont
  {Notarnicola}}, \bibinfo {author} {\bibfnamefont {Mario}\ \bibnamefont
  {Collura}}, \ and\ \bibinfo {author} {\bibfnamefont {Simone}\ \bibnamefont
  {Montangero}},\ }\bibfield  {title} {\enquote {\bibinfo {title}
  {Real-time-dynamics quantum simulation of $(1+1)\text{-dimensional}$ lattice
  {QED} with {R}ydberg atoms},}\ }\href {\doibase
  10.1103/PhysRevResearch.2.013288} {\bibfield  {journal} {\bibinfo  {journal}
  {Phys. Rev. Res.}\ }\textbf {\bibinfo {volume} {2}},\ \bibinfo {pages}
  {013288} (\bibinfo {year} {2020})}\BibitemShut {NoStop}%
\bibitem [{\citenamefont {Verresen}\ \emph {et~al.}(2021)\citenamefont
  {Verresen}, \citenamefont {Lukin},\ and\ \citenamefont
  {Vishwanath}}]{VerresenVishwanath2021}%
  \BibitemOpen
  \bibfield  {author} {\bibinfo {author} {\bibfnamefont {Ruben}\ \bibnamefont
  {Verresen}}, \bibinfo {author} {\bibfnamefont {Mikhail~D.}\ \bibnamefont
  {Lukin}}, \ and\ \bibinfo {author} {\bibfnamefont {Ashvin}\ \bibnamefont
  {Vishwanath}},\ }\bibfield  {title} {\enquote {\bibinfo {title} {Prediction
  of toric code topological order from {R}ydberg blockade},}\ }\href {\doibase
  10.1103/PhysRevX.11.031005} {\bibfield  {journal} {\bibinfo  {journal} {Phys.
  Rev. X}\ }\textbf {\bibinfo {volume} {11}},\ \bibinfo {pages} {031005}
  (\bibinfo {year} {2021})}\BibitemShut {NoStop}%
\bibitem [{\citenamefont {Giudice}\ \emph {et~al.}(2022)\citenamefont
  {Giudice}, \citenamefont {Surace}, \citenamefont {Pichler},\ and\
  \citenamefont {Giudici}}]{GiudiceGiudici2022}%
  \BibitemOpen
  \bibfield  {author} {\bibinfo {author} {\bibfnamefont {Giacomo}\ \bibnamefont
  {Giudice}}, \bibinfo {author} {\bibfnamefont {Federica~Maria}\ \bibnamefont
  {Surace}}, \bibinfo {author} {\bibfnamefont {Hannes}\ \bibnamefont
  {Pichler}}, \ and\ \bibinfo {author} {\bibfnamefont {Giuliano}\ \bibnamefont
  {Giudici}},\ }\bibfield  {title} {\enquote {\bibinfo {title} {Trimer states
  with $\mathbb{Z}_3$ topological order in {R}ydberg atom arrays},}\
  }\href@noop {} {\bibfield  {journal} {\bibinfo  {journal} {arXiv preprint
  arXiv:2205.10387}\ } (\bibinfo {year} {2022})}\BibitemShut {NoStop}%
\bibitem [{\citenamefont {Dalmonte}\ and\ \citenamefont
  {Montangero}(2016)}]{DalmonteMontangero2016}%
  \BibitemOpen
  \bibfield  {author} {\bibinfo {author} {\bibfnamefont {M.}~\bibnamefont
  {Dalmonte}}\ and\ \bibinfo {author} {\bibfnamefont {S.}~\bibnamefont
  {Montangero}},\ }\bibfield  {title} {\enquote {\bibinfo {title} {Lattice
  gauge theory simulations in the quantum information era},}\ }\href {\doibase
  10.1080/00107514.2016.1151199} {\bibfield  {journal} {\bibinfo  {journal}
  {Contemporary Physics}\ }\textbf {\bibinfo {volume} {57}},\ \bibinfo {pages}
  {388--412} (\bibinfo {year} {2016})}\BibitemShut {NoStop}%
\bibitem [{\citenamefont {Banuls}\ \emph {et~al.}(2020)\citenamefont {Banuls},
  \citenamefont {Blatt}, \citenamefont {Catani}, \citenamefont {Celi},
  \citenamefont {Cirac}, \citenamefont {Dalmonte}, \citenamefont {Fallani},
  \citenamefont {Jansen}, \citenamefont {Lewenstein}, \citenamefont
  {Montangero} \emph {et~al.}}]{BanulsZoller2020}%
  \BibitemOpen
  \bibfield  {author} {\bibinfo {author} {\bibfnamefont {Mari~Carmen}\
  \bibnamefont {Banuls}}, \bibinfo {author} {\bibfnamefont {Rainer}\
  \bibnamefont {Blatt}}, \bibinfo {author} {\bibfnamefont {Jacopo}\
  \bibnamefont {Catani}}, \bibinfo {author} {\bibfnamefont {Alessio}\
  \bibnamefont {Celi}}, \bibinfo {author} {\bibfnamefont {Juan~Ignacio}\
  \bibnamefont {Cirac}}, \bibinfo {author} {\bibfnamefont {Marcello}\
  \bibnamefont {Dalmonte}}, \bibinfo {author} {\bibfnamefont {Leonardo}\
  \bibnamefont {Fallani}}, \bibinfo {author} {\bibfnamefont {Karl}\
  \bibnamefont {Jansen}}, \bibinfo {author} {\bibfnamefont {Maciej}\
  \bibnamefont {Lewenstein}}, \bibinfo {author} {\bibfnamefont {Simone}\
  \bibnamefont {Montangero}},  \emph {et~al.},\ }\bibfield  {title} {\enquote
  {\bibinfo {title} {Simulating lattice gauge theories within quantum
  technologies},}\ }\href@noop {} {\bibfield  {journal} {\bibinfo  {journal}
  {The European physical journal D}\ }\textbf {\bibinfo {volume} {74}},\
  \bibinfo {pages} {1--42} (\bibinfo {year} {2020})}\BibitemShut {NoStop}%
\bibitem [{\citenamefont {Satzinger}\ \emph {et~al.}(2021)\citenamefont
  {Satzinger}, \citenamefont {Liu}, \citenamefont {Smith}, \citenamefont
  {Knapp}, \citenamefont {Newman}, \citenamefont {Jones}, \citenamefont {Chen},
  \citenamefont {Quintana}, \citenamefont {Mi}, \citenamefont {Dunsworth} \emph
  {et~al.}}]{SatzingerRoushan2021}%
  \BibitemOpen
  \bibfield  {author} {\bibinfo {author} {\bibfnamefont {KJ}~\bibnamefont
  {Satzinger}}, \bibinfo {author} {\bibfnamefont {Y-J}\ \bibnamefont {Liu}},
  \bibinfo {author} {\bibfnamefont {A}~\bibnamefont {Smith}}, \bibinfo {author}
  {\bibfnamefont {C}~\bibnamefont {Knapp}}, \bibinfo {author} {\bibfnamefont
  {M}~\bibnamefont {Newman}}, \bibinfo {author} {\bibfnamefont {C}~\bibnamefont
  {Jones}}, \bibinfo {author} {\bibfnamefont {Z}~\bibnamefont {Chen}}, \bibinfo
  {author} {\bibfnamefont {C}~\bibnamefont {Quintana}}, \bibinfo {author}
  {\bibfnamefont {X}~\bibnamefont {Mi}}, \bibinfo {author} {\bibfnamefont
  {A}~\bibnamefont {Dunsworth}},  \emph {et~al.},\ }\bibfield  {title}
  {\enquote {\bibinfo {title} {Realizing topologically ordered states on a
  quantum processor},}\ }\href@noop {} {\bibfield  {journal} {\bibinfo
  {journal} {Science}\ }\textbf {\bibinfo {volume} {374}},\ \bibinfo {pages}
  {1237--1241} (\bibinfo {year} {2021})}\BibitemShut {NoStop}%
\bibitem [{\citenamefont {Pisarski}(2021)}]{Pisarski2021}%
  \BibitemOpen
  \bibfield  {author} {\bibinfo {author} {\bibfnamefont {Robert~D.}\
  \bibnamefont {Pisarski}},\ }\bibfield  {title} {\enquote {\bibinfo {title}
  {Remarks on nuclear matter: How an ${\ensuremath{\omega}}_{0}$ condensate can
  spike the speed of sound, and a model of ${Z(3)}$ baryons},}\ }\href
  {\doibase 10.1103/PhysRevD.103.L071504} {\bibfield  {journal} {\bibinfo
  {journal} {Phys. Rev. D}\ }\textbf {\bibinfo {volume} {103}},\ \bibinfo
  {pages} {L071504} (\bibinfo {year} {2021})}\BibitemShut {NoStop}%
\bibitem [{\citenamefont {Gazit}\ \emph {et~al.}(2017)\citenamefont {Gazit},
  \citenamefont {Randeria},\ and\ \citenamefont
  {Vishwanath}}]{GazitVishwanath2017}%
  \BibitemOpen
  \bibfield  {author} {\bibinfo {author} {\bibfnamefont {Snir}\ \bibnamefont
  {Gazit}}, \bibinfo {author} {\bibfnamefont {Mohit}\ \bibnamefont {Randeria}},
  \ and\ \bibinfo {author} {\bibfnamefont {Ashvin}\ \bibnamefont
  {Vishwanath}},\ }\bibfield  {title} {\enquote {\bibinfo {title} {Emergent
  {D}irac fermions and broken symmetries in confined and deconfined phases of
  {$Z_2$} gauge theories},}\ }\href@noop {} {\bibfield  {journal} {\bibinfo
  {journal} {Nature Physics}\ }\textbf {\bibinfo {volume} {13}},\ \bibinfo
  {pages} {484--490} (\bibinfo {year} {2017})}\BibitemShut {NoStop}%
\bibitem [{\citenamefont {Gazit}\ \emph {et~al.}(2018)\citenamefont {Gazit},
  \citenamefont {Assaad}, \citenamefont {Sachdev}, \citenamefont {Vishwanath},\
  and\ \citenamefont {Wang}}]{GazitWang2018}%
  \BibitemOpen
  \bibfield  {author} {\bibinfo {author} {\bibfnamefont {Snir}\ \bibnamefont
  {Gazit}}, \bibinfo {author} {\bibfnamefont {Fakher~F}\ \bibnamefont
  {Assaad}}, \bibinfo {author} {\bibfnamefont {Subir}\ \bibnamefont {Sachdev}},
  \bibinfo {author} {\bibfnamefont {Ashvin}\ \bibnamefont {Vishwanath}}, \ and\
  \bibinfo {author} {\bibfnamefont {Chong}\ \bibnamefont {Wang}},\ }\bibfield
  {title} {\enquote {\bibinfo {title} {Confinement transition of ${Z}_{2}$
  gauge theories coupled to massless fermions: Emergent quantum chromodynamics
  and {SO(5)} symmetry},}\ }\href@noop {} {\bibfield  {journal} {\bibinfo
  {journal} {Proceedings of the National Academy of Sciences}\ }\textbf
  {\bibinfo {volume} {115}},\ \bibinfo {pages} {E6987--E6995} (\bibinfo {year}
  {2018})}\BibitemShut {NoStop}%
\bibitem [{\citenamefont {Haase}\ \emph {et~al.}(2021)\citenamefont {Haase},
  \citenamefont {Dellantonio}, \citenamefont {Celi}, \citenamefont {Paulson},
  \citenamefont {Kan}, \citenamefont {Jansen},\ and\ \citenamefont
  {Muschik}}]{HaaseDellantonio2021}%
  \BibitemOpen
  \bibfield  {author} {\bibinfo {author} {\bibfnamefont {Jan~F}\ \bibnamefont
  {Haase}}, \bibinfo {author} {\bibfnamefont {Luca}\ \bibnamefont
  {Dellantonio}}, \bibinfo {author} {\bibfnamefont {Alessio}\ \bibnamefont
  {Celi}}, \bibinfo {author} {\bibfnamefont {Danny}\ \bibnamefont {Paulson}},
  \bibinfo {author} {\bibfnamefont {Angus}\ \bibnamefont {Kan}}, \bibinfo
  {author} {\bibfnamefont {Karl}\ \bibnamefont {Jansen}}, \ and\ \bibinfo
  {author} {\bibfnamefont {Christine~A}\ \bibnamefont {Muschik}},\ }\bibfield
  {title} {\enquote {\bibinfo {title} {A resource efficient approach for
  quantum and classical simulations of gauge theories in particle physics},}\
  }\href@noop {} {\bibfield  {journal} {\bibinfo  {journal} {Quantum}\ }\textbf
  {\bibinfo {volume} {5}},\ \bibinfo {pages} {393} (\bibinfo {year}
  {2021})}\BibitemShut {NoStop}%
\bibitem [{\citenamefont {Einhorn}\ \emph {et~al.}(1980)\citenamefont
  {Einhorn}, \citenamefont {Savit},\ and\ \citenamefont
  {Rabinovici}}]{EinhornRabinovici1980}%
  \BibitemOpen
  \bibfield  {author} {\bibinfo {author} {\bibfnamefont {Martin~B}\
  \bibnamefont {Einhorn}}, \bibinfo {author} {\bibfnamefont {Robert}\
  \bibnamefont {Savit}}, \ and\ \bibinfo {author} {\bibfnamefont {Eliezer}\
  \bibnamefont {Rabinovici}},\ }\bibfield  {title} {\enquote {\bibinfo {title}
  {A physical picture for the phase transitions in {$Z_N$} symmetric models},}\
  }\href@noop {} {\bibfield  {journal} {\bibinfo  {journal} {Nuclear Physics
  B}\ }\textbf {\bibinfo {volume} {170}},\ \bibinfo {pages} {16--31} (\bibinfo
  {year} {1980})}\BibitemShut {NoStop}%
\bibitem [{\citenamefont {Jos\'e}\ \emph {et~al.}(1977)\citenamefont {Jos\'e},
  \citenamefont {Kadanoff}, \citenamefont {Kirkpatrick},\ and\ \citenamefont
  {Nelson}}]{JoseNelson1977}%
  \BibitemOpen
  \bibfield  {author} {\bibinfo {author} {\bibfnamefont {Jorge~V.}\
  \bibnamefont {Jos\'e}}, \bibinfo {author} {\bibfnamefont {Leo~P.}\
  \bibnamefont {Kadanoff}}, \bibinfo {author} {\bibfnamefont {Scott}\
  \bibnamefont {Kirkpatrick}}, \ and\ \bibinfo {author} {\bibfnamefont
  {David~R.}\ \bibnamefont {Nelson}},\ }\bibfield  {title} {\enquote {\bibinfo
  {title} {Renormalization, vortices, and symmetry-breaking perturbations in
  the two-dimensional planar model},}\ }\href {\doibase
  10.1103/PhysRevB.16.1217} {\bibfield  {journal} {\bibinfo  {journal} {Phys.
  Rev. B}\ }\textbf {\bibinfo {volume} {16}},\ \bibinfo {pages} {1217--1241}
  (\bibinfo {year} {1977})}\BibitemShut {NoStop}%
\bibitem [{\citenamefont {Patil}\ \emph {et~al.}(2021)\citenamefont {Patil},
  \citenamefont {Shao},\ and\ \citenamefont {Sandvik}}]{PatilSadvik2021}%
  \BibitemOpen
  \bibfield  {author} {\bibinfo {author} {\bibfnamefont {Pranay}\ \bibnamefont
  {Patil}}, \bibinfo {author} {\bibfnamefont {Hui}\ \bibnamefont {Shao}}, \
  and\ \bibinfo {author} {\bibfnamefont {Anders~W.}\ \bibnamefont {Sandvik}},\
  }\bibfield  {title} {\enquote {\bibinfo {title} {Unconventional {U(1)} to
  ${Z}_{q}$ crossover in quantum and classical $q$-state clock models},}\
  }\href {\doibase 10.1103/PhysRevB.103.054418} {\bibfield  {journal} {\bibinfo
   {journal} {Phys. Rev. B}\ }\textbf {\bibinfo {volume} {103}},\ \bibinfo
  {pages} {054418} (\bibinfo {year} {2021})}\BibitemShut {NoStop}%
\bibitem [{foo()}]{footnoteFirstOrder}%
  \BibitemOpen
  \href@noop {} {}\bibinfo {note} {The case $N = 3$ is special: Here a first
  order transition generally occurs, as can be easily seen by considering a
  complex $\vert \phi \vert^4$ theory perturbed by the $\mathbb{Z}_3$ symmetric
  potential $\phi^3 + {\phi^*}^3$.}\BibitemShut {Stop}%
\bibitem [{\citenamefont {Altshuler}\ \emph {et~al.}(1994)\citenamefont
  {Altshuler}, \citenamefont {Ioffe},\ and\ \citenamefont
  {Millis}}]{AltshulerMillis1994}%
  \BibitemOpen
  \bibfield  {author} {\bibinfo {author} {\bibfnamefont {B.~L.}\ \bibnamefont
  {Altshuler}}, \bibinfo {author} {\bibfnamefont {L.~B.}\ \bibnamefont
  {Ioffe}}, \ and\ \bibinfo {author} {\bibfnamefont {A.~J.}\ \bibnamefont
  {Millis}},\ }\bibfield  {title} {\enquote {\bibinfo {title} {Low-energy
  properties of fermions with singular interactions},}\ }\href {\doibase
  10.1103/PhysRevB.50.14048} {\bibfield  {journal} {\bibinfo  {journal} {Phys.
  Rev. B}\ }\textbf {\bibinfo {volume} {50}},\ \bibinfo {pages} {14048--14064}
  (\bibinfo {year} {1994})}\BibitemShut {NoStop}%
\bibitem [{\citenamefont {Kim}\ \emph {et~al.}(1994)\citenamefont {Kim},
  \citenamefont {Furusaki}, \citenamefont {Wen},\ and\ \citenamefont
  {Lee}}]{KimLee1994}%
  \BibitemOpen
  \bibfield  {author} {\bibinfo {author} {\bibfnamefont {Yong~Baek}\
  \bibnamefont {Kim}}, \bibinfo {author} {\bibfnamefont {Akira}\ \bibnamefont
  {Furusaki}}, \bibinfo {author} {\bibfnamefont {Xiao-Gang}\ \bibnamefont
  {Wen}}, \ and\ \bibinfo {author} {\bibfnamefont {Patrick~A.}\ \bibnamefont
  {Lee}},\ }\bibfield  {title} {\enquote {\bibinfo {title} {Gauge-invariant
  response functions of fermions coupled to a gauge field},}\ }\href {\doibase
  10.1103/PhysRevB.50.17917} {\bibfield  {journal} {\bibinfo  {journal} {Phys.
  Rev. B}\ }\textbf {\bibinfo {volume} {50}},\ \bibinfo {pages} {17917--17932}
  (\bibinfo {year} {1994})}\BibitemShut {NoStop}%
\bibitem [{\citenamefont {Hermele}\ \emph {et~al.}(2004)\citenamefont
  {Hermele}, \citenamefont {Senthil}, \citenamefont {Fisher}, \citenamefont
  {Lee}, \citenamefont {Nagaosa},\ and\ \citenamefont {Wen}}]{HermeleWen2004}%
  \BibitemOpen
  \bibfield  {author} {\bibinfo {author} {\bibfnamefont {Michael}\ \bibnamefont
  {Hermele}}, \bibinfo {author} {\bibfnamefont {T.}~\bibnamefont {Senthil}},
  \bibinfo {author} {\bibfnamefont {Matthew P.~A.}\ \bibnamefont {Fisher}},
  \bibinfo {author} {\bibfnamefont {Patrick~A.}\ \bibnamefont {Lee}}, \bibinfo
  {author} {\bibfnamefont {Naoto}\ \bibnamefont {Nagaosa}}, \ and\ \bibinfo
  {author} {\bibfnamefont {Xiao-Gang}\ \bibnamefont {Wen}},\ }\bibfield
  {title} {\enquote {\bibinfo {title} {Stability of {$U(1)$} spin liquids in
  two dimensions},}\ }\href {\doibase 10.1103/PhysRevB.70.214437} {\bibfield
  {journal} {\bibinfo  {journal} {Phys. Rev. B}\ }\textbf {\bibinfo {volume}
  {70}},\ \bibinfo {pages} {214437} (\bibinfo {year} {2004})}\BibitemShut
  {NoStop}%
\bibitem [{\citenamefont {Nave}\ \emph {et~al.}(2007)\citenamefont {Nave},
  \citenamefont {Lee},\ and\ \citenamefont {Lee}}]{NaveLee2007}%
  \BibitemOpen
  \bibfield  {author} {\bibinfo {author} {\bibfnamefont {Cody~P.}\ \bibnamefont
  {Nave}}, \bibinfo {author} {\bibfnamefont {Sung-Sik}\ \bibnamefont {Lee}}, \
  and\ \bibinfo {author} {\bibfnamefont {Patrick~A.}\ \bibnamefont {Lee}},\
  }\bibfield  {title} {\enquote {\bibinfo {title} {Susceptibility of a spinon
  {F}ermi surface coupled to a {$U(1)$} gauge field},}\ }\href {\doibase
  10.1103/PhysRevB.76.165104} {\bibfield  {journal} {\bibinfo  {journal} {Phys.
  Rev. B}\ }\textbf {\bibinfo {volume} {76}},\ \bibinfo {pages} {165104}
  (\bibinfo {year} {2007})}\BibitemShut {NoStop}%
\bibitem [{\citenamefont {Lee}(2009)}]{Lee2009}%
  \BibitemOpen
  \bibfield  {author} {\bibinfo {author} {\bibfnamefont {Sung-Sik}\
  \bibnamefont {Lee}},\ }\bibfield  {title} {\enquote {\bibinfo {title}
  {Low-energy effective theory of {F}ermi surface coupled with {$U(1)$} gauge
  field in $2+1$ dimensions},}\ }\href {\doibase 10.1103/PhysRevB.80.165102}
  {\bibfield  {journal} {\bibinfo  {journal} {Phys. Rev. B}\ }\textbf {\bibinfo
  {volume} {80}},\ \bibinfo {pages} {165102} (\bibinfo {year}
  {2009})}\BibitemShut {NoStop}%
\bibitem [{\citenamefont {Mross}\ \emph {et~al.}(2010)\citenamefont {Mross},
  \citenamefont {McGreevy}, \citenamefont {Liu},\ and\ \citenamefont
  {Senthil}}]{MrossSenthil2010}%
  \BibitemOpen
  \bibfield  {author} {\bibinfo {author} {\bibfnamefont {David~F.}\
  \bibnamefont {Mross}}, \bibinfo {author} {\bibfnamefont {John}\ \bibnamefont
  {McGreevy}}, \bibinfo {author} {\bibfnamefont {Hong}\ \bibnamefont {Liu}}, \
  and\ \bibinfo {author} {\bibfnamefont {T.}~\bibnamefont {Senthil}},\
  }\bibfield  {title} {\enquote {\bibinfo {title} {Controlled expansion for
  certain non-fermi-liquid metals},}\ }\href {\doibase
  10.1103/PhysRevB.82.045121} {\bibfield  {journal} {\bibinfo  {journal} {Phys.
  Rev. B}\ }\textbf {\bibinfo {volume} {82}},\ \bibinfo {pages} {045121}
  (\bibinfo {year} {2010})}\BibitemShut {NoStop}%
\bibitem [{\citenamefont {Magnifico}\ \emph {et~al.}(2019)\citenamefont
  {Magnifico}, \citenamefont {Vodola}, \citenamefont {Ercolessi}, \citenamefont
  {Kumar}, \citenamefont {M\"uller},\ and\ \citenamefont
  {Bermudez}}]{MagnificoBermudez2019}%
  \BibitemOpen
  \bibfield  {author} {\bibinfo {author} {\bibfnamefont {G.}~\bibnamefont
  {Magnifico}}, \bibinfo {author} {\bibfnamefont {D.}~\bibnamefont {Vodola}},
  \bibinfo {author} {\bibfnamefont {E.}~\bibnamefont {Ercolessi}}, \bibinfo
  {author} {\bibfnamefont {S.~P.}\ \bibnamefont {Kumar}}, \bibinfo {author}
  {\bibfnamefont {M.}~\bibnamefont {M\"uller}}, \ and\ \bibinfo {author}
  {\bibfnamefont {A.}~\bibnamefont {Bermudez}},\ }\bibfield  {title} {\enquote
  {\bibinfo {title} {$\mathbb{Z}_{N}$ gauge theories coupled to topological
  fermions: {${\mathrm{QED}}_{2}$} with a quantum mechanical
  $\ensuremath{\theta}$ angle},}\ }\href {\doibase 10.1103/PhysRevB.100.115152}
  {\bibfield  {journal} {\bibinfo  {journal} {Phys. Rev. B}\ }\textbf {\bibinfo
  {volume} {100}},\ \bibinfo {pages} {115152} (\bibinfo {year}
  {2019})}\BibitemShut {NoStop}%
\bibitem [{\citenamefont {Nyhegn}\ \emph {et~al.}(2021)\citenamefont {Nyhegn},
  \citenamefont {Chung},\ and\ \citenamefont {Burrello}}]{NyhegnBurello2021}%
  \BibitemOpen
  \bibfield  {author} {\bibinfo {author} {\bibfnamefont {Jens}\ \bibnamefont
  {Nyhegn}}, \bibinfo {author} {\bibfnamefont {Chia-Min}\ \bibnamefont
  {Chung}}, \ and\ \bibinfo {author} {\bibfnamefont {Michele}\ \bibnamefont
  {Burrello}},\ }\bibfield  {title} {\enquote {\bibinfo {title}
  {$\mathbb{Z}_{N}$ lattice gauge theory in a ladder geometry},}\ }\href
  {\doibase 10.1103/PhysRevResearch.3.013133} {\bibfield  {journal} {\bibinfo
  {journal} {Phys. Rev. Research}\ }\textbf {\bibinfo {volume} {3}},\ \bibinfo
  {pages} {013133} (\bibinfo {year} {2021})}\BibitemShut {NoStop}%
\bibitem [{\citenamefont {Pradhan}\ \emph {et~al.}(2022)\citenamefont
  {Pradhan}, \citenamefont {Maroncelli},\ and\ \citenamefont
  {Ercolessi}}]{PradhanErcolessi2022}%
  \BibitemOpen
  \bibfield  {author} {\bibinfo {author} {\bibfnamefont {Sunny}\ \bibnamefont
  {Pradhan}}, \bibinfo {author} {\bibfnamefont {Andrea}\ \bibnamefont
  {Maroncelli}}, \ and\ \bibinfo {author} {\bibfnamefont {Elisa}\ \bibnamefont
  {Ercolessi}},\ }\bibfield  {title} {\enquote {\bibinfo {title} {Discrete
  abelian lattice gauge theories on a ladder and their dualities with clock
  models},}\ }\href@noop {} {\bibfield  {journal} {\bibinfo  {journal} {arXiv
  preprint arXiv:2208.04182}\ } (\bibinfo {year} {2022})}\BibitemShut {NoStop}%
\bibitem [{\citenamefont {Xu}\ \emph {et~al.}(2022)\citenamefont {Xu},
  \citenamefont {Wu}, \citenamefont {Ye}, \citenamefont {Luo}, \citenamefont
  {Jian},\ and\ \citenamefont {Xu}}]{XuXu2022}%
  \BibitemOpen
  \bibfield  {author} {\bibinfo {author} {\bibfnamefont {Yichen}\ \bibnamefont
  {Xu}}, \bibinfo {author} {\bibfnamefont {Xiao-Chuan}\ \bibnamefont {Wu}},
  \bibinfo {author} {\bibfnamefont {Mengxing}\ \bibnamefont {Ye}}, \bibinfo
  {author} {\bibfnamefont {Zhu-Xi}\ \bibnamefont {Luo}}, \bibinfo {author}
  {\bibfnamefont {Chao-Ming}\ \bibnamefont {Jian}}, \ and\ \bibinfo {author}
  {\bibfnamefont {Cenke}\ \bibnamefont {Xu}},\ }\bibfield  {title} {\enquote
  {\bibinfo {title} {Interaction-driven metal-insulator transition with charge
  fractionalization},}\ }\href {\doibase 10.1103/PhysRevX.12.021067} {\bibfield
   {journal} {\bibinfo  {journal} {Phys. Rev. X}\ }\textbf {\bibinfo {volume}
  {12}},\ \bibinfo {pages} {021067} (\bibinfo {year} {2022})}\BibitemShut
  {NoStop}%
\bibitem [{\citenamefont {Vaezi}(2014)}]{Vaezi2014}%
  \BibitemOpen
  \bibfield  {author} {\bibinfo {author} {\bibfnamefont {Abolhassan}\
  \bibnamefont {Vaezi}},\ }\bibfield  {title} {\enquote {\bibinfo {title}
  {${Z}_{3}$ generalization of the {K}itaev's spin-1/2 model},}\ }\href
  {\doibase 10.1103/PhysRevB.90.075106} {\bibfield  {journal} {\bibinfo
  {journal} {Phys. Rev. B}\ }\textbf {\bibinfo {volume} {90}},\ \bibinfo
  {pages} {075106} (\bibinfo {year} {2014})}\BibitemShut {NoStop}%
\bibitem [{\citenamefont {Barkeshli}\ \emph {et~al.}(2015)\citenamefont
  {Barkeshli}, \citenamefont {Jiang}, \citenamefont {Thomale},\ and\
  \citenamefont {Qi}}]{BarkeshliQi2015}%
  \BibitemOpen
  \bibfield  {author} {\bibinfo {author} {\bibfnamefont {Maissam}\ \bibnamefont
  {Barkeshli}}, \bibinfo {author} {\bibfnamefont {Hong-Chen}\ \bibnamefont
  {Jiang}}, \bibinfo {author} {\bibfnamefont {Ronny}\ \bibnamefont {Thomale}},
  \ and\ \bibinfo {author} {\bibfnamefont {Xiao-Liang}\ \bibnamefont {Qi}},\
  }\bibfield  {title} {\enquote {\bibinfo {title} {Generalized {K}itaev models
  and extrinsic non-{A}belian twist defects},}\ }\href {\doibase
  10.1103/PhysRevLett.114.026401} {\bibfield  {journal} {\bibinfo  {journal}
  {Phys. Rev. Lett.}\ }\textbf {\bibinfo {volume} {114}},\ \bibinfo {pages}
  {026401} (\bibinfo {year} {2015})}\BibitemShut {NoStop}%
\bibitem [{\citenamefont {Maciejko}\ \emph {et~al.}(2012)\citenamefont
  {Maciejko}, \citenamefont {Qi}, \citenamefont {Karch},\ and\ \citenamefont
  {Zhang}}]{MaciejkoZhang2012}%
  \BibitemOpen
  \bibfield  {author} {\bibinfo {author} {\bibfnamefont {Joseph}\ \bibnamefont
  {Maciejko}}, \bibinfo {author} {\bibfnamefont {Xiao-Liang}\ \bibnamefont
  {Qi}}, \bibinfo {author} {\bibfnamefont {Andreas}\ \bibnamefont {Karch}}, \
  and\ \bibinfo {author} {\bibfnamefont {Shou-Cheng}\ \bibnamefont {Zhang}},\
  }\bibfield  {title} {\enquote {\bibinfo {title} {Models of three-dimensional
  fractional topological insulators},}\ }\href {\doibase
  10.1103/PhysRevB.86.235128} {\bibfield  {journal} {\bibinfo  {journal} {Phys.
  Rev. B}\ }\textbf {\bibinfo {volume} {86}},\ \bibinfo {pages} {235128}
  (\bibinfo {year} {2012})}\BibitemShut {NoStop}%
\bibitem [{\citenamefont {Florens}\ and\ \citenamefont
  {Georges}(2004)}]{FlorensGeorges2004}%
  \BibitemOpen
  \bibfield  {author} {\bibinfo {author} {\bibfnamefont {Serge}\ \bibnamefont
  {Florens}}\ and\ \bibinfo {author} {\bibfnamefont {Antoine}\ \bibnamefont
  {Georges}},\ }\bibfield  {title} {\enquote {\bibinfo {title} {Slave-rotor
  mean-field theories of strongly correlated systems and the {M}ott transition
  in finite dimensions},}\ }\href {\doibase 10.1103/PhysRevB.70.035114}
  {\bibfield  {journal} {\bibinfo  {journal} {Phys. Rev. B}\ }\textbf {\bibinfo
  {volume} {70}},\ \bibinfo {pages} {035114} (\bibinfo {year}
  {2004})}\BibitemShut {NoStop}%
\bibitem [{\citenamefont {Coleman}(1984)}]{Coleman1984}%
  \BibitemOpen
  \bibfield  {author} {\bibinfo {author} {\bibfnamefont {Piers}\ \bibnamefont
  {Coleman}},\ }\bibfield  {title} {\enquote {\bibinfo {title} {New approach to
  the mixed-valence problem},}\ }\href {\doibase 10.1103/PhysRevB.29.3035}
  {\bibfield  {journal} {\bibinfo  {journal} {Phys. Rev. B}\ }\textbf {\bibinfo
  {volume} {29}},\ \bibinfo {pages} {3035--3044} (\bibinfo {year}
  {1984})}\BibitemShut {NoStop}%
\bibitem [{\citenamefont {Kotliar}\ and\ \citenamefont
  {Ruckenstein}(1986)}]{KotliarRuckenstein1986}%
  \BibitemOpen
  \bibfield  {author} {\bibinfo {author} {\bibfnamefont {Gabriel}\ \bibnamefont
  {Kotliar}}\ and\ \bibinfo {author} {\bibfnamefont {Andrei~E.}\ \bibnamefont
  {Ruckenstein}},\ }\bibfield  {title} {\enquote {\bibinfo {title} {New
  functional integral approach to strongly correlated {F}ermi systems: The
  {G}utzwiller approximation as a saddle point},}\ }\href {\doibase
  10.1103/PhysRevLett.57.1362} {\bibfield  {journal} {\bibinfo  {journal}
  {Phys. Rev. Lett.}\ }\textbf {\bibinfo {volume} {57}},\ \bibinfo {pages}
  {1362--1365} (\bibinfo {year} {1986})}\BibitemShut {NoStop}%
\bibitem [{\citenamefont {Read}\ and\ \citenamefont
  {Sachdev}(1991)}]{SachdevRead1991}%
  \BibitemOpen
  \bibfield  {author} {\bibinfo {author} {\bibfnamefont {N.}~\bibnamefont
  {Read}}\ and\ \bibinfo {author} {\bibfnamefont {Subir}\ \bibnamefont
  {Sachdev}},\ }\bibfield  {title} {\enquote {\bibinfo {title} {Large-{N}
  expansion for frustrated quantum antiferromagnets},}\ }\href {\doibase
  10.1103/PhysRevLett.66.1773} {\bibfield  {journal} {\bibinfo  {journal}
  {Phys. Rev. Lett.}\ }\textbf {\bibinfo {volume} {66}},\ \bibinfo {pages}
  {1773--1776} (\bibinfo {year} {1991})}\BibitemShut {NoStop}%
\bibitem [{\citenamefont {Yu}\ and\ \citenamefont {Si}(2012)}]{YuSi2012}%
  \BibitemOpen
  \bibfield  {author} {\bibinfo {author} {\bibfnamefont {Rong}\ \bibnamefont
  {Yu}}\ and\ \bibinfo {author} {\bibfnamefont {Qimiao}\ \bibnamefont {Si}},\
  }\bibfield  {title} {\enquote {\bibinfo {title} {{$U(1)$} slave-spin theory
  and its application to {M}ott transition in a multiorbital model for iron
  pnictides},}\ }\href {\doibase 10.1103/PhysRevB.86.085104} {\bibfield
  {journal} {\bibinfo  {journal} {Phys. Rev. B}\ }\textbf {\bibinfo {volume}
  {86}},\ \bibinfo {pages} {085104} (\bibinfo {year} {2012})}\BibitemShut
  {NoStop}%
\bibitem [{\citenamefont {Bonetti}\ and\ \citenamefont
  {Metzner}(2022)}]{BonettiMetzner2022}%
  \BibitemOpen
  \bibfield  {author} {\bibinfo {author} {\bibfnamefont {Pietro~M.}\
  \bibnamefont {Bonetti}}\ and\ \bibinfo {author} {\bibfnamefont {Walter}\
  \bibnamefont {Metzner}},\ }\bibfield  {title} {\enquote {\bibinfo {title}
  {{SU(2)} gauge theory of the pseudogap phase in the two-dimensional {H}ubbard
  model},}\ }\href {\doibase 10.1103/PhysRevB.106.205152} {\bibfield  {journal}
  {\bibinfo  {journal} {Phys. Rev. B}\ }\textbf {\bibinfo {volume} {106}},\
  \bibinfo {pages} {205152} (\bibinfo {year} {2022})}\BibitemShut {NoStop}%
\bibitem [{\citenamefont {Gu}\ \emph {et~al.}(2014)\citenamefont {Gu},
  \citenamefont {Wang},\ and\ \citenamefont {Wen}}]{GuWen2014}%
  \BibitemOpen
  \bibfield  {author} {\bibinfo {author} {\bibfnamefont {Zheng-Cheng}\
  \bibnamefont {Gu}}, \bibinfo {author} {\bibfnamefont {Zhenghan}\ \bibnamefont
  {Wang}}, \ and\ \bibinfo {author} {\bibfnamefont {Xiao-Gang}\ \bibnamefont
  {Wen}},\ }\bibfield  {title} {\enquote {\bibinfo {title} {Lattice model for
  fermionic toric code},}\ }\href {\doibase 10.1103/PhysRevB.90.085140}
  {\bibfield  {journal} {\bibinfo  {journal} {Phys. Rev. B}\ }\textbf {\bibinfo
  {volume} {90}},\ \bibinfo {pages} {085140} (\bibinfo {year}
  {2014})}\BibitemShut {NoStop}%
\bibitem [{\citenamefont {Borla}\ \emph {et~al.}(2022)\citenamefont {Borla},
  \citenamefont {Jeevanesan}, \citenamefont {Pollmann},\ and\ \citenamefont
  {Moroz}}]{BorlaMoroz2022}%
  \BibitemOpen
  \bibfield  {author} {\bibinfo {author} {\bibfnamefont {Umberto}\ \bibnamefont
  {Borla}}, \bibinfo {author} {\bibfnamefont {Bhilahari}\ \bibnamefont
  {Jeevanesan}}, \bibinfo {author} {\bibfnamefont {Frank}\ \bibnamefont
  {Pollmann}}, \ and\ \bibinfo {author} {\bibfnamefont {Sergej}\ \bibnamefont
  {Moroz}},\ }\bibfield  {title} {\enquote {\bibinfo {title} {Quantum phases of
  two-dimensional $\mathbb{Z}_{2}$ gauge theory coupled to single-component
  fermion matter},}\ }\href {\doibase 10.1103/PhysRevB.105.075132} {\bibfield
  {journal} {\bibinfo  {journal} {Phys. Rev. B}\ }\textbf {\bibinfo {volume}
  {105}},\ \bibinfo {pages} {075132} (\bibinfo {year} {2022})}\BibitemShut
  {NoStop}%
\bibitem [{\citenamefont {Emonts}\ \emph {et~al.}(2023)\citenamefont {Emonts},
  \citenamefont {Kelman}, \citenamefont {Borla}, \citenamefont {Moroz},
  \citenamefont {Gazit},\ and\ \citenamefont {Zohar}}]{EmontsMoroz2023}%
  \BibitemOpen
  \bibfield  {author} {\bibinfo {author} {\bibfnamefont {Patrick}\ \bibnamefont
  {Emonts}}, \bibinfo {author} {\bibfnamefont {Ariel}\ \bibnamefont {Kelman}},
  \bibinfo {author} {\bibfnamefont {Umberto}\ \bibnamefont {Borla}}, \bibinfo
  {author} {\bibfnamefont {Sergej}\ \bibnamefont {Moroz}}, \bibinfo {author}
  {\bibfnamefont {Snir}\ \bibnamefont {Gazit}}, \ and\ \bibinfo {author}
  {\bibfnamefont {Erez}\ \bibnamefont {Zohar}},\ }\bibfield  {title} {\enquote
  {\bibinfo {title} {Finding the ground state of a lattice gauge theory with
  fermionic tensor networks: A $2+1\mathrm{D}$ $\mathbb{Z}_2$ demonstration},}\
  }\href {\doibase 10.1103/PhysRevD.107.014505} {\bibfield  {journal} {\bibinfo
   {journal} {Phys. Rev. D}\ }\textbf {\bibinfo {volume} {107}},\ \bibinfo
  {pages} {014505} (\bibinfo {year} {2023})}\BibitemShut {NoStop}%
\bibitem [{\citenamefont {Nandkishore}\ \emph {et~al.}(2012)\citenamefont
  {Nandkishore}, \citenamefont {Metlitski},\ and\ \citenamefont
  {Senthil}}]{NandkishoreSenthil2012}%
  \BibitemOpen
  \bibfield  {author} {\bibinfo {author} {\bibfnamefont {Rahul}\ \bibnamefont
  {Nandkishore}}, \bibinfo {author} {\bibfnamefont {Max~A.}\ \bibnamefont
  {Metlitski}}, \ and\ \bibinfo {author} {\bibfnamefont {T.}~\bibnamefont
  {Senthil}},\ }\bibfield  {title} {\enquote {\bibinfo {title} {Orthogonal
  metals: The simplest non-{F}ermi liquids},}\ }\href {\doibase
  10.1103/PhysRevB.86.045128} {\bibfield  {journal} {\bibinfo  {journal} {Phys.
  Rev. B}\ }\textbf {\bibinfo {volume} {86}},\ \bibinfo {pages} {045128}
  (\bibinfo {year} {2012})}\BibitemShut {NoStop}%
\bibitem [{\citenamefont {Zhong}\ \emph {et~al.}(2013)\citenamefont {Zhong},
  \citenamefont {Wang},\ and\ \citenamefont {Luo}}]{ZhongLuo2013}%
  \BibitemOpen
  \bibfield  {author} {\bibinfo {author} {\bibfnamefont {Yin}\ \bibnamefont
  {Zhong}}, \bibinfo {author} {\bibfnamefont {Yu-Feng}\ \bibnamefont {Wang}}, \
  and\ \bibinfo {author} {\bibfnamefont {Hong-Gang}\ \bibnamefont {Luo}},\
  }\bibfield  {title} {\enquote {\bibinfo {title} {${Z}_{2}$ fractionalized
  {C}hern/topological insulators in an exactly soluble correlated model},}\
  }\href {\doibase 10.1103/PhysRevB.88.045109} {\bibfield  {journal} {\bibinfo
  {journal} {Phys. Rev. B}\ }\textbf {\bibinfo {volume} {88}},\ \bibinfo
  {pages} {045109} (\bibinfo {year} {2013})}\BibitemShut {NoStop}%
\bibitem [{\citenamefont {Hohenadler}\ and\ \citenamefont
  {Assaad}(2019)}]{HohenadlerAssaad2019}%
  \BibitemOpen
  \bibfield  {author} {\bibinfo {author} {\bibfnamefont {Martin}\ \bibnamefont
  {Hohenadler}}\ and\ \bibinfo {author} {\bibfnamefont {Fakher~F.}\
  \bibnamefont {Assaad}},\ }\bibfield  {title} {\enquote {\bibinfo {title}
  {Orthogonal metal in the {H}ubbard model with liberated slave spins},}\
  }\href {\doibase 10.1103/PhysRevB.100.125133} {\bibfield  {journal} {\bibinfo
   {journal} {Phys. Rev. B}\ }\textbf {\bibinfo {volume} {100}},\ \bibinfo
  {pages} {125133} (\bibinfo {year} {2019})}\BibitemShut {NoStop}%
\bibitem [{\citenamefont {R\"uegg}\ \emph {et~al.}(2010)\citenamefont
  {R\"uegg}, \citenamefont {Huber},\ and\ \citenamefont
  {Sigrist}}]{RueggSigrist2010}%
  \BibitemOpen
  \bibfield  {author} {\bibinfo {author} {\bibfnamefont {A.}~\bibnamefont
  {R\"uegg}}, \bibinfo {author} {\bibfnamefont {S.~D.}\ \bibnamefont {Huber}},
  \ and\ \bibinfo {author} {\bibfnamefont {M.}~\bibnamefont {Sigrist}},\
  }\bibfield  {title} {\enquote {\bibinfo {title} {$\mathbb{Z}_2$-slave-spin
  theory for strongly correlated fermions},}\ }\href {\doibase
  10.1103/PhysRevB.81.155118} {\bibfield  {journal} {\bibinfo  {journal} {Phys.
  Rev. B}\ }\textbf {\bibinfo {volume} {81}},\ \bibinfo {pages} {155118}
  (\bibinfo {year} {2010})}\BibitemShut {NoStop}%
\bibitem [{\citenamefont {Paramekanti}\ and\ \citenamefont
  {Vishwanath}(2004)}]{ParamekantiVishwanath2004}%
  \BibitemOpen
  \bibfield  {author} {\bibinfo {author} {\bibfnamefont {Arun}\ \bibnamefont
  {Paramekanti}}\ and\ \bibinfo {author} {\bibfnamefont {Ashvin}\ \bibnamefont
  {Vishwanath}},\ }\bibfield  {title} {\enquote {\bibinfo {title} {Extending
  luttinger's theorem to ${Z}_{2}$ fractionalized phases of matter},}\ }\href
  {\doibase 10.1103/PhysRevB.70.245118} {\bibfield  {journal} {\bibinfo
  {journal} {Phys. Rev. B}\ }\textbf {\bibinfo {volume} {70}},\ \bibinfo
  {pages} {245118} (\bibinfo {year} {2004})}\BibitemShut {NoStop}%
\bibitem [{\citenamefont {Scheurer}\ \emph {et~al.}(2018)\citenamefont
  {Scheurer}, \citenamefont {Chatterjee}, \citenamefont {Wu}, \citenamefont
  {Ferrero}, \citenamefont {Georges},\ and\ \citenamefont
  {Sachdev}}]{ScheurerSachdev2018}%
  \BibitemOpen
  \bibfield  {author} {\bibinfo {author} {\bibfnamefont {Mathias~S}\
  \bibnamefont {Scheurer}}, \bibinfo {author} {\bibfnamefont {Shubhayu}\
  \bibnamefont {Chatterjee}}, \bibinfo {author} {\bibfnamefont {Wei}\
  \bibnamefont {Wu}}, \bibinfo {author} {\bibfnamefont {Michel}\ \bibnamefont
  {Ferrero}}, \bibinfo {author} {\bibfnamefont {Antoine}\ \bibnamefont
  {Georges}}, \ and\ \bibinfo {author} {\bibfnamefont {Subir}\ \bibnamefont
  {Sachdev}},\ }\bibfield  {title} {\enquote {\bibinfo {title} {Topological
  order in the pseudogap metal},}\ }\href@noop {} {\bibfield  {journal}
  {\bibinfo  {journal} {Proceedings of the National Academy of Sciences}\
  }\textbf {\bibinfo {volume} {115}},\ \bibinfo {pages} {E3665--E3672}
  (\bibinfo {year} {2018})}\BibitemShut {NoStop}%
\bibitem [{\citenamefont {K\"onig}\ \emph {et~al.}(2020)\citenamefont
  {K\"onig}, \citenamefont {Coleman},\ and\ \citenamefont
  {Tsvelik}}]{KoenigTsvelik2020}%
  \BibitemOpen
  \bibfield  {author} {\bibinfo {author} {\bibfnamefont {Elio~J.}\ \bibnamefont
  {K\"onig}}, \bibinfo {author} {\bibfnamefont {Piers}\ \bibnamefont
  {Coleman}}, \ and\ \bibinfo {author} {\bibfnamefont {Alexei~M.}\ \bibnamefont
  {Tsvelik}},\ }\bibfield  {title} {\enquote {\bibinfo {title} {Soluble limit
  and criticality of fermions in $\mathbb{Z}_{2}$ gauge theories},}\ }\href
  {\doibase 10.1103/PhysRevB.102.155143} {\bibfield  {journal} {\bibinfo
  {journal} {Phys. Rev. B}\ }\textbf {\bibinfo {volume} {102}},\ \bibinfo
  {pages} {155143} (\bibinfo {year} {2020})}\BibitemShut {NoStop}%
\bibitem [{\citenamefont {Luttinger}(1960)}]{Luttinger1960}%
  \BibitemOpen
  \bibfield  {author} {\bibinfo {author} {\bibfnamefont {J.~M.}\ \bibnamefont
  {Luttinger}},\ }\bibfield  {title} {\enquote {\bibinfo {title} {{F}ermi
  surface and some simple equilibrium properties of a system of interacting
  fermions},}\ }\href {\doibase 10.1103/PhysRev.119.1153} {\bibfield  {journal}
  {\bibinfo  {journal} {Phys. Rev.}\ }\textbf {\bibinfo {volume} {119}},\
  \bibinfo {pages} {1153--1163} (\bibinfo {year} {1960})}\BibitemShut {NoStop}%
\bibitem [{\citenamefont {Oshikawa}(2000)}]{Oshikawa_2000}%
  \BibitemOpen
  \bibfield  {author} {\bibinfo {author} {\bibfnamefont {Masaki}\ \bibnamefont
  {Oshikawa}},\ }\bibfield  {title} {\enquote {\bibinfo {title} {Topological
  approach to {L}uttinger{\textquotesingle}s theorem and the {F}ermi surface of
  a {K}ondo lattice},}\ }\href {\doibase 10.1103/physrevlett.84.3370}
  {\bibfield  {journal} {\bibinfo  {journal} {Physical Review Letters}\
  }\textbf {\bibinfo {volume} {84}},\ \bibinfo {pages} {3370--3373} (\bibinfo
  {year} {2000})}\BibitemShut {NoStop}%
\bibitem [{\citenamefont {Proust}\ and\ \citenamefont
  {Taillefer}(2019)}]{ProustTaillefer2019}%
  \BibitemOpen
  \bibfield  {author} {\bibinfo {author} {\bibfnamefont {Cyril}\ \bibnamefont
  {Proust}}\ and\ \bibinfo {author} {\bibfnamefont {Louis}\ \bibnamefont
  {Taillefer}},\ }\bibfield  {title} {\enquote {\bibinfo {title} {The
  remarkable underlying ground states of cuprate superconductors},}\
  }\href@noop {} {\bibfield  {journal} {\bibinfo  {journal} {Annual Review of
  Condensed Matter Physics}\ }\textbf {\bibinfo {volume} {10}},\ \bibinfo
  {pages} {409--429} (\bibinfo {year} {2019})}\BibitemShut {NoStop}%
\bibitem [{\citenamefont {Si}\ and\ \citenamefont
  {Steglich}(2010)}]{SiSteglich2010}%
  \BibitemOpen
  \bibfield  {author} {\bibinfo {author} {\bibfnamefont {Qimiao}\ \bibnamefont
  {Si}}\ and\ \bibinfo {author} {\bibfnamefont {Frank}\ \bibnamefont
  {Steglich}},\ }\bibfield  {title} {\enquote {\bibinfo {title} {Heavy fermions
  and quantum phase transitions},}\ }\href {\doibase 10.1126/science.1191195}
  {\bibfield  {journal} {\bibinfo  {journal} {Science}\ }\textbf {\bibinfo
  {volume} {329}},\ \bibinfo {pages} {1161--1166} (\bibinfo {year}
  {2010})}\BibitemShut {NoStop}%
\bibitem [{\citenamefont {Xiang}\ \emph {et~al.}(2018)\citenamefont {Xiang},
  \citenamefont {Kasahara}, \citenamefont {Asaba}, \citenamefont {Lawson},
  \citenamefont {Tinsman}, \citenamefont {Chen}, \citenamefont {Sugimoto},
  \citenamefont {Kawaguchi}, \citenamefont {Sato}, \citenamefont {Li},
  \citenamefont {Yao}, \citenamefont {Chen}, \citenamefont {Iga}, \citenamefont
  {Singleton}, \citenamefont {Matsuda},\ and\ \citenamefont
  {Li}}]{XiangLi2018}%
  \BibitemOpen
  \bibfield  {author} {\bibinfo {author} {\bibfnamefont {Z.}~\bibnamefont
  {Xiang}}, \bibinfo {author} {\bibfnamefont {Y.}~\bibnamefont {Kasahara}},
  \bibinfo {author} {\bibfnamefont {T.}~\bibnamefont {Asaba}}, \bibinfo
  {author} {\bibfnamefont {B.}~\bibnamefont {Lawson}}, \bibinfo {author}
  {\bibfnamefont {C.}~\bibnamefont {Tinsman}}, \bibinfo {author} {\bibfnamefont
  {Lu}~\bibnamefont {Chen}}, \bibinfo {author} {\bibfnamefont {K.}~\bibnamefont
  {Sugimoto}}, \bibinfo {author} {\bibfnamefont {S.}~\bibnamefont {Kawaguchi}},
  \bibinfo {author} {\bibfnamefont {Y.}~\bibnamefont {Sato}}, \bibinfo {author}
  {\bibfnamefont {G.}~\bibnamefont {Li}}, \bibinfo {author} {\bibfnamefont
  {S.}~\bibnamefont {Yao}}, \bibinfo {author} {\bibfnamefont {Y.~L.}\
  \bibnamefont {Chen}}, \bibinfo {author} {\bibfnamefont {F.}~\bibnamefont
  {Iga}}, \bibinfo {author} {\bibfnamefont {John}\ \bibnamefont {Singleton}},
  \bibinfo {author} {\bibfnamefont {Y.}~\bibnamefont {Matsuda}}, \ and\
  \bibinfo {author} {\bibfnamefont {Lu}~\bibnamefont {Li}},\ }\bibfield
  {title} {\enquote {\bibinfo {title} {Quantum oscillations of electrical
  resistivity in an insulator},}\ }\href {\doibase 10.1126/science.aap9607}
  {\bibfield  {journal} {\bibinfo  {journal} {Science}\ }\textbf {\bibinfo
  {volume} {362}},\ \bibinfo {pages} {65--69} (\bibinfo {year}
  {2018})}\BibitemShut {NoStop}%
\bibitem [{\citenamefont {Li}\ \emph {et~al.}(2014)\citenamefont {Li},
  \citenamefont {Xiang}, \citenamefont {Yu}, \citenamefont {Asaba},
  \citenamefont {Lawson}, \citenamefont {Cai}, \citenamefont {Tinsman},
  \citenamefont {Berkley}, \citenamefont {Wolgast}, \citenamefont {Eo},
  \citenamefont {Kim}, \citenamefont {Kurdak}, \citenamefont {Allen},
  \citenamefont {Sun}, \citenamefont {Chen}, \citenamefont {Wang},
  \citenamefont {Fisk},\ and\ \citenamefont {Li}}]{LiLi2014}%
  \BibitemOpen
  \bibfield  {author} {\bibinfo {author} {\bibfnamefont {G.}~\bibnamefont
  {Li}}, \bibinfo {author} {\bibfnamefont {Z.}~\bibnamefont {Xiang}}, \bibinfo
  {author} {\bibfnamefont {F.}~\bibnamefont {Yu}}, \bibinfo {author}
  {\bibfnamefont {T.}~\bibnamefont {Asaba}}, \bibinfo {author} {\bibfnamefont
  {B.}~\bibnamefont {Lawson}}, \bibinfo {author} {\bibfnamefont
  {P.}~\bibnamefont {Cai}}, \bibinfo {author} {\bibfnamefont {C.}~\bibnamefont
  {Tinsman}}, \bibinfo {author} {\bibfnamefont {A.}~\bibnamefont {Berkley}},
  \bibinfo {author} {\bibfnamefont {S.}~\bibnamefont {Wolgast}}, \bibinfo
  {author} {\bibfnamefont {Y.~S.}\ \bibnamefont {Eo}}, \bibinfo {author}
  {\bibfnamefont {Dae-Jeong}\ \bibnamefont {Kim}}, \bibinfo {author}
  {\bibfnamefont {C.}~\bibnamefont {Kurdak}}, \bibinfo {author} {\bibfnamefont
  {J.~W.}\ \bibnamefont {Allen}}, \bibinfo {author} {\bibfnamefont
  {K.}~\bibnamefont {Sun}}, \bibinfo {author} {\bibfnamefont {X.~H.}\
  \bibnamefont {Chen}}, \bibinfo {author} {\bibfnamefont {Y.~Y.}\ \bibnamefont
  {Wang}}, \bibinfo {author} {\bibfnamefont {Z.}~\bibnamefont {Fisk}}, \ and\
  \bibinfo {author} {\bibfnamefont {Lu}~\bibnamefont {Li}},\ }\bibfield
  {title} {\enquote {\bibinfo {title} {Two-dimensional {F}ermi surfaces in
  {K}ondo insulator {S}m{B}$_6$},}\ }\href {\doibase 10.1126/science.1250366}
  {\bibfield  {journal} {\bibinfo  {journal} {Science}\ }\textbf {\bibinfo
  {volume} {346}},\ \bibinfo {pages} {1208--1212} (\bibinfo {year}
  {2014})}\BibitemShut {NoStop}%
\bibitem [{\citenamefont {Tan}\ \emph {et~al.}(2015)\citenamefont {Tan},
  \citenamefont {Hsu}, \citenamefont {Zeng}, \citenamefont {Hatnean},
  \citenamefont {Harrison}, \citenamefont {Zhu}, \citenamefont {Hartstein},
  \citenamefont {Kiourlappou}, \citenamefont {Srivastava}, \citenamefont
  {Johannes}, \citenamefont {Murphy}, \citenamefont {Park}, \citenamefont
  {Balicas}, \citenamefont {Lonzarich}, \citenamefont {Balakrishnan},\ and\
  \citenamefont {Sebastian}}]{TanSebastian2015}%
  \BibitemOpen
  \bibfield  {author} {\bibinfo {author} {\bibfnamefont {B.~S.}\ \bibnamefont
  {Tan}}, \bibinfo {author} {\bibfnamefont {Y.-T.}\ \bibnamefont {Hsu}},
  \bibinfo {author} {\bibfnamefont {B.}~\bibnamefont {Zeng}}, \bibinfo {author}
  {\bibfnamefont {M.~Ciomaga}\ \bibnamefont {Hatnean}}, \bibinfo {author}
  {\bibfnamefont {N.}~\bibnamefont {Harrison}}, \bibinfo {author}
  {\bibfnamefont {Z.}~\bibnamefont {Zhu}}, \bibinfo {author} {\bibfnamefont
  {M.}~\bibnamefont {Hartstein}}, \bibinfo {author} {\bibfnamefont
  {M.}~\bibnamefont {Kiourlappou}}, \bibinfo {author} {\bibfnamefont
  {A.}~\bibnamefont {Srivastava}}, \bibinfo {author} {\bibfnamefont {M.~D.}\
  \bibnamefont {Johannes}}, \bibinfo {author} {\bibfnamefont {T.~P.}\
  \bibnamefont {Murphy}}, \bibinfo {author} {\bibfnamefont {J.-H.}\
  \bibnamefont {Park}}, \bibinfo {author} {\bibfnamefont {L.}~\bibnamefont
  {Balicas}}, \bibinfo {author} {\bibfnamefont {G.~G.}\ \bibnamefont
  {Lonzarich}}, \bibinfo {author} {\bibfnamefont {G.}~\bibnamefont
  {Balakrishnan}}, \ and\ \bibinfo {author} {\bibfnamefont {Suchitra~E.}\
  \bibnamefont {Sebastian}},\ }\bibfield  {title} {\enquote {\bibinfo {title}
  {Unconventional {F}ermi surface in an insulating state},}\ }\href {\doibase
  10.1126/science.aaa7974} {\bibfield  {journal} {\bibinfo  {journal}
  {Science}\ }\textbf {\bibinfo {volume} {349}},\ \bibinfo {pages} {287--290}
  (\bibinfo {year} {2015})}\BibitemShut {NoStop}%
\bibitem [{\citenamefont {Czajka}\ \emph {et~al.}(2021)\citenamefont {Czajka},
  \citenamefont {Gao}, \citenamefont {Hirschberger}, \citenamefont
  {Lampen-Kelley}, \citenamefont {Banerjee}, \citenamefont {Yan}, \citenamefont
  {Mandrus}, \citenamefont {Nagler},\ and\ \citenamefont
  {Ong}}]{CzajkaOng2021}%
  \BibitemOpen
  \bibfield  {author} {\bibinfo {author} {\bibfnamefont {Peter}\ \bibnamefont
  {Czajka}}, \bibinfo {author} {\bibfnamefont {Tong}\ \bibnamefont {Gao}},
  \bibinfo {author} {\bibfnamefont {Max}\ \bibnamefont {Hirschberger}},
  \bibinfo {author} {\bibfnamefont {Paula}\ \bibnamefont {Lampen-Kelley}},
  \bibinfo {author} {\bibfnamefont {Arnab}\ \bibnamefont {Banerjee}}, \bibinfo
  {author} {\bibfnamefont {Jiaqiang}\ \bibnamefont {Yan}}, \bibinfo {author}
  {\bibfnamefont {David~G}\ \bibnamefont {Mandrus}}, \bibinfo {author}
  {\bibfnamefont {Stephen~E}\ \bibnamefont {Nagler}}, \ and\ \bibinfo {author}
  {\bibfnamefont {NP}~\bibnamefont {Ong}},\ }\bibfield  {title} {\enquote
  {\bibinfo {title} {Oscillations of the thermal conductivity in the
  spin-liquid state of $\alpha$-{R}u{C}l$_3$},}\ }\href@noop {} {\bibfield
  {journal} {\bibinfo  {journal} {Nature Physics}\ }\textbf {\bibinfo {volume}
  {17}},\ \bibinfo {pages} {915--919} (\bibinfo {year} {2021})}\BibitemShut
  {NoStop}%
\bibitem [{\citenamefont {Dzyaloshinskii}(2003)}]{Dzyaloshinskii2003}%
  \BibitemOpen
  \bibfield  {author} {\bibinfo {author} {\bibfnamefont {Igor}\ \bibnamefont
  {Dzyaloshinskii}},\ }\bibfield  {title} {\enquote {\bibinfo {title} {Some
  consequences of the {L}uttinger theorem: The {L}uttinger surfaces in
  non-{F}ermi liquids and {M}ott insulators},}\ }\href {\doibase
  10.1103/PhysRevB.68.085113} {\bibfield  {journal} {\bibinfo  {journal} {Phys.
  Rev. B}\ }\textbf {\bibinfo {volume} {68}},\ \bibinfo {pages} {085113}
  (\bibinfo {year} {2003})}\BibitemShut {NoStop}%
\bibitem [{\citenamefont {Fabrizio}(2023)}]{Fabrizio2023}%
  \BibitemOpen
  \bibfield  {author} {\bibinfo {author} {\bibfnamefont {Michele}\ \bibnamefont
  {Fabrizio}},\ }\bibfield  {title} {\enquote {\bibinfo {title} {Spin-liquid
  insulators can be {L}andau's {F}ermi liquids},}\ }\href {\doibase
  10.1103/PhysRevLett.130.156702} {\bibfield  {journal} {\bibinfo  {journal}
  {Phys. Rev. Lett.}\ }\textbf {\bibinfo {volume} {130}},\ \bibinfo {pages}
  {156702} (\bibinfo {year} {2023})}\BibitemShut {NoStop}%
\bibitem [{\citenamefont {Wagner}\ \emph {et~al.}(2023)\citenamefont {Wagner},
  \citenamefont {Crippa}, \citenamefont {Amaricci}, \citenamefont {Hansmann},
  \citenamefont {Klett}, \citenamefont {K{\"o}nig}, \citenamefont
  {Sch{\"a}fer}, \citenamefont {Di~Sante}, \citenamefont {Cano}, \citenamefont
  {Millis} \emph {et~al.}}]{WagnerSangiovanni2023}%
  \BibitemOpen
  \bibfield  {author} {\bibinfo {author} {\bibfnamefont {Niklas}\ \bibnamefont
  {Wagner}}, \bibinfo {author} {\bibfnamefont {Lorenzo}\ \bibnamefont
  {Crippa}}, \bibinfo {author} {\bibfnamefont {Adriano}\ \bibnamefont
  {Amaricci}}, \bibinfo {author} {\bibfnamefont {Philipp}\ \bibnamefont
  {Hansmann}}, \bibinfo {author} {\bibfnamefont {Marcel}\ \bibnamefont
  {Klett}}, \bibinfo {author} {\bibfnamefont {Elio}\ \bibnamefont {K{\"o}nig}},
  \bibinfo {author} {\bibfnamefont {Thomas}\ \bibnamefont {Sch{\"a}fer}},
  \bibinfo {author} {\bibfnamefont {Domenico}\ \bibnamefont {Di~Sante}},
  \bibinfo {author} {\bibfnamefont {Jennifer}\ \bibnamefont {Cano}}, \bibinfo
  {author} {\bibfnamefont {Andrew}\ \bibnamefont {Millis}},  \emph {et~al.},\
  }\bibfield  {title} {\enquote {\bibinfo {title} {{M}ott insulators with
  boundary zeros},}\ }\href@noop {} {\bibfield  {journal} {\bibinfo  {journal}
  {arXiv preprint arXiv:2301.05588}\ } (\bibinfo {year} {2023})}\BibitemShut
  {NoStop}%
\bibitem [{\citenamefont {Setty}\ \emph {et~al.}(2023)\citenamefont {Setty},
  \citenamefont {Sur}, \citenamefont {Chen}, \citenamefont {Xie}, \citenamefont
  {Hu}, \citenamefont {Paschen}, \citenamefont {Cano},\ and\ \citenamefont
  {Si}}]{SettySi2023}%
  \BibitemOpen
  \bibfield  {author} {\bibinfo {author} {\bibfnamefont {Chandan}\ \bibnamefont
  {Setty}}, \bibinfo {author} {\bibfnamefont {Shouvik}\ \bibnamefont {Sur}},
  \bibinfo {author} {\bibfnamefont {Lei}\ \bibnamefont {Chen}}, \bibinfo
  {author} {\bibfnamefont {Fang}\ \bibnamefont {Xie}}, \bibinfo {author}
  {\bibfnamefont {Haoyu}\ \bibnamefont {Hu}}, \bibinfo {author} {\bibfnamefont
  {Silke}\ \bibnamefont {Paschen}}, \bibinfo {author} {\bibfnamefont
  {Jennifer}\ \bibnamefont {Cano}}, \ and\ \bibinfo {author} {\bibfnamefont
  {Qimiao}\ \bibnamefont {Si}},\ }\bibfield  {title} {\enquote {\bibinfo
  {title} {Symmetry constraints and spectral crossing in a {M}ott insulator
  with {G}reen's function zeros},}\ }\href@noop {} {\bibfield  {journal}
  {\bibinfo  {journal} {arXiv preprint arXiv:2301.13870}\ } (\bibinfo {year}
  {2023})}\BibitemShut {NoStop}%
\bibitem [{\citenamefont {Blason}\ and\ \citenamefont
  {Fabrizio}(2023)}]{BlasonFabrizio2023}%
  \BibitemOpen
  \bibfield  {author} {\bibinfo {author} {\bibfnamefont {Andrea}\ \bibnamefont
  {Blason}}\ and\ \bibinfo {author} {\bibfnamefont {Michele}\ \bibnamefont
  {Fabrizio}},\ }\bibfield  {title} {\enquote {\bibinfo {title} {Unified role
  of {G}reen's function poles and zeros in topological insulators},}\
  }\href@noop {} {\bibfield  {journal} {\bibinfo  {journal} {arXiv preprint
  arXiv:2304.08180}\ } (\bibinfo {year} {2023})}\BibitemShut {NoStop}%
\bibitem [{\citenamefont {Wen}(2003)}]{Wen2003}%
  \BibitemOpen
  \bibfield  {author} {\bibinfo {author} {\bibfnamefont {Xiao-Gang}\
  \bibnamefont {Wen}},\ }\bibfield  {title} {\enquote {\bibinfo {title}
  {Quantum orders in an exact soluble model},}\ }\href {\doibase
  10.1103/PhysRevLett.90.016803} {\bibfield  {journal} {\bibinfo  {journal}
  {Phys. Rev. Lett.}\ }\textbf {\bibinfo {volume} {90}},\ \bibinfo {pages}
  {016803} (\bibinfo {year} {2003})}\BibitemShut {NoStop}%
\bibitem [{\citenamefont {Bullock}\ and\ \citenamefont
  {Brennen}(2007)}]{BullockBrennen2007}%
  \BibitemOpen
  \bibfield  {author} {\bibinfo {author} {\bibfnamefont {Stephen~S}\
  \bibnamefont {Bullock}}\ and\ \bibinfo {author} {\bibfnamefont {Gavin~K}\
  \bibnamefont {Brennen}},\ }\bibfield  {title} {\enquote {\bibinfo {title}
  {Qudit surface codes and gauge theory with finite cyclic groups},}\
  }\href@noop {} {\bibfield  {journal} {\bibinfo  {journal} {Journal of Physics
  A: Mathematical and Theoretical}\ }\textbf {\bibinfo {volume} {40}},\
  \bibinfo {pages} {3481} (\bibinfo {year} {2007})}\BibitemShut {NoStop}%
\bibitem [{\citenamefont {Schulz}\ \emph {et~al.}(2012)\citenamefont {Schulz},
  \citenamefont {Dusuel}, \citenamefont {Orus}, \citenamefont {Vidal},\ and\
  \citenamefont {Schmidt}}]{SchulzSchmidt2012}%
  \BibitemOpen
  \bibfield  {author} {\bibinfo {author} {\bibfnamefont {Marc~Daniel}\
  \bibnamefont {Schulz}}, \bibinfo {author} {\bibfnamefont {S{\'e}bastien}\
  \bibnamefont {Dusuel}}, \bibinfo {author} {\bibfnamefont {Roman}\
  \bibnamefont {Orus}}, \bibinfo {author} {\bibfnamefont {Julien}\ \bibnamefont
  {Vidal}}, \ and\ \bibinfo {author} {\bibfnamefont {Kai~Phillip}\ \bibnamefont
  {Schmidt}},\ }\bibfield  {title} {\enquote {\bibinfo {title} {Breakdown of a
  perturbed topological phase},}\ }\href@noop {} {\bibfield  {journal}
  {\bibinfo  {journal} {New Journal of Physics}\ }\textbf {\bibinfo {volume}
  {14}},\ \bibinfo {pages} {025005} (\bibinfo {year} {2012})}\BibitemShut
  {NoStop}%
\bibitem [{\citenamefont {Zou}\ and\ \citenamefont {Haah}(2016)}]{ZouHaah2016}%
  \BibitemOpen
  \bibfield  {author} {\bibinfo {author} {\bibfnamefont {Liujun}\ \bibnamefont
  {Zou}}\ and\ \bibinfo {author} {\bibfnamefont {Jeongwan}\ \bibnamefont
  {Haah}},\ }\bibfield  {title} {\enquote {\bibinfo {title} {Spurious
  long-range entanglement and replica correlation length},}\ }\href {\doibase
  10.1103/PhysRevB.94.075151} {\bibfield  {journal} {\bibinfo  {journal} {Phys.
  Rev. B}\ }\textbf {\bibinfo {volume} {94}},\ \bibinfo {pages} {075151}
  (\bibinfo {year} {2016})}\BibitemShut {NoStop}%
\bibitem [{\citenamefont {Pachos}(2012)}]{PachosBook}%
  \BibitemOpen
  \bibfield  {author} {\bibinfo {author} {\bibfnamefont {Jiannis~K}\
  \bibnamefont {Pachos}},\ }\href@noop {} {\emph {\bibinfo {title}
  {Introduction to topological quantum computation}}}\ (\bibinfo  {publisher}
  {Cambridge University Press},\ \bibinfo {year} {2012})\BibitemShut {NoStop}%
\bibitem [{\citenamefont {Watanabe}\ \emph {et~al.}(2023)\citenamefont
  {Watanabe}, \citenamefont {Cheng},\ and\ \citenamefont
  {Fuji}}]{WatanabeFuji2023}%
  \BibitemOpen
  \bibfield  {author} {\bibinfo {author} {\bibfnamefont {Haruki}\ \bibnamefont
  {Watanabe}}, \bibinfo {author} {\bibfnamefont {Meng}\ \bibnamefont {Cheng}},
  \ and\ \bibinfo {author} {\bibfnamefont {Yohei}\ \bibnamefont {Fuji}},\
  }\bibfield  {title} {\enquote {\bibinfo {title} {{Ground state degeneracy on
  torus in a family of ${Z}_{N}$ toric code}},}\ }\href {\doibase
  10.1063/5.0134010} {\bibfield  {journal} {\bibinfo  {journal} {Journal of
  Mathematical Physics}\ }\textbf {\bibinfo {volume} {64}},\ \bibinfo {pages}
  {051901} (\bibinfo {year} {2023})}\BibitemShut {NoStop}%
\bibitem [{\citenamefont {Fradkin}\ and\ \citenamefont
  {Shenker}(1979)}]{FradkinShenker1979}%
  \BibitemOpen
  \bibfield  {author} {\bibinfo {author} {\bibfnamefont {Eduardo}\ \bibnamefont
  {Fradkin}}\ and\ \bibinfo {author} {\bibfnamefont {Stephen~H.}\ \bibnamefont
  {Shenker}},\ }\bibfield  {title} {\enquote {\bibinfo {title} {Phase diagrams
  of lattice gauge theories with {H}iggs fields},}\ }\href {\doibase
  10.1103/PhysRevD.19.3682} {\bibfield  {journal} {\bibinfo  {journal} {Phys.
  Rev. D}\ }\textbf {\bibinfo {volume} {19}},\ \bibinfo {pages} {3682--3697}
  (\bibinfo {year} {1979})}\BibitemShut {NoStop}%
\bibitem [{\citenamefont {Tupitsyn}\ \emph {et~al.}(2010)\citenamefont
  {Tupitsyn}, \citenamefont {Kitaev}, \citenamefont {Prokof'ev},\ and\
  \citenamefont {Stamp}}]{TupitsynStamp2010}%
  \BibitemOpen
  \bibfield  {author} {\bibinfo {author} {\bibfnamefont {I.~S.}\ \bibnamefont
  {Tupitsyn}}, \bibinfo {author} {\bibfnamefont {A.}~\bibnamefont {Kitaev}},
  \bibinfo {author} {\bibfnamefont {N.~V.}\ \bibnamefont {Prokof'ev}}, \ and\
  \bibinfo {author} {\bibfnamefont {P.~C.~E.}\ \bibnamefont {Stamp}},\
  }\bibfield  {title} {\enquote {\bibinfo {title} {Topological multicritical
  point in the phase diagram of the toric code model and three-dimensional
  lattice gauge {H}iggs model},}\ }\href {\doibase 10.1103/PhysRevB.82.085114}
  {\bibfield  {journal} {\bibinfo  {journal} {Phys. Rev. B}\ }\textbf {\bibinfo
  {volume} {82}},\ \bibinfo {pages} {085114} (\bibinfo {year}
  {2010})}\BibitemShut {NoStop}%
\bibitem [{\citenamefont {Hasegawa}\ \emph {et~al.}(1989)\citenamefont
  {Hasegawa}, \citenamefont {Lederer}, \citenamefont {Rice},\ and\
  \citenamefont {Wiegmann}}]{HasegawaWiegmann1989}%
  \BibitemOpen
  \bibfield  {author} {\bibinfo {author} {\bibfnamefont {Y.}~\bibnamefont
  {Hasegawa}}, \bibinfo {author} {\bibfnamefont {P.}~\bibnamefont {Lederer}},
  \bibinfo {author} {\bibfnamefont {T.~M.}\ \bibnamefont {Rice}}, \ and\
  \bibinfo {author} {\bibfnamefont {P.~B.}\ \bibnamefont {Wiegmann}},\
  }\bibfield  {title} {\enquote {\bibinfo {title} {Theory of electronic
  diamagnetism in two-dimensional lattices},}\ }\href {\doibase
  10.1103/PhysRevLett.63.907} {\bibfield  {journal} {\bibinfo  {journal} {Phys.
  Rev. Lett.}\ }\textbf {\bibinfo {volume} {63}},\ \bibinfo {pages} {907--910}
  (\bibinfo {year} {1989})}\BibitemShut {NoStop}%
\bibitem [{\citenamefont {Prosko}\ \emph {et~al.}(2017)\citenamefont {Prosko},
  \citenamefont {Lee},\ and\ \citenamefont {Maciejko}}]{ProskoMaciejko2017}%
  \BibitemOpen
  \bibfield  {author} {\bibinfo {author} {\bibfnamefont {Christian}\
  \bibnamefont {Prosko}}, \bibinfo {author} {\bibfnamefont {Shu-Ping}\
  \bibnamefont {Lee}}, \ and\ \bibinfo {author} {\bibfnamefont {Joseph}\
  \bibnamefont {Maciejko}},\ }\bibfield  {title} {\enquote {\bibinfo {title}
  {Simple $\mathbb{Z}_{2}$ lattice gauge theories at finite fermion density},}\
  }\href {\doibase 10.1103/PhysRevB.96.205104} {\bibfield  {journal} {\bibinfo
  {journal} {Phys. Rev. B}\ }\textbf {\bibinfo {volume} {96}},\ \bibinfo
  {pages} {205104} (\bibinfo {year} {2017})}\BibitemShut {NoStop}%
\bibitem [{\citenamefont {Hazra}\ and\ \citenamefont
  {Coleman}(2021)}]{HazraColeman2021}%
  \BibitemOpen
  \bibfield  {author} {\bibinfo {author} {\bibfnamefont {Tamaghna}\
  \bibnamefont {Hazra}}\ and\ \bibinfo {author} {\bibfnamefont {Piers}\
  \bibnamefont {Coleman}},\ }\bibfield  {title} {\enquote {\bibinfo {title}
  {Luttinger sum rules and spin fractionalization in the {SU($N$)} kondo
  lattice},}\ }\href {\doibase 10.1103/PhysRevResearch.3.033284} {\bibfield
  {journal} {\bibinfo  {journal} {Phys. Rev. Res.}\ }\textbf {\bibinfo {volume}
  {3}},\ \bibinfo {pages} {033284} (\bibinfo {year} {2021})}\BibitemShut
  {NoStop}%
\bibitem [{\citenamefont {Knolle}\ \emph {et~al.}(2014)\citenamefont {Knolle},
  \citenamefont {Kovrizhin}, \citenamefont {Chalker},\ and\ \citenamefont
  {Moessner}}]{KnolleMoessner2014}%
  \BibitemOpen
  \bibfield  {author} {\bibinfo {author} {\bibfnamefont {J.}~\bibnamefont
  {Knolle}}, \bibinfo {author} {\bibfnamefont {D.~L.}\ \bibnamefont
  {Kovrizhin}}, \bibinfo {author} {\bibfnamefont {J.~T.}\ \bibnamefont
  {Chalker}}, \ and\ \bibinfo {author} {\bibfnamefont {R.}~\bibnamefont
  {Moessner}},\ }\bibfield  {title} {\enquote {\bibinfo {title} {Dynamics of a
  two-dimensional quantum spin liquid: Signatures of emergent {M}ajorana
  fermions and fluxes},}\ }\href {\doibase 10.1103/PhysRevLett.112.207203}
  {\bibfield  {journal} {\bibinfo  {journal} {Phys. Rev. Lett.}\ }\textbf
  {\bibinfo {volume} {112}},\ \bibinfo {pages} {207203} (\bibinfo {year}
  {2014})}\BibitemShut {NoStop}%
\bibitem [{\citenamefont {K\"onig}\ \emph {et~al.}(2021)\citenamefont
  {K\"onig}, \citenamefont {Coleman},\ and\ \citenamefont
  {Komijani}}]{KoenigKomijani2021}%
  \BibitemOpen
  \bibfield  {author} {\bibinfo {author} {\bibfnamefont {Elio~J.}\ \bibnamefont
  {K\"onig}}, \bibinfo {author} {\bibfnamefont {Piers}\ \bibnamefont
  {Coleman}}, \ and\ \bibinfo {author} {\bibfnamefont {Yashar}\ \bibnamefont
  {Komijani}},\ }\bibfield  {title} {\enquote {\bibinfo {title} {Frustrated
  {K}ondo impurity triangle: A simple model of deconfinement},}\ }\href
  {\doibase 10.1103/PhysRevB.104.115103} {\bibfield  {journal} {\bibinfo
  {journal} {Phys. Rev. B}\ }\textbf {\bibinfo {volume} {104}},\ \bibinfo
  {pages} {115103} (\bibinfo {year} {2021})}\BibitemShut {NoStop}%
\bibitem [{\citenamefont {Anderson}(1967)}]{Anderson1967}%
  \BibitemOpen
  \bibfield  {author} {\bibinfo {author} {\bibfnamefont {P.~W.}\ \bibnamefont
  {Anderson}},\ }\bibfield  {title} {\enquote {\bibinfo {title} {Infrared
  catastrophe in {F}ermi gases with local scattering potentials},}\ }\href
  {\doibase 10.1103/PhysRevLett.18.1049} {\bibfield  {journal} {\bibinfo
  {journal} {Phys. Rev. Lett.}\ }\textbf {\bibinfo {volume} {18}},\ \bibinfo
  {pages} {1049--1051} (\bibinfo {year} {1967})}\BibitemShut {NoStop}%
\bibitem [{\citenamefont {Gogolin}\ \emph {et~al.}(2004)\citenamefont
  {Gogolin}, \citenamefont {Nersesian},\ and\ \citenamefont
  {Tsvelik}}]{GogolinTsvelikBook}%
  \BibitemOpen
  \bibfield  {author} {\bibinfo {author} {\bibfnamefont {A.~O.}\ \bibnamefont
  {Gogolin}}, \bibinfo {author} {\bibfnamefont {A.~A.}\ \bibnamefont
  {Nersesian}}, \ and\ \bibinfo {author} {\bibfnamefont {A.~M.}\ \bibnamefont
  {Tsvelik}},\ }\href@noop {} {\emph {\bibinfo {title} {{Bosonization and
  strongly correlated systems}}}}\ (\bibinfo {year} {2004})\BibitemShut
  {NoStop}%
\bibitem [{\citenamefont {Nozieres}\ and\ \citenamefont
  {De~Dominicis}(1969)}]{NozieresdeDominicis1969}%
  \BibitemOpen
  \bibfield  {author} {\bibinfo {author} {\bibfnamefont {P.}~\bibnamefont
  {Nozieres}}\ and\ \bibinfo {author} {\bibfnamefont {C.~T.}\ \bibnamefont
  {De~Dominicis}},\ }\bibfield  {title} {\enquote {\bibinfo {title}
  {Singularities in the x-ray absorption and emission of metals. iii. one-body
  theory exact solution},}\ }\href {\doibase 10.1103/PhysRev.178.1097}
  {\bibfield  {journal} {\bibinfo  {journal} {Phys. Rev.}\ }\textbf {\bibinfo
  {volume} {178}},\ \bibinfo {pages} {1097--1107} (\bibinfo {year}
  {1969})}\BibitemShut {NoStop}%
\bibitem [{\citenamefont {{Gogolin}}(1993)}]{Gogolin1993}%
  \BibitemOpen
  \bibfield  {author} {\bibinfo {author} {\bibfnamefont {A.~O.}\ \bibnamefont
  {{Gogolin}}},\ }\bibfield  {title} {\enquote {\bibinfo {title} {{Effect of
  the van {H}ove singularities on the x-ray absorption and emission in
  metal}},}\ }\href@noop {} {\bibfield  {journal} {\bibinfo  {journal} {ZhETF
  Pisma Redaktsiiu}\ }\textbf {\bibinfo {volume} {57}},\ \bibinfo {pages} {300}
  (\bibinfo {year} {1993})}\BibitemShut {NoStop}%
\bibitem [{\citenamefont {Polyakov}(1987)}]{Polyakov1987}%
  \BibitemOpen
  \bibfield  {author} {\bibinfo {author} {\bibfnamefont
  {Aleksandr~Michajlovi{\v{c}}}\ \bibnamefont {Polyakov}},\ }\href@noop {}
  {\emph {\bibinfo {title} {Gauge fields and strings}}}\ (\bibinfo  {publisher}
  {Taylor \& Francis},\ \bibinfo {year} {1987})\BibitemShut {NoStop}%
\end{thebibliography}%

\end{document}